\documentclass[twocolumn,aps,noshowpacs]{revtex4upd}

\usepackage{graphicx}
\usepackage{dcolumn}
\usepackage{multirow}
\usepackage{bm}
\usepackage{times}
\usepackage{color}
\usepackage{amstext}
\usepackage{amsmath}
\usepackage{amsfonts}
\usepackage[bookmarks=true,colorlinks=true,pdfpagemode=UseNone,pdfstartview=FitH,linkcolor=myrefblue,urlcolor=myrefblue,citecolor=myrefblue]{hyperref}
\definecolor{lightblue}{rgb}{0.6,0.9,1}
\definecolor{myrefblue}{rgb}{0.1,0.6,1}
\definecolor{myblue}{rgb}{0,0,0}
\definecolor{nmat}{rgb}{0.7,0.04,0.26}

\newcommand{\lfaof}{\mbox{LaFeAsO$_{1-x}$F$_{x}$}}
\newcommand{\nfaof}{\mbox{NdFeAsO$_{1-x}$F$_{x}$}}

\newcommand{\cfaof}{\mbox{CeFeAsO$_{1-x}$F$_{x}$}}
\newcommand{\thcrsi}{\mbox{ThCr$_2$Si$_2$}}
\newcommand{\bfa}{\mbox{BaFe$_2$As$_2$}}
\newcommand{\cfa}{\mbox{CaFe$_2$As$_2$}}
\newcommand{\sfa}{\mbox{SrFe$_2$As$_2$}}
\newcommand{\efa}{\mbox{EuFe$_2$As$_2$}}
\newcommand{\bfp}{\mbox{BaFe$_2$P$_2$}}

\newcommand{\lfa}{\mbox{LiFeAs}}
\newcommand{\bkfa}{\mbox{Ba$_{1-x}$K$_{x}$Fe$_2$As$_2$}}
\newcommand{\bkfaopt}{\mbox{Ba$_{0.6}$K$_{0.4}$Fe$_2$As$_2$}}
\newcommand{\bkfaoptmpi}{\mbox{Ba$_{0.68}$K$_{0.32}$Fe$_2$As$_2$}}

\newcommand{\bfra}{\mbox{Ba(Fe$_{1-x}$Ru$_x$)$_2$As$_2$}}
\newcommand{\bfap}{\mbox{BaFe$_2$(As$_{1-x}$P$_x$)$_2$}}
\newcommand{\bfca}{\mbox{Ba(Fe$_{1-x}$Co$_x$)$_2$As$_2$}}
\newcommand{\cfca}{\mbox{Ca(Fe$_{1-x}$Co$_x$)$_2$As$_2$}}

\newcommand{\bfna}{\mbox{Ba(Fe$_{1-x}$Ni$_x$)$_2$As$_2$}}
\newcommand{\bfcua}{\mbox{Ba(Fe$_{1-x}$Cu$_x$)$_2$As$_2$}}
\newcommand{\rfs}{\mbox{Rb$_x$Fe$_{2-y}$Se$_2$}}
\newcommand{\kfs}{\mbox{K$_x$Fe$_{2-y}$Se$_2$}}

\newcommand{\afs}{\mbox{$A_x$Fe$_{2-y}$Se$_2$} ($A=$K, Rb, Cs)}
\newcommand{\afsgen}{\mbox{$A_x$Fe$_{2-y}$Se$_2$}}
\newcommand{\fetese}{\mbox{FeTe$_{1-x}$Se$_x$}}

\begin{document}

\renewcommand{\arraystretch}{1.5}

\title{Optical conductivity of iron-based superconductors}

\author{Aliaksei Charnukha}
\affiliation{Leibniz Institute for Solid State and Materials Research, IFW, D-01069 Dresden, Germany}

\begin{abstract}
The new family of unconventional iron-based superconductors discovered in 2006 immediately relieved their copper-based high-temperature predecessors as the most actively studied superconducting compounds in the world. The experimental and theoretical effort made in order to unravel the mechanism of superconductivity in these materials has been overwhelming. Although our understanding of their microscopic properties has been improving steadily, the pairing mechanism giving rise to superconducting transition temperatures up to $55\ \textrm{K}$ remains elusive. And yet the hope is strong that these materials, which possess a drastically different electronic structure but similarly high transition temperatures compared to the copper-based compounds, will shed essential new light onto the several-decade-old problem of unconventional superconductivity. In this work we review the current understanding of the itinerant-charge-carrier dynamics in the iron-based superconductors and parent compounds largely based on the optical-conductivity data the community has gleaned over the past seven years using such experimental techniques as reflectivity, ellipsometry, and terahertz transmission measurements and analyze the implications of these studies for the microscopic properties of the iron-based materials as well as the mechanism of superconductivity therein.
\end{abstract}

\maketitle

\tableofcontents

\section{Introduction}

In year 2006 the group of Hideo Hosono discovered that the quaternary compound LaFePO becomes superconducting at about $4 \textrm{K}$~\cite{kamihara_original_LFPO}. At the time, this observation generated little interest in the condensed-matter community due to the relatively low superconducting transition temperature of this compound, comparable to that of many conventional superconductors~\cite{RevModPhys.62.1027}, its very different crystallographic structure notwithstanding. Nevertheless, Hosono and collaborators kept faith in this material and continued searching for ways to enhance its superconducting transition temperature. This search bore fruit when they discovered an order-of-magnitude higher transition temperature in a closely related LaFeAsO doped with fluorine~\cite{kamihara_original_LFPO}, which immediately propelled the iron-based superconductors into the elite group of the very few with superconducting transition temperatures above $25\ \textrm{K}$, alongside with high-temperature cuprate superconductors~\cite{RevModPhys.77.721,RevModPhys.62.1027,RevModPhys.78.17}, $\textrm{Ba}_{1-x}\textrm{K}_x\textrm{BiO}_3$~\cite{Cava_Battlog_BKBO_discovery_1988}, select fullerenes~\cite{RevModPhys.69.575}, and $\textrm{MgB}_2$~\cite{0953-2048-14-11-201,Canfield_MgB2_2003}. This discovery led to explosive development of the field, even more so when only several months later a twice higher superconducting transition temperature was reported in a samarium-based sister material~\cite{Ren_discovery_Sm1111}.

\begin{figure}[t!]
\includegraphics[width=\columnwidth]{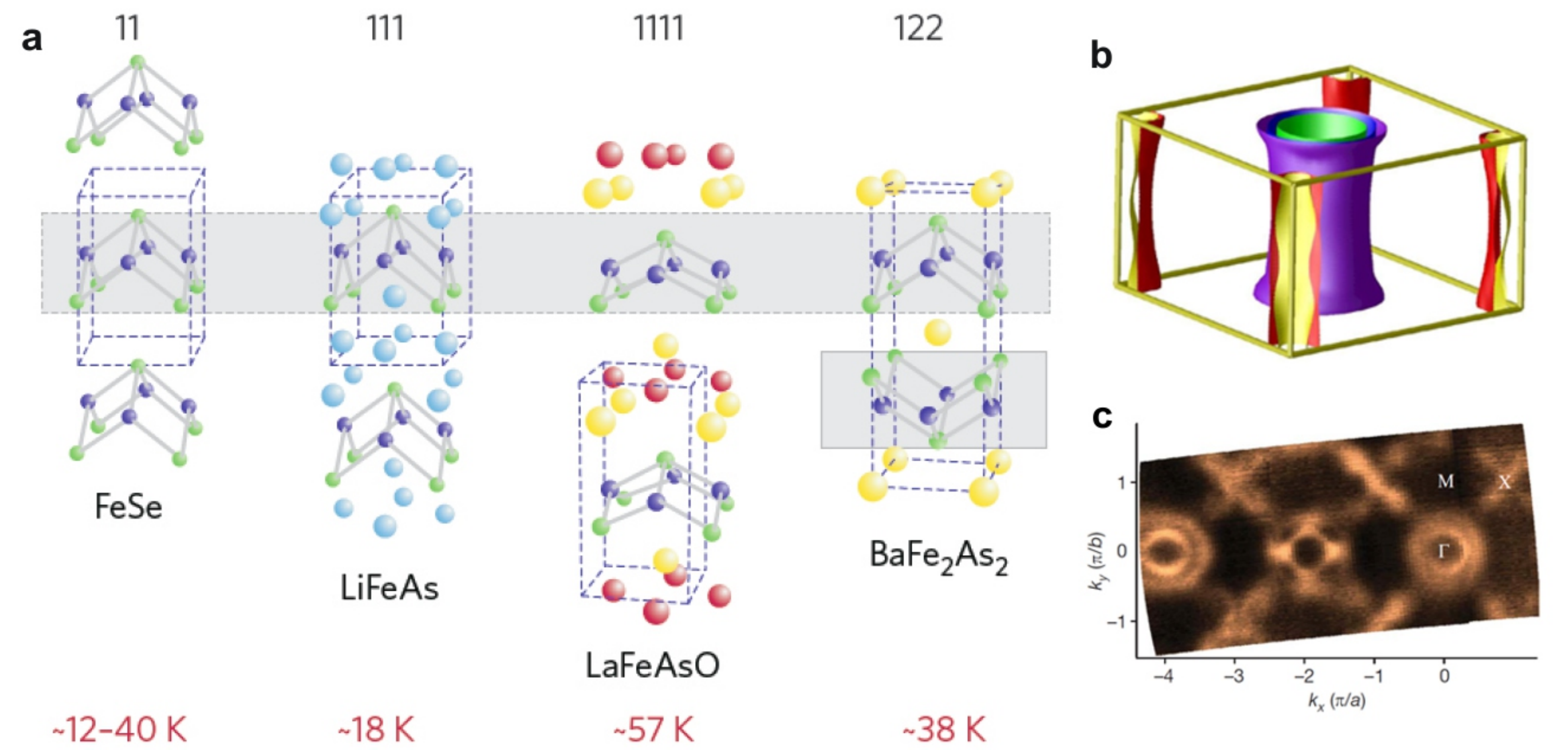}
\caption{\label{fig:intro}~\textbf{a},~Several representative classes of the iron-based materials. The numbers at the top denote the chemical formula of the compound underneath (for instance, LiFeAs is a 111 compound), while the temperatures at the bottom specify either the superconducting transition temperature of the material itself or the highest $T_{\mathrm{c}}$ obtained by doping or substitution in a material of that type. The common structural unit of all iron-based compounds is the Fe-As tetrahedral layer (gray areas), with the adjacent layers either co-aligned out-of-plane or alternating in orientation. Adapted by permission from Macmillan Publishers Ltd: Nature Physics Ref.~\onlinecite{Chu_Pnictides_Classes_2009}, copyright (2009).~\textbf{b},~Fermi surface of \bkfaopt\ predicted by {\it ab initio} calculations. Reprinted from Ref.~\onlinecite{2009PhyC469614M}, copyright (2009), with permission from Elsevier.~\textbf{c},~Fermi surface of \bkfaopt\ obtained by means of angle-resolved photoemission spectroscopy. Reprinted by permission from Macmillan Publishers Ltd: Nature Ref.~\onlinecite{Borisenko_BKFA_FS_NormState2009}, copyright (2009).}
\end{figure}

The crystallographic structure of these iron-based compounds was found to belong to the tetragonal $P4/mmm$ space group symmetry at ambient conditions and feature conductive layers of tetrahedrally coordinated Fe and As ions intercalated with rare-earth and oxygen ions, as shown in Fig.~\ref{fig:intro}a. The layered structure has drawn comparisons to the high-temperature copper-oxide superconductors, in which similar layers are formed by copper and oxygen ions, albeit octahedrally coordinated. It was soon realized that, unlike in the copper-based materials, the presence of oxygen in iron pnictides is inessential for the occurrence of superconductivity. Transition temperatures up to about $40\ \textrm{K}$ were reported upon intercalation of the iron-pnictogen layers with a single-type alkaline earth metal ion with subsequent alio- or isovalent substitution or application of external pressure~\cite{Johnston_Review_2010,0953-8984-22-20-203203,RevModPhys.83.1589}. Somewhat lower but nevertheless substantial superconducting transition temperatures could be attained without any intercalation at all in the binary PbO-type $\alpha$--FeSe iron chalcogenide~\cite{Hsu23092008}. It is now well-established that tetrahedral coordination of iron ions via a pnictogen or a chalcogen is essential for iron-based superconductivity~\cite{Johnston_Review_2010,RevModPhys.83.1589}. The iron-pnictogen bond angle or, similarly, the pntictogen or chalcogen ion height over the plane formed by iron ions, characterizing the tetrahedral coordination, was found to show a certain correlation with the superconducting transition temperature~\cite{Dai_TcvsPnictogenBondAngle,JPSJ.77.083704,0953-2048-23-5-054013,PhysRevB.81.205119,Johnston_Review_2010}.

In order to gain a more fundamental physical understanding of superconductivity and its dependence on the structure and doping in the iron-based materials, the knowledge of their electronic structure is indispensable. Already very early {\it ab initio} calculations have shown that the low-energy band structure of these compounds is formed predominantly by {Fe-3d} orbitals, very weakly hybridized with {As-4p} orbitals, which commonly produce a rather complex Fermi surface with up to five sheets of different electronic character~\cite{2009PhyC469614M}. These calculations further revealed that, as can be expected from their layered structure, the iron-based materials possess a quasi-two-dimensional Fermi surface (see Fig.~\ref{fig:intro}b), with seemingly very good nesting (good geometrical overlap of superimposed portions of the Fermi surface) between the holelike and electronlike Fermi-surface sheets in the center and in the corner of the Brillouin zone, respectively, which would enhance susceptibility to electronic instabilities leading to the corresponding reconstruction of the Brillouin zone. Neutron-scattering measurements have indeed identified long-range antiferromagnetic order at low temperatures in the overwhelming majority of the parent (undoped) iron-based materials at a wave vector consistent with the nesting vector predicted by theoretical calculations~\cite{Johnston_Review_2010,RevModPhys.83.1589} (with very few exceptions such as, e.g., LiFeAs superconductor possessing no magnetism or nesting whatsoever~\cite{PhysRevB.78.060505,Chu2009326,PhysRevB.79.052508,PhysRevLett.105.067002} and FeTe, in which the antiferromagnetic wave vector is very different from the nesting vector~\cite{NatPhys_Dai_Dagotto_ReviewMagnPnictides}). This very good agreement between theory and experiment, along with the fact that iron-based materials remain metallic in the magnetic state, led the community to believe that antiferromagnetism in these compounds is largely of itinerant nesting-enhanced spin-density--wave type~\cite{Johnston_Review_2010}. However, further extensive systematic neutron-scattering studies of most of the known relevant iron-based materials have revealed that their antiferromagnetism bears essentially dual, simultaneously local and itinerant, character~\cite{NatPhys_Dai_Dagotto_ReviewMagnPnictides}.

The observation that the superconducting transition temperature in the iron-based materials is maximized when the antiferromagnetism of the parent compounds has been largely or entirely suppressed~\cite{Johnston_Review_2010,0953-8984-22-20-203203} and the theoretical prediction of electron-phonon interaction too small to account for the superconducting transition temperatures found in these compounds~\cite{boeri:026403,Boeri2009628,PhysRevB.82.020506} has led to the suggestion that superconductivity is mediated by nesting-enhanced antiferromagnetic spin fluctuations. As a result, the superconducting order parameter is expected to possess an $s$-wave symmetry~\cite{Annett_symmetry_SC_1990} but exhibit a sign change between the Fermi surfaces connected by the antiferromagnetic/nesting wave vector (the so-called ``extended $s$-wave'', or $s_\pm$ symmetry) due to the repulsive character of the mediated interaction~\cite{2009PhyC469614M}. An experimental test of this hypothesis is problematic in this case, unlike the historic confirmation of the sign-changing, d-wave, character of the superconducting order parameter in the high-temperature cuprate superconductors~\cite{RevModPhys.67.515}, because the proposed extended (sign-changing) s-wave symmetry preserves the tetragonal symmetry of the underlying lattice. Nevertheless, several proposals for a phase-sensitive test of the symmetry of the superconducting order paramter have been put forward~\cite{golubov032601,2013arXiv1312.5930B} and even realized~\cite{Science328.474}.

It is important to note that, although the nesting scenario seemed to provide a plausible explanation for the observed ground state of the parent and superconducting iron-based materials at the dawn of their era~\cite{Johnston_Review_2010,RevModPhys.83.1589,NatPhys_Dai_Dagotto_ReviewMagnPnictides}, extensive angle-resolved photoemission experiments have revealed an unusual propellerlike shape of the electron sheets of the Fermi surface located in the corners of the Brillouin zone, as shown in Fig.~\ref{fig:intro}c, in a number of iron-based compounds~\cite{Borisenko_BKFA_FS_NormState2009}, with no significant nesting between the electron- and holelike sheets of the Fermi surface. These measurements have further found that the theoretically predicted electronic band structure can largely be brought into agreement with experiment if the former is renormalized (in certain materials in a band/orbital-dependent way) by a factor of about $2$--$3$ up to a certain intermediate energy~\cite{PhysRevLett.105.067002,2011JPSJ80k3707S,0953-8984-23-13-135701,PhysRevB.86.155143,PhysRevLett.110.067003,2013arXiv1307.1280M} as well as if a band/orbital-dependent Fermi-level shifts are introduced~\cite{PhysRevLett.105.067002,PhysRevLett.109.177001,PhysRevB.89.064514,PhysRevLett.110.167002} (in the \bfca\ compound a strong temperature dependence of such a shift has been identified in Ref.~\onlinecite{PhysRevLett.110.167002}). Similar shifts were found necessary to reconcile {\it ab initio} calculations with the results of quantum-oscillation measurements on many different iron-based compounds~\cite{0034-4885-74-12-124507}.

These observations appear to challenge the importance of nesting and instead emphasize the role of the orbital character of the Fermi surface for superconducting pairing~\cite{PhysRevB.89.064514}, therefore, lending some support to theories of orbital-fluctuation--mediated pairing~\cite{PhysRevLett.104.157001,0295-5075-93-5-57003,PhysRevB.83.140512,2013arXiv1307.6119O}. However, while the bandwidth renormalization can be explained by the existence of relatively high-energy electronic correlations, consistent with the observation of ubiquitous high-energy spectral weight in the spin susceptibility via inelastic neutron scattering~\cite{HighEnergySF_Dai_2012}, the band/orbital-dependent shifts have been argued to be a natural consequence of the strong interband interaction in the presence of particle-hole asymmetry in the iron-based compounds~\cite{Cappelluti2010S508}. In this picture, the experimentally observed complex topology of the Fermi surface could be obtained from the underlying well-nested {\it bare} Fermi surface (as obtained in {\it ab initio} calculations), once the aforementioned high-energy band renormalization and the band/orbital-dependent shifts have been taken into account.

Angle-resolved photoemission experiments have further uncovered the proximity of one or more bands to a band-edge singularity, a type of van Hove singularity, in most iron-based superconductors with a relatively high superconducting transition temperature~\cite{PhysRevB.88.134501}. The intensity and number of such van Hove singularities seem to correlate positively with the magnitude of the latter.

Optical studies of charge dynamics have been instrumental in the investigation of the superconducting and normal-state properties of the iron-based compounds. They have provided a number of quantities indispensable for the elucidation of the mechanism of superconductivity in these materials, such as the number and magnitude of the superconducting gaps, certain characteristics of the bosonic excitations mediating superconducting pairing, dynamic characteristics of quasiparticles (plasma frequency and scattering rate), the energy scales involved in the antiferromagnetic and superconducting phase transitions; explored possible quantum phase transitions and exotic electronic states, identified various manifestations of the inherently multiband character of iron-based materials, and provided valuable insights into the importance of electronic correlations for superconductivity.

In this work we will review the contribution of optical-conductivity investigations using such experimental techniques as reflectivity, ellipsometry, and terahertz transmission measurements to the current understanding of the physics of superconductivity and magnetism in the iron-based materials. The structure of this review is as follows. In Sec.~\ref{sec:phasediagram} we will introduce the general phase diagram of iron-based compounds and outline their commonalities and peculiarities manifested in the topology of the phase diagram as well as in the phenomena of coexistence/phase separation of the superconducting and antiferromagnetic phases, existence of a quantum critical point beneath the superconducting dome, occurrence of nematicity, and pseudogap. Section~\ref{sec:parents} will digest the optical-conductivity studies on the parent compounds of the iron-based family in the normal and antiferromagnetic state, as well as the phenomenon of nematicity. Section~\ref{sec:superconductors} presents and analyzes the superconducting properties accessible to optical measurements as well as models used to extract such information and their applicability to the case of iron-based superconductors. It then discusses the effect of correlations on the optical conductivity and their importance for superconductivity and considers the issue of quantum criticality. In every section we try to present experimental results for all classes of the iron-based materials, when available, as well as juxtapose and compare them in the given context. The only exception are the iron-selenide \afs\ compounds, in which phase separation plays an important role and which appear to have somewhat different superconducting properties as compared to the other iron-based superconductors. These iron selenides are reviewed in Sec.~\ref{sec:selenides}. Finally, in Sec.~\ref{sec:conclusions}, main experimental observations based on optical-conductivity measurements of the iron-based materials are briefly summarized and some concluding remarks are made.

\section{\label{sec:phasediagram}Phase diagram}

\begin{figure}[t!]
\includegraphics[width=\columnwidth]{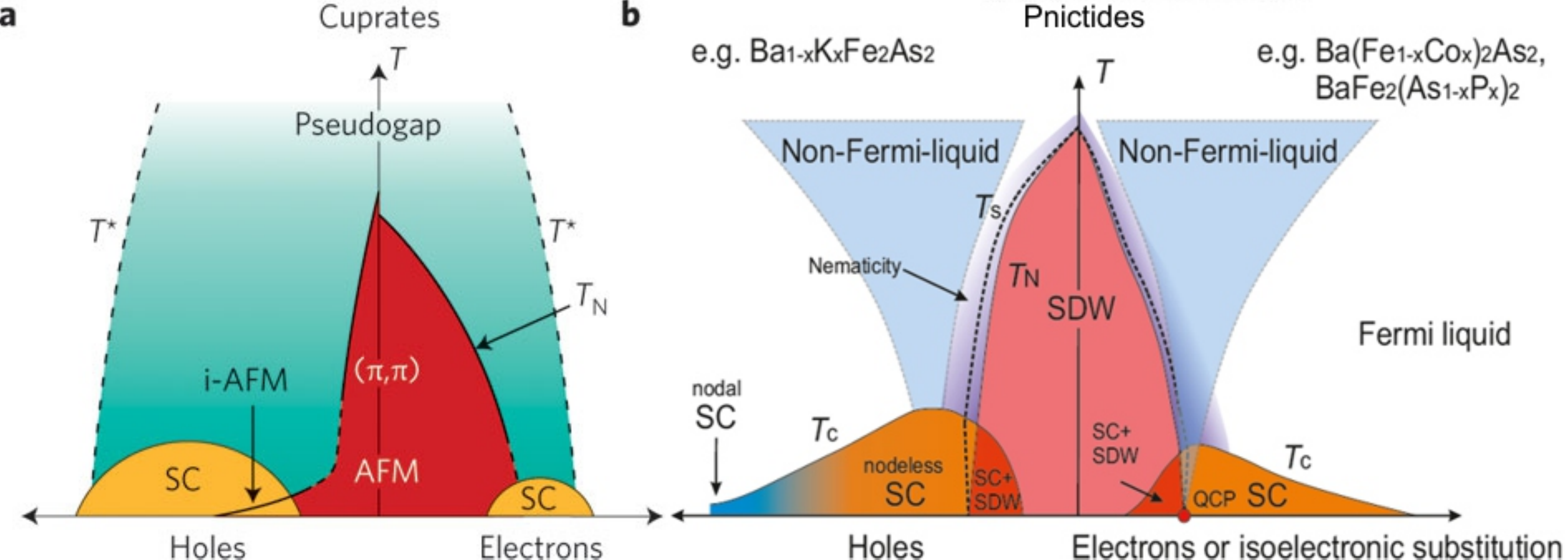}
\caption{\label{fig:phasediagcomb}{\bf a},~Phase diagram of the electron- and hole-doped cuprate superconductors. The parent compounds are Mott insulators and show local antiferromagnetism with a wave vector of $(\pi,\pi)$. The antiferromagnetic phase region (red) extends into that of superconductivity (yellow) and in the overlap region the two either coexist or compete with each other. Above the N\'eel and/or superconducting transition temperature the so-called pseudogap phase is present, which breaks the tetragonal symmetry of the high-temperature normal state reducing it to $C_2$~\cite{DavisTakagi_PseudogapSTM_2012}.~{\bf b},~Phase diagram of the electron- and hole-doped iron-based superconductors. The general features are similar to those of the cuprate superconductors: superconductivity in otherwise non-superconducting antiferromagnetic parent compounds can be induced by doping charge carriers of either sign, which results in a weakening of the itinerant antiferromagnetic phase (pink) and appearance of a dome-shaped superconducting phase region (orange) with coexistence of the superconducting and antiferromagnetic phase at certain doping levels. On the hole-doped side of the phase diagram, e.g. in the \bkfa\ series of compounds, superconductivity has been found to occur continuously up to the complete substitution of potassium for barium~\cite{Rotter_Johrendt_BKFA_phasediagram2008,PhysRevB.85.184507}, with a crossover (possibly with a phase transition)~\cite{PhysRevB.87.144511} from a nodeless superconducting order parameter in the underdoped, optimally doped and slightly overdoped regions of the phase diagram~\cite{0295-5075-83-4-47001,1367-2630-11-5-055069,PhysRevB.79.054517,PhysRevB.83.060510} to a nodal one in the extremely overdoped regime~\cite{PhysRevLett.109.087001,0953-2048-25-8-084013,Okazaki_Eisaki_Matsuda_Octet_s-wave_KFA_2012}. Nematic and non-Fermi-liquid behavior of the iron-based materials will be discussed in detail in the following sections. Panel {\bf a} adapted by permission from Macmillan Publishers Ltd: Nature Physics Ref.~\onlinecite{BasovChubukov_NatPhys2011}, copyright (2011).}
\end{figure}

The highest superconducting transition temperature observed in a bulk iron-based superconductor ($55\ \textrm{K}$ in $\textrm{SmFeAsO}_{1-x}\textrm{F}_x$~\cite{Ren_discovery_Sm1111}) is second only to that of high-temperature superconducting cuprate materials. Therefore, a juxtaposition of these two families appears instructive. As mentioned above, already at the structural level these materials have much in common: they consist of conductive layers of iron and pnictogen/chalcogen in the iron-based materials as compared to those of copper and oxygen in their cuprate counterparts. However, whereas the former remain conductive at all attainable temperatures, pressures, and doping levels~\cite{Johnston_Review_2010,RevModPhys.83.1589}, the latter are driven insulating by strong electronic correlations~\cite{RevModPhys.61.433,RevModPhys.66.763}. Superconductivity in the materials from both families can be induced by providing additional charge carriers via doping, which gives rise to very similar phase diagrams (shown schematically in Fig.~\ref{fig:phasediagcomb}). In the parent state (without doping) both types of materials experience a phase transition into an antiferromagnetic state, with a concomitant structural distortion at the same or slightly higher temperature in the case of the iron-based materials. Charge doping of either sign results in a gradual suppression of antiferromagnetism and eventually leads to the appearance of superconductivity in a dome-shaped region of the phase diagram. In the cuprates a high-temeprature suppression of the density of states near the Fermi level (called ``pseudogap'') accompanied by the spontaneous breaking of the four-fold rotational symmetry within the unit cell exists over a relatively broad doping range~\cite{0034-4885-62-1-002,Varma_Nature_PhaseDiagramCuprates_2010,DavisTakagi_PseudogapSTM_2012}. The existence of a pseudogap has also been reported for some of the iron-based compounds~\cite{boris:027001,PhysRevLett.109.027006}.

Consideration of these general trends together with the energetics of the collective ground states of the iron- and copper-based materials~\cite{BasovChubukov_NatPhys2011} suggests that several factors are of importance for high-temperature superconductivity. First of all, low dimensionality or, equivalently, strong anisotropy appears to play an important role by reducing the total kinetic energy of itinerant charge carriers through confinement and thus tilting the balance towards higher condensation energy (difference in the kinetic and potential energy above and below the phase transition) of electronic collective ground states driven by the reduction of potential energy, as is the case in, e.g., the conventional BCS theory of superconductivity~\cite{Tinkham_superconductivity_1995_articlestyle}. Secondly, sizable electronic correlations also seem to be indispensable. These conditions, however, foster a variety of collective ground states, all of which except for superconductivity are localizing, as evidenced by the complex phase diagram of the cuprate (Fig.~\ref{fig:phasediagcomb}a and Refs.~\onlinecite{DavisTakagi_PseudogapSTM_2012,Ghiringhelli17082012,Forgan_Hayden_YBCO_CDW_2012,PhysRevLett.110.137004,PhysRevLett.110.187001}) and iron-based (Fig.~\ref{fig:phasediagcomb}b and the discussion below) superconductors. Such competing states must, therefore, be eliminated by reducing the Fermi-surface nesting and/or the strength of electronic correlations driving them, in order to obtain high superconducting transition temperatures.

While the gross features of the phase diagram of the iron-based materials comply quite well with the generic shape presented in Fig.~\ref{fig:phasediagcomb}b, its detailed structure does depend significantly on the particular material class. The phase diagrams of different classes of iron-based superconductors have been extensively compared in excellent previous reviews on the topic~\cite{Johnston_Review_2010,RevModPhys.83.1589,0953-8984-22-20-203203}. Here we would like to emphasize the most recent extensions of these phase diagrams.

As can be seen in Fig.~\ref{fig:phasediagcomb}b, the antiferromagnetic and superconducting regions of the phase diagram overlap significantly at low doping levels, raising the question of how these two phases coexist with each other. At the beginning of the iron-pnictide research it was believed, for instance, that some materials of the \thcrsi\ type (or 122 type, shown in Fig.~\ref{fig:intro}a), such as \bfca, show microscopic coexistence of antiferromagnetism and superconductivity~\cite{PhysRevLett.105.057001}, whereas others, such as \bkfa, undergo intrinsic phase separation~\cite{PhysRevLett.102.117006}. Recently, experimental evidence in favor of microscopic phase coexistence in \bkfa\ as well as in \bfra\ single crystals of sufficiently high quality has emerged~\cite{PhysRevB.86.180501,PhysRevB.85.184507,PhysRevB.88.184514,PhysRevLett.109.197002}, proving that phase separation is an extrinsic effect, at least in the 122 iron pnictides.

Such coexistence of superconductivity and antiferromagnetism implies the occurrence of an antiferromagnetic quantum critical point~\cite{KeimerSachdev_QC_2011,Sachdev_QPT_2011_articlestyle,CarringtonMatsuda_review_QCP_Pnictides2014} underneath the superconducting dome. Given that microscopically coexisting phases are bound to be strongly intertwined, it may be possible to study the properties of one via the other. This approach has indeed been realized in a recent study of the London penetration depth, which uncovered that the latter is sharply peaked at the optimal doping level of superconducting \bfap~\cite{Hashimoto_Prozorov_QCPinBFAP_2012}, thus unambiguously proving the existence of a quantum critical point with respect to doping in this material~\cite{PhysRevLett.110.177003}.

Recent experiments have further demonstrated the existence of a nematic electronic state, which spontaneously breaks the four-fold rotational symmetry of the underlying lattice and induces the structural transition, in several materials of the \thcrsi\ type~\cite{Jiun-HawChu08132010,Fisher_BFCA_Nematicity_2012,Kasahara_nematicity_BFAP_2012}. Whether the nematic order parameter is driven by spin-nematic or orbital fluctuations remains unclear~\cite{PhysRevLett.111.137001,PhysRevLett.111.217001,NematicityPnictides_Fernandes_NatPhys2014,Nature_magnetism_pnictides_2010}, which suggests an intimate interplay of spin and orbital degrees of freedom in the iron-based superconductors. This interpretation will gain further support in the following sections based on the results of systematic measurements of the optical conductivity. 

\begin{figure*}[t!]
\includegraphics[width=175mm]{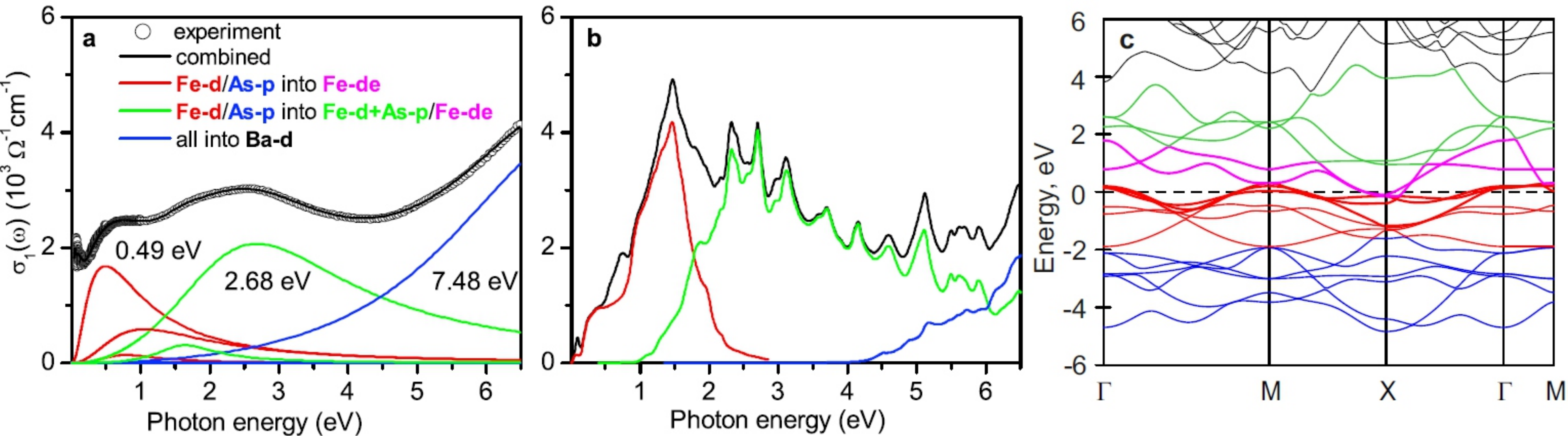}
\caption{\label{fig:bkfaoptlda}\textbf{a},~Real part of the optical conductivity of $\textrm{Ba}_{0.68}\textrm{K}_{0.32}\textrm{Fe}_2\textrm{As}_2$ and contributing interband transitions determined by a dispersion analysis. \textbf{b},~Corresponding LDA calculation with a breakdown into separate orbital contributions described in the legend~\textbf{a}. \textbf{c},~Band structure from the same LDA calculation. Color coding of the dispersion curves corresponds to the \textit{text} color in legend~\textbf{a}. Figure adapted by permission from Macmillan Publishers Ltd: Nature Communications Ref.~\onlinecite{CharnukhaNatCommun2011}, copyright (2011).}
\end{figure*}

Yet another aspect of the phase diagram of iron-based superconductors that has received considerable attention in the last several years is the continuity of the superconducting phase region in hole-doped compounds (left side of the phase diagram in Fig.~\ref{fig:phasediagcomb}b) throughout the overdoped side up to the terminal hole doping, at which the alkaline earth metal of the parent is completely substituted by an alkali metal thus producing a hole doping of 0.5 electrons per iron ion. These extremely hole-overdoped materials were found to display very different superconducting and normal-state properties compared to their underdoped and optimally doped counterparts, such as the scaling of the specific-heat jump upon entering the superconducting state with respect to the superconducting transition temperature and thermal transport strongly reminiscent of d-wave superconductors~\cite{PhysRevB.87.100509,PhysRevB.89.060504,PhysRevLett.109.087001}, as well as a very large renormalization of the effective mass~\cite{JPSJ.79.053702,PhysRevLett.111.027002,PhysRevB.87.224512}. If, as it appears to be very likely, the symmetry of the superconducting order parameter in these materials is indeed very different from the extended s-wave symmetry widely believed to be possessed by the underdoped and optimally doped 122 materials and given that superconductivity does not vanish upon overdoping~\cite{Rotter_Johrendt_BKFA_phasediagram2008,0295-5075-85-1-17006,PhysRevB.85.184507,PhysRevB.87.100509}, a phase transition between the superconducting ground states exhibiting these two different symmetries must occur at intermediate doping levels and might entail a time-reversal--symmetry breaking~\cite{0953-2048-25-8-084013,PhysRevB.87.144511}, which has a large potential for industrial applications. Yet a recent muon-spin--rotation investigation of polycrystalline \bkfa\ materials for $0.5\leq x\leq0.9$ failed to detect any time-reversal--symmetry breaking to the accuracy of the experiment~\cite{PhysRevB.89.020502}. In order to shed definitive light on this issue, detailed systematic investigation of an entire doping series on a fine doping grid in the overdoped regime is required. Unfortunately, the synthesis of single-crystalline 122 materials on the hole-overdoped side of the phase diagram has proven challenging~\cite{PhysRevB.87.100509} and will require further efforts of the community.

\section{\label{sec:parents}Antiferromagnetic parent compounds}

We now turn to the discussion of the charge dynamics of the iron-based materials across their phase diagram. Given that superconductivity in these compounds emerges upon alio-, isovalent substitution, or application of external pressure to the parents, a deep fundamental understanding of the physical properties of the latter in the normal as well as in the low-temperature antiferromagnetic state is essential in order to elucidate the microscopic mechanism of superconductivity.

\begin{figure}[b!]
\includegraphics[width=\columnwidth]{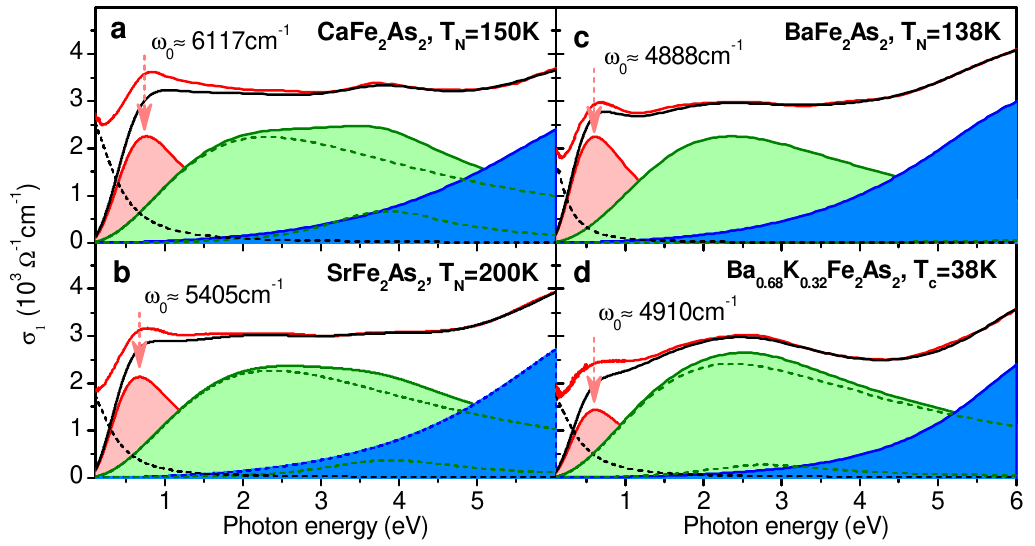}
\caption{\label{fig:comp122interband}~\textbf{a}--\textbf{d}, Real part of the optical conductivity of \cfa, \sfa, \bfa, and \bkfaoptmpi\, respectively, at $300\ \textrm{K}$ (red line) and the corresponding interband (red, green, and blue areas) and itinerant (black dashed lines) contributions obtained in a Drude-Lorentz dispersion analysis as described in the text. The lowest clearly identifiable absorption band is indicated with pink arrows. Major partial Lorentz contributions to the absorption band in the visible (green area) are plotted as green dashed lines. The total contribution of all interband transitions (black line). Panels~\textbf{a}--\textbf{c} and~\textbf{d} adapted with permission from Ref.~\onlinecite{PhysRevB.88.184511} and Ref.~\onlinecite{PhysRevB.84.174511}, respectively. Copyright (2011, 2013) by the American Physical Society.}
\end{figure}

The iron-based materials of the \thcrsi\ type are arguably the most extensively studied, largely due to the early discovery~\cite{PhysRevB.78.020503,PhysRevLett.101.107006} and synthesis of large high-quality single-crystalline compounds of this class~\cite{PhysRevLett.101.117004,PhysRevB.79.014506,PhysRevLett.102.117005,PhysRevLett.102.187004}, thus allowing the entire arsenal of experimental condensed-matter techniques to be applied to them virtually simultaneously, giving rise to explosive progress in this area of research. Therefore, the experimental optical investigation of these 122 compounds will set the tone for this and the following sections, with the relatively scarce data available on other materials of the iron-based family provided for comparison where available.

\subsection{Normal state}

According to the predictions of {\it ab initio} electronic band-structure calculations the intermediate-energy electronic band structure (within $\approx\pm 3\ \textrm{eV}$ from the Fermi level) of virtually all iron-based materials is dominated by the more or less hybridized 3d orbitals of iron ions and 4p orbitals of the pnictogen/chalcogen~\cite{2009PhyC469614M}. This suggests that all of these materials should have quite similar overall structure of the interband optical transitions in this energy range even upon substitution. Optical investigations on a large subset of the iron-based compounds have indeed confirmed this conclusion. Figure~\ref{fig:bkfaoptlda} shows a representative spectrum of the optical conductivity of a hole-doped \bkfaoptmpi, the theoretical prediction, as well as the corresponding electronic band structure obtained for this material in the local-density approximation~\cite{CharnukhaNatCommun2011}. The Drude-Lorentz analysis of the experimental data allows one to clearly identify three main contributions to the interband optical conductivity (red, green, and blue Lorentzians in Fig.~\ref{fig:bkfaoptlda}a). A comparison with the theoretical prediction and the underlying electronic band structure plotted in Figs.~\ref{fig:bkfaoptlda}b~and~c allows one to assign these different contributions to the following transitions:
\begin{itemize}
\item the lowest-energy interband contribution (red line in Figs.~\ref{fig:bkfaoptlda}a~and~b) originates in transitions from weakly hybridized Fe-d (red lines in Fig.~\ref{fig:bkfaoptlda}c) and As-p (blue lines in Fig.~\ref{fig:bkfaoptlda}c) states below the Fermi level to Fe-d states of electronic character in the vicinity of the Fermi level (purple lines in Fig.~\ref{fig:bkfaoptlda}c);
\item the intermediate-energy interband contribution (green line in Figs.~\ref{fig:bkfaoptlda}a~and~b) stems from transitions from the same initial states to strongly hybridized Fe-d and As-p states above the Fermi level (green lines in Fig.~\ref{fig:bkfaoptlda}c);
\item finally, the higher-energy contribution (blue line in Figs.~\ref{fig:bkfaoptlda}a~and~b) comes predominantly from transitions from all of the above initial states to Ba-d states above the Fermi level.
\end{itemize}
This general structure of the interband optical conductivity is largely independent of the intercalating atom, as demonstrated in Ref.~\onlinecite{PhysRevB.88.184511} and summarized in Fig.~\ref{fig:comp122interband}.

It is evident from the comparison in Fig.~\ref{fig:bkfaoptlda} that the agreement between between experiment and theory is rather satisfactory in this intermediate energy range except for the significant softening of the $1.5\ \textrm{eV}$ band predicted by {\it ab initio} calculations to about $0.6$--$0.75\ \textrm{eV}$ inferred from experimental data (peak positions of the red lines in Fig.~\ref{fig:bkfaoptlda}a~and~b), i.e. by a factor of about $2.5$. According to the assignment above, this shift must come predominantly from the bandwidth renormalization of the Fe-d states below the Fermi level (red lines in Fig.~\ref{fig:bkfaoptlda}c). Such a renormalization has been observed consistently in many different iron-based materials using angle-resolved photoemission spectroscopy~\cite{PhysRevLett.105.067002,2011JPSJ80k3707S,0953-8984-23-13-135701,PhysRevB.86.155143,PhysRevLett.110.067003,2013arXiv1307.1280M}. The experimentally observed location of the absorption band at about $0.6$--$0.75\ \textrm{eV}$ has been reproduced by theoretical calculations in the dynamic mean-field theory~\cite{PhysRevB.82.045105,Yin_NatPhys_BFA_2011}, which identified the origin of the aforementioned renormalization in substantial Hund's-coupling correlations between the Fe-d electrons in the iron-based compounds.

The occurrence of numerous Fe-d bands very close to the Fermi level results in the concentration of sizable spectral weight in the mid-infrared optical band at quite low energies. As a result, this absorption band makes an unusually large contribution to the total zero-frequency interband permittivity $\Delta\varepsilon_{\mathrm{tot}}$ of the order of 60, which leads to a drastic reduction of the plasma frequency of itinerant charge carriers by a factor of $\sim8$ with respect to its bare value~\cite{CharnukhaNatCommun2011,PhysRevB.84.174511,PhysRevB.88.184511}.

\begin{figure}[t!]
\includegraphics[width=\columnwidth]{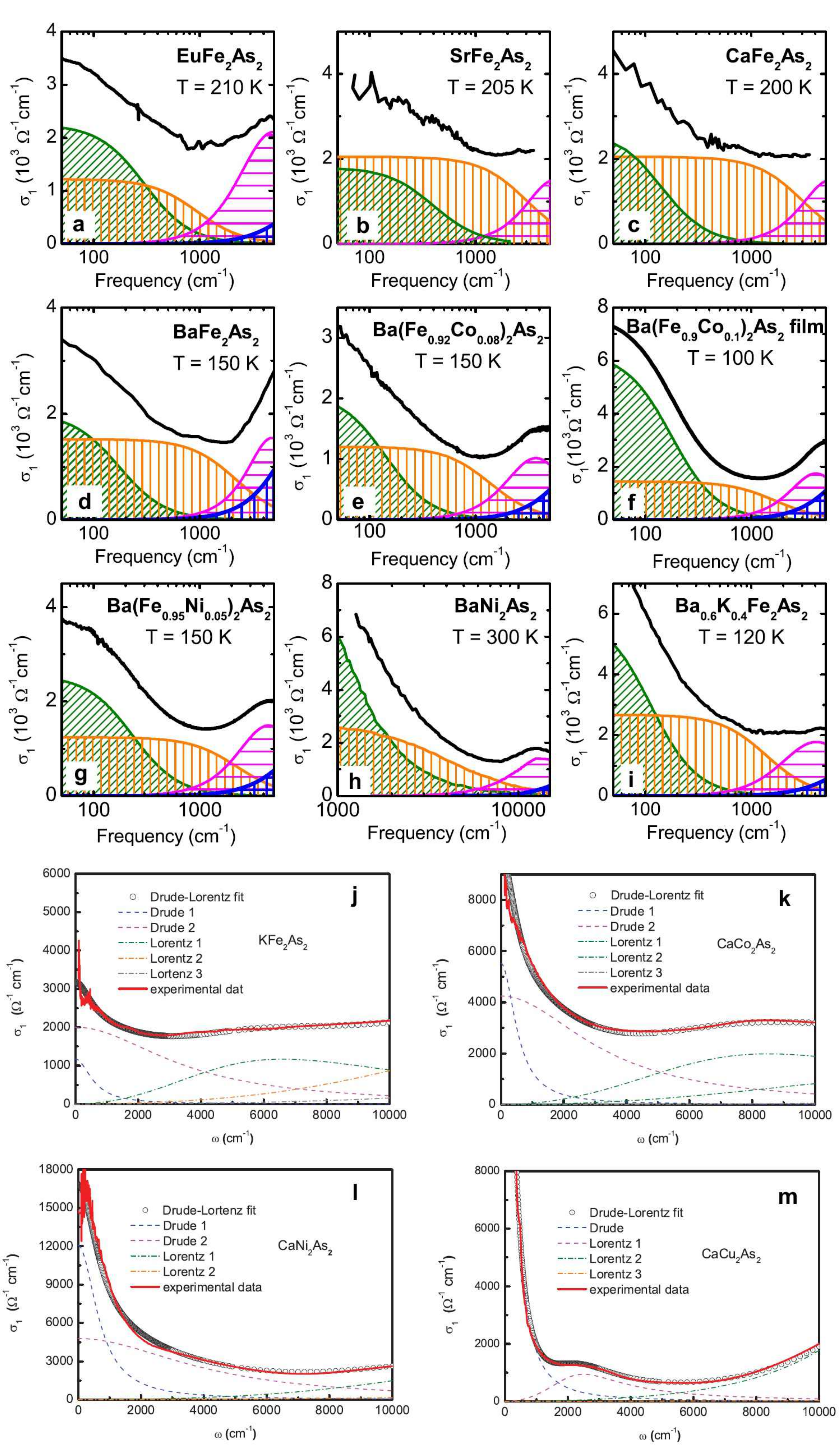}
\caption{\label{fig:twodrude} Real part of the optical conductivity of various isostructural \thcrsi-type compounds and the corresponding Drude-Lorentz fits to the experimental data. Panels~\textbf{a}--\textbf{i} and~\textbf{j}--\textbf{m} reprinted with permission from Ref.~\onlinecite{PhysRevB.82.054518} and Ref.~\onlinecite{PhysRevB.86.134503}, respectively. Copyright (2010, 2012) by the American Physical Society.}
\end{figure}

As already mentioned, all single-phase iron-based materials remain conductive at all attainable temperatures, doping levels, and pressures. It implies that the low-energy, far-infrared, optical conductivity can be expected to be dominated by the itinerant-charge-carrier response. Such behavior has indeed been observed in all studied compounds of the iron-based family, as will be shown below. However, unlike in conventional metals, the infrared charge dynamics of these materials was found to have pronounced bad-metal characteristics, exhibiting a long high-frequency tail along with a coherent low-frequency response. This ambivalent response can be modeled with two independent Drudelike contributions to the optical conductivity: a very narrow, ``coherent'', and a broad, ``incoherent'' term, as first demonstrated in Ref.~\onlinecite{PhysRevB.81.100512} for a representative set of 122 compounds (see Fig.~\ref{fig:twodrude}a--f) as well as in Refs.~\onlinecite{PhysRevB.81.104528,1367-2630-12-7-073036} for \bfca\ and later found in other classes of the iron-based family~\cite{PhysRevB.81.180508,1367-2630-15-7-073029} and different isostructural compounds (see Fig.~\ref{fig:twodrude}g--i). While the precise properties of the two types of itinerant response to some extent depend on the material composition, the breakdown into a coherent and incoherent contribution is universal. It is, therefore, important to understand the origin of these two types of response and their relation to itinerant charge carriers. The most straightforward explanation would be to remember that all iron-based materials are inherently multiband compounds and thus possess multiple electronic subsystems according to the number of Fermi-surface sheets. In the case of weakly interacting quasiparticles the width of the free-charge-carrier response would be dominated by the rate of scattering from impurities in the crystal. This argument would then lead one to conclude that there are two types of charge carriers in the system: one that experiences very strong impurity scattering, giving rise to the incoherent part of the itinerant response, while the other is subject to much weaker scattering, thus producing a very narrow, coherent, Drudelike term in the optical conductivity. Since in this case all of the scattering comes from randomly distributed impurities in the crystal, one may extract the quasiparticle mean-free path from the experimentally obtained scattering rate. In iron-based compounds the latter is on the order of $2500\ \textrm{cm}^{-1}$~\cite{PhysRevB.88.184511}. Assuming the Fermi velocity of about $0.4\ \textrm{eV}/$\AA\ one obtains the mean-free path of $0.8$~\AA, much smaller than the in-plane lattice constant of about $3.7$~\AA\ and even smaller than the shortest interionic distance~\cite{Johnston_Review_2010}. This situation is clearly unphysical and suggests that the incoherent term might be unrelated to the itinerant response in the iron-based compounds but rather be a manifestation of low-lying interband transitions, indeed predicted by {\it ab initio} calculations~\cite{PhysRevB.83.224514,PhysRevB.87.075136} and recently identified experimentally~\cite{PhysRevB.88.180508}. However, the total spectral weight of these low-energy transitions (both in theory and in experiment) is several orders of magnitude smaller than that of the incoherent conductivity term. Moreover, the detailed analysis of the optical spectral weight in \bfca\ has revealed that only by assigning most of the spectral weight at energies up to $1500\ \textrm{cm}^{-1}$ to the response of itinerant charge carriers does one find consistency with band-structure renormalization revealed by other probes~\cite{PhysRevLett.108.147002}.

\begin{figure}[t!]
\includegraphics[width=\columnwidth]{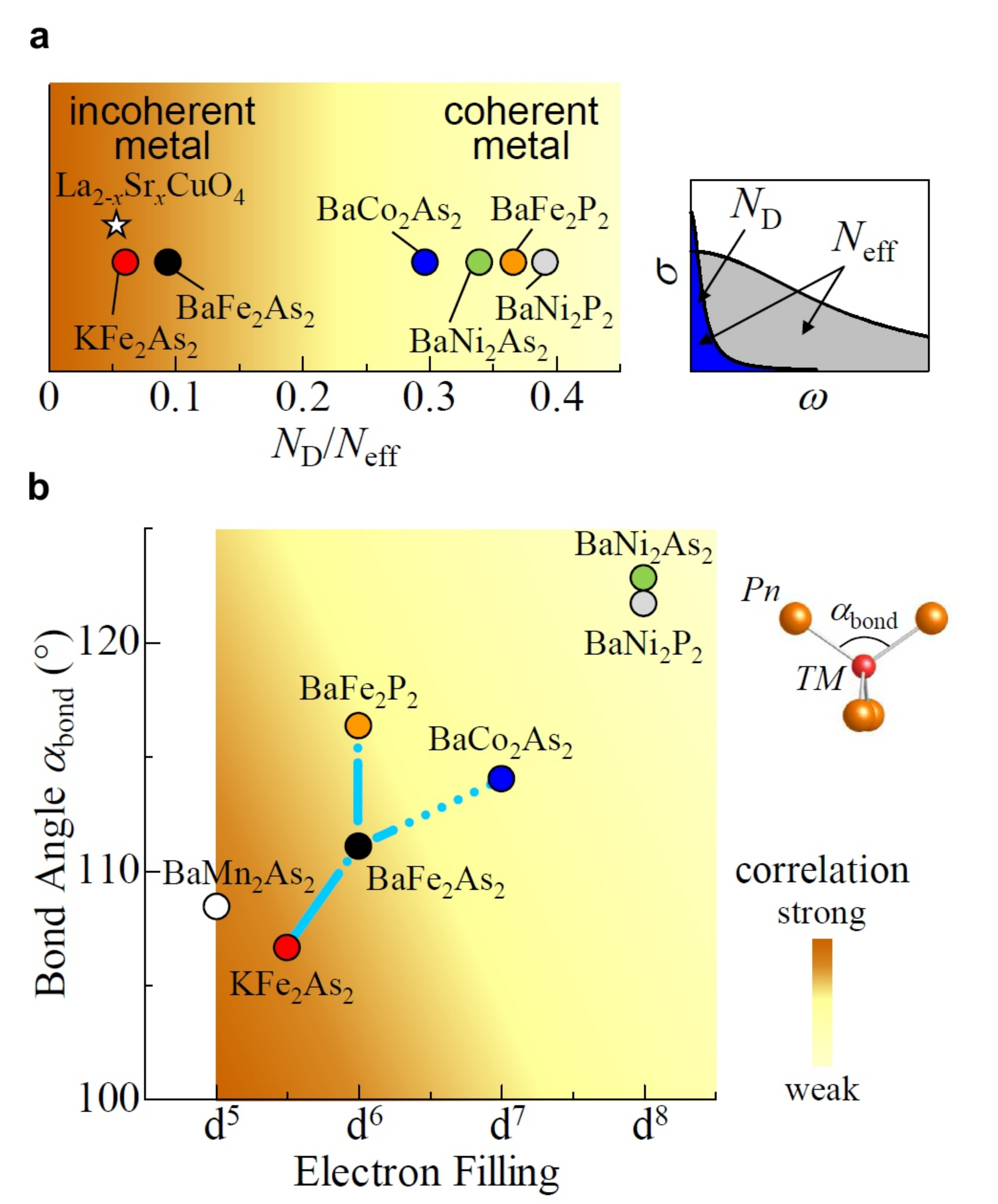}
\caption{\label{fig:correlations}~\textbf{a}, Fraction of the coherent Drude spectral weight $N_{\mathrm{D}}/N_{\mathrm{eff}}$ for various transition-metal iron pnictides with the same crystal structure, including one typical underdoped cuprate.~\textbf{b}, Transition-metal pnictides mapped on the diagram of pnictogen-transition metal-pnictogen bond angle $\alpha_{\mathrm{bond}}$ as a function of the electron filling of transition-metal d orbitals in the $2+$ state. Blue dotted and solid lines are trajectories of three types of doping into \bfa, with superconductivity observed in the regions indicated by the solid blue lines. Smaller $\alpha_{\mathrm{bond}}$ and lower electron filling towards $d^5$ (half filling) indicate stronger electronic correlations. Correlation strength is shown schematically as a color gradient. Reprinted figure and adapted caption with permission from Ref.~\onlinecite{2013arXiv1308.6113N}.}
\end{figure}

The above difficulty with the unphysically small mean-free path inferred from the very large quasiparticle scattering rate of the incoherent term can be completely resolved if one considers interacting quasiparticles. In this case, the total scattering is a combination of the elastic scattering from impurities in the crystal mentioned above and the inelastic scattering mediated by various bosonic excitation coupled to the itinerant charge carriers (such as phonons, spin fluctuations, orbital fluctuations, etc.)~\cite{Shulga1991266,A1995RX14100023}. The inelastic scattering rate is determined by the strength of the electron-boson interaction and unrelated to the lattice and impurities in the crystal. Good nesting between the hole- and electronlike sheets of the Fermi surface originally identified in the iron-based compounds from {\it ab initio} calculations could in principle provide a large enough density of states for the corresponding bosonic excitations to give rise to the observed magnitude of the quasiparticle scattering rate of the incoherent itinerant response. However, as has been mentioned in the introduction, the experimentally determined Fermi surface topology essentially excludes any effective nesting in iron-based compounds. Additionally, the observation of very similar incoherent response in iron-based materials with quite different composition and, as a result, drastically different Fermi surfaces and types of dominating charge carriers~\cite{Johnston_Review_2010,RevModPhys.83.1589} implies that it is unrelated to the specific structure of the Fermi surface but rather is a universal feature in this family of compounds~\cite{PhysRevB.86.134503}. One of the plausible candidates for the source of the strong interactions underlying the incoherent term in the optical conductivity seem to be Hund's-coupling orbital correlations between Fe-d electrons, predicted to have a sizable intensity and play an important role in the low-energy electrodynamics of the iron-based materials~\cite{Yin_NatPhys_BFA_2011}. Recent detailed optical study of several isostructural 122 compounds has revealed the presence of the incoherent term in the optical conductivity of various 3d-transition-metal--based materials in stark contrast to its complete absence in $\textrm{CaCu}_2\textrm{As}_2$  (see Fig.~\ref{fig:twodrude}), a 4s metal with a fully filled 3d shell, thus entirely lacking Hund's-coupling correlations. A detailed analysis of the relative intensity of the coherent and incoherent contributions to the optical conductivity as a function of doping has revealed a pronounced dependence of the degree of incoherence on the pnictogen-transition metal-pnictogen bond angle and the electron filling of the transition-metal 3d orbitals (Fig.~\ref{fig:correlations}), both of which provide a measure of electronic correlations. An angle-resolved photoemission spectroscopy study of the fully cobalt-substituted $\textrm{BaCo}_2\textrm{As}_2$ has likewise revealed significantly weaker electronic correlations compared to its iron-based counterpart due to the higher filling of the Co-3d orbitals in the presence of strong Hund's-coupling correlations~\cite{PhysRevX.3.011006}. The degree of electronic correlations can further be assessed by comparing the spectral weight of itinerant charge carriers (related to their kinetic energy~\cite{PhysRevB.16.2437,Hirsch1992305}) extracted from the optical conductivity obtained experimentally and in {\it ab initio} band-structure calculations and has been found to be quite substantial~\cite{PhysRevLett.108.147002}. At the same time, numerous previous studies of the iron-based materials have revealed a trend towards the maximization of the superconducting transition temperature close to the ideal tetrahedral coordination ($\alpha_{\mathrm{bond}}\approx109.5^\circ$) of the transition metal and pnictogen ions~\cite{Johnston_Review_2010,PhysRevB.85.184507}. It, therefore, appears that strong electronic correlations are an essential ingredient of high-temperature superconductivity in the iron-based materials. The importance of orbital physics in the iron-based superconductors is further reflected in the correlation between the magnitude of the superconducting energy gap and the orbital character of the Fermi-surface sheets on which it occurs, identified by means of angle-resolved photoemission spectroscopy~\cite{PhysRevB.89.064514}.

In the case of interacting quasiparticles the traditional Drude expression for the optical conductivity of itinerant charge carriers can be generalized to the so-called extended Drude model~\cite{RevModPhys.77.721}. The complex optical conductivity $\sigma(\omega)=\sigma_1(\omega)+i\sigma_2(\omega)$, determined by two real-valued functions of frequency $\sigma_1(\omega)$ and $\sigma_2(\omega)$, is uniquely expressed in this model in terms of two different functions: frequency-dependent quasiparticle scattering rate $\gamma(\omega)$ and mass renormalization $m(\omega)/m_{\mathrm{band}}$, where $m_{\mathrm{band}}$ is the bare quasiparticle effective mass. Since such a mapping always exists for an arbitrary shape of the optical conductivity, whether it truly represents itinerant-charge-carrier response or not, extreme care must be taken in the interpretation of the results of such a data analysis. As it so happens, several properties of the iron-based materials eloquently highlight the main difficulties of the extended-Drude approach. 

\begin{figure}[t!]
\includegraphics[width=\columnwidth]{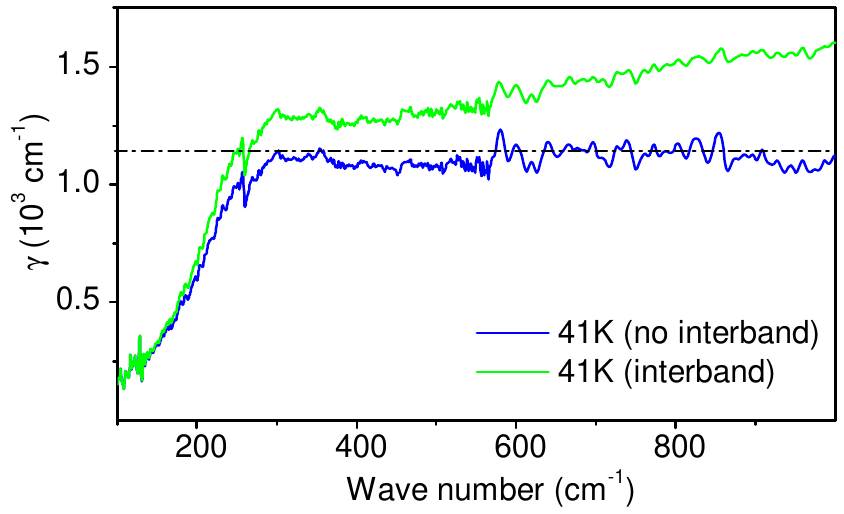}
\caption{\label{fig:extDrude} Optical scattering rate obtained from the experimental optical conductivity of \bkfaoptmpi\ at $41\ \textrm{K}$ in the framework of the extended-Drude model, with the contribution of the interband transitions subtracted (blue line) and without subtraction (green line). Dash-dotted line indicates the saturation level of the high-energy optical scattering rate. Reprinted with permission from Ref.~\onlinecite{PhysRevB.84.174511}. Copyright (2011) by the American Physical Society.}
\end{figure}

The first complication comes from the relatively low-lying intense interband transitions. A typical extended Drude analysis of a correlated material would be confined to the frequencies of the order of the largest quasiparticle scattering rate or less, in the pnictides --- usually up to around $1000$--$2500\ \textrm{cm}^{-1}$. As was discussed above, the lowest-energy pronounced optical absorption band is located around $0.6$--$0.75\ \textrm{eV}\approx5000$--$6000\ \textrm{cm}^{-1}$, i.e. well beyond the frequency range of interest for the extended Drude analysis and thus of little concern. However, this analysis essentially depends not only on the real part of the optical conductivity but also on its imaginary part. While the influence of the $0.6\ \textrm{eV}$ band on the former is negligible, that on the latter is substantial {\it in spite} of the fact that it lies well beyond the spectral range of interest because it makes a very large contribution to the total zero-frequency interband permittivity of the order of 60. The importance of the constant contribution of each interband transition to the real part of the dielectric function (or, equivalently, to the imaginary part of the optical conductivity: $\varepsilon(\omega)=1+4\pi i\sigma(\omega)/\omega)$ well below the resonance frequency of that transition is often forgotten because in the overwhelming majority of materials this contribution is very small, adding up to a value of order unity. The iron-based compounds are one of the prominent exceptions. Figure~\ref{fig:extDrude} demonstrates the effect of unsubtracted interband transitions above $0.6\ \textrm{eV}$ on the extended Drude analysis below $1000\ \textrm{cm}^{-1}$ in the low-temperature normal state of superconducting \bkfaoptmpi. It is clear that, with all interband transitions eliminated, the quasiparticle scattering rate saturates above $300\ \textrm{cm}^{-1}$. Quite on the contrary, if the extended Drude analysis is carried out on the raw data then the quasiparticle scattering rate shows a linear increase up to very high frequencies. Albeit such a linear frequency dependence is a mere artifact of the data analysis it can be mistaken for the signatures of strong electronic correlations or a quantum critical point~\cite{0034-4885-66-8-202,PhysRevB.84.174511,PhysRevB.83.224514}. In light of the recently reported observation of low-energy absorption bands around $1000\ \textrm{cm}^{-1}$ and $2300\ \textrm{cm}^{-1}$ in \bfca\ (albeit of rather low intensity)~\cite{PhysRevB.88.180508} in line with the theoretical predictions~\cite{PhysRevB.83.224514,PhysRevB.87.075136} even more care should be exercised in order not to misinterpret the extrinsic effect of interband transition as intrinsic properties of itinerant charge carriers.

The second complication in the extended-Drude analysis of the optical conductivity of the iron-based materials comes from the inherently multiband character of the latter, which implies the existence of multiple electronic subsystems, each with its own quasiparticle properties. In the case of a single-band material with weakly interacting quasiparticles the extended Drude analysis of the optical conductivity provides physically correct result: constant quasiparticle scattering rate equal to the elastic scattering rate from random impurities in the crystal, as well as unity effective-mass renormalization. In the multiband case, however, even if the quasiparticles in all of the bands are weakly interacting, a simple application of the extended Drude analysis to the total optical conductivity of all bands will result in a frequency-dependent quasiparticle scattering rate and effective-mass renormalization. Even though the procedure is mathematically correct and the result is unique, its interpretation is difficult, unless all bands have very similar properties, which is not the case for the iron-based compounds. If, in addition, every band features (strongly) interacting quasiparticles the rationality of this analysis appears dubious.

\begin{figure}[t!]
\includegraphics[width=\columnwidth]{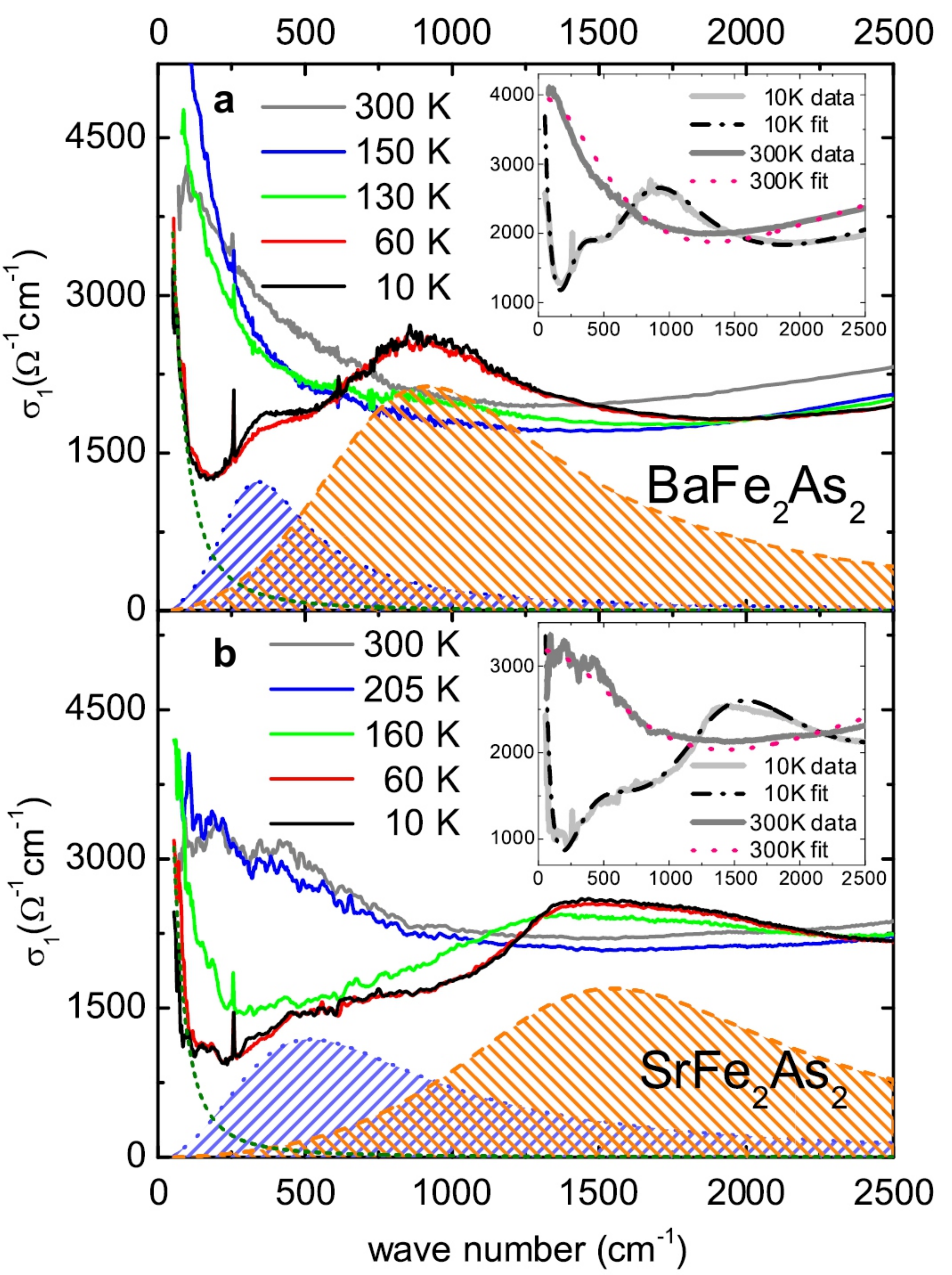}
\caption{\label{fig:sdw} Real part of the complex optical conductivity of $\textrm{BaFe}_2\textrm{As}_2$~\textbf{a} and $\textrm{SrFe}_2\textrm{As}_2$~\textbf{b} at several temperatures in the normal and antiferromagnetic state. Instets show the experimental data along with the result of a Drude-Lorentz fit. Reprinted with permission from Ref.~\onlinecite{PhysRevLett.101.257005}. Copyright (2008) by the American Physical Society.}
\end{figure}

It is, therefore, more reliable to start with a certain microscopic model of the normal-state multiband optical conductivity, such as the Eliashberg theory~\cite{RevModPhys.62.1027}, constrained by a large body of experimental data obtained with other techniques and {\it fit} the frequency-dependent quasiparticle scattering rate and effective-mass renormalization obtained in this model to those extracted via the effective Drude analysis of the experimental optical conductivity data~\cite{PhysRevB.84.174511}. In this approach the interpretation of the latter is irrelevant and it is used as a sheer mathematical transformation of the original complex optical conductivity, which might provide certain insights into some characteristics of the underlying physical phenomena. If, on the contrary, the multiband extended-Drude data are fitted or inverted in a single-band theory one might still be able to get approximate qualitative information about the energy scales of relevant interactions~\cite{PhysRevB.82.144519}. However, there is  a significant risk of mistaking multiband effects for intrinsic properties of quasiparticle interactions.

\subsection{Antiferromagnetic state}

Upon lowering the temperature, essentially all of the iron-based compounds undergo a sequence of a structural and a magnetic phase transition~\cite{Johnston_Review_2010,RevModPhys.83.1589}. At the structural transition the $C4$ rotational symmetry of the tetragonal high-temperature state is reduced to the orthorhombic $C2$ symmetry in a weakly or strongly first-order phase transition~\cite{PhysRevB.83.134505}, with the new orthorhombic crystallographic axes rotated by $45^\circ$ with respect to their tetragonal counterparts. This process is accompanied by the formation of twin domains, oriented perpendicularly to each other due to the two possible directions for the longer orthorhombic axis. At a somewhat lower temperature the materials enter an antiferromagnetically ordered phase that exhibits a finite magnetic moment on iron ions and a corresponding reconstruction of the Brillouin zone~\cite{PhysRevB.82.134503} but retains metallic properties. The latter, together with the early observation of a relatively small ordered moment, suggested that the antiferromagnetic state in the iron-based materials is of itinerant character and thus is driven by the nesting instability of the Fermi surface at a commensurate nesting vector connecting the hole and the electron sheets of the Fermi surface predicted by {\it ab initio} calculations~\cite{2009PhyC469614M}. However, as already mentioned above, extensive ARPES~\cite{Borisenko_BKFA_FS_NormState2009} and neutron-scattering~\cite{NatPhys_Dai_Dagotto_ReviewMagnPnictides} investigations of various iron-based materials have identified virtually absent nesting and an essentially dual, simultaneously itinerant and local, character of antiferromagnetism in these compounds, respectively.

\begin{figure*}[t!]
\includegraphics[width=\textwidth]{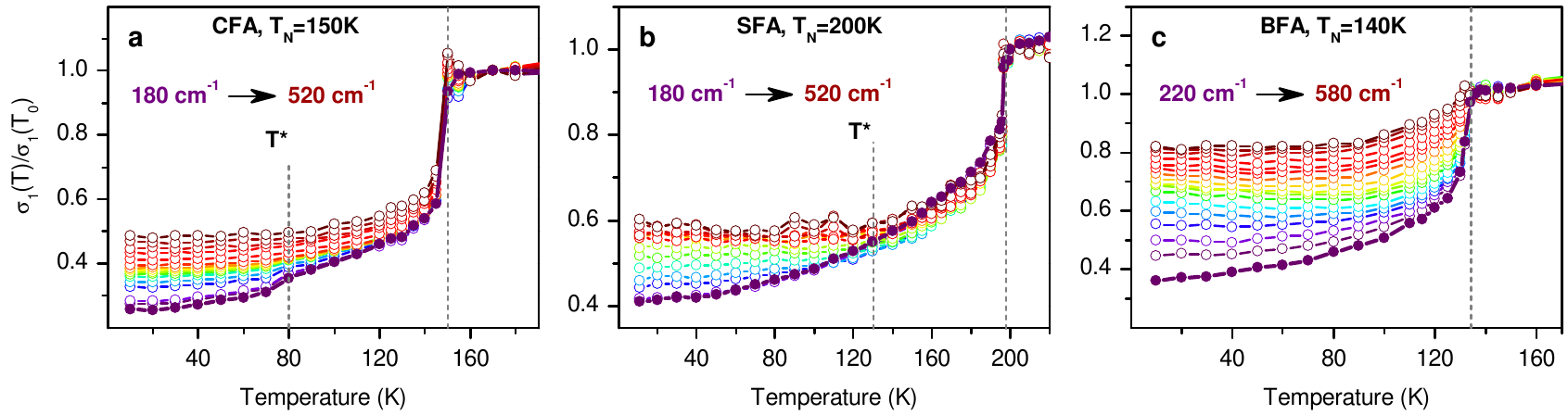}
\caption{\label{fig:sdwgaps} Detailed temperature dependence of the real part of the complex optical conductivity of \textbf{a}~$\textrm{CaFe}_2\textrm{As}_2$, \textbf{b}~$\textrm{SrFe}_2\textrm{As}_2$, and \textbf{c}~$\textrm{BaFe}_2\textrm{As}_2$ on a fine grid of wavenumbers in the far-infrared spectral range, normalized to the value of the optical conductivity at $T=T_{\mathrm{0}}$ close to $T_{\mathrm{N}}$ ($T_{\mathrm{0}}=170\ \textrm{K},\ 200\ \textrm{K},\ \textrm{and}\ 140\ \textrm{K}$ for the Ca-, Sr-, and Ba-based compounds, respectively). The grid of wavenumbers runs from $180\ \textrm{cm}^{-1}$ (blue colors) to $520\ \textrm{cm}^{-1}$ (red colors) in steps of $20\ \textrm{cm}^{-1}$ with the exception of the spectral window of the $260\ \textrm{cm}^{-1}$ phonon for the Ca- and Sr-based compounds, as well as from $220\ \textrm{cm}^{-1}$ to $580\ \textrm{cm}^{-1}$ for $\textrm{BaFe}_2\textrm{As}_2$. Spin-density--wave--related features in the temperature dependence of the optical conductivity (vertical dashed lines), for interpretation see text. The lowest-frequency temperature dependence is plotted as filled circles unlike the rest (open circles) to emphasize the change in the temperature dependence of the optical conductivity. Figure, caption, and the corresponding discussion in the text reprinted with permission from Ref.~\onlinecite{PhysRevB.88.184511}. Copyright (2013) by the American Physical Society.}
\end{figure*}

The low-energy modification of the electronic band structure induced by the antiferromagnetic phase transition can be probed effectively by studying its effect on the charge dynamics reflected in the optical conductivity. Already very first investigations of the latter in many different iron-based parent compounds revealed a strong suppression of the infrared conductivity at the N\'eel temperature below a certain energy~\cite{PhysRevLett.101.257005,PhysRevB.81.100512,PhysRevB.81.104528,PhysRevB.81.205114,1367-2630-12-7-073036,2013arXiv1303.4182O,boris:027001,detwinned_BFCA_lobo_2011,Hu2009545}, see Fig.~\ref{fig:sdw}. Due to the conservation of the total number of electrons, the missing area under the conductivity curve below this energy is transferred to higher energies, which results in the formation of a ``hump'' structure~\cite{JPSJ.12.570,PhysRevB.66.134510}. In the case of itinerant magnetism the above suppression in the optical conductivity results from the opening of the so-called spin-density--wave energy gap at the hot spots of the Fermi surface (portions of the Fermi surface connected by the nesting vector). In the most common case of imperfect nesting the hot spots and, consequently, the spin-density--wave gap do not cover the entire Fermi surface. Therefore, the material remains metallic in the low-symmetry state, with the optical conductivity non-zero at all frequencies. This circumstance makes the determination of the optical spin-density--wave energy gap $2\Delta_{\mathrm{SDW}}$ from the conductivity spectra difficult but it can be approximated by the energy, at which the low-temperature conductivity spectrum crosses the one at the N\'eel temperature for the first time~\cite{PhysRevB.66.134510}. Having estimated the energy gap one may proceed to calculate the corresponding gap ratio $2\Delta_{\mathrm{SDW}}/k_{\mathrm{B}}T_{\mathrm{N}}$, i.e. the ratio of the optical energy gap $2\Delta_{\mathrm{SDW}}$ and the N\'eel temperature in energy units $k_{\mathrm{B}}T_{\mathrm{N}}$, where $k_{\mathrm{B}}$ is the Boltzmann constant. The gap ratio provides a rough estimate of the coupling strength between electrons giving rise to a given electronic instability, such as superconductivity, spin-density wave etc.~\cite{RevModPhys.62.1027,Gruner_Density_waves_in_solids_2000_articlestyle}. A systematic comparison of the gap ratios of three common \thcrsi-type iron-based parent compounds was given in Ref.~\onlinecite{PhysRevB.88.184511}. This study found a clear trend towards stronger coupling with decreasing atomic number of the intercalating atom: from barium via strontium to calcium. A similar trend was found in the total plasma frequency of the itinerant charge carriers, with \cfa\ being significantly more metallic than the other two materials.

As can be seen in Fig.~\ref{fig:sdw}, the optical conductivity of \bfa\ and \sfa\ reveals a change in its shape at lowest frequencies at $10\ \textrm{K}$ as compared to intermediate temperatures below $T_{\mathrm{N}}$, which indicates the presence of another spin-density--wave gap, with a smaller magnitude. This feature has also been observed in \bfa\ and \cfa\ compounds in Ref.~\onlinecite{PhysRevLett.108.147002} and Ref.~\onlinecite{PhysRevB.88.184511}, respectively, and associated with a second spin-density--wave energy gap. The existence of two separate spin-density--wave energy gaps raises a question as to the degree of interaction between the corresponding electronic subsystems. This question becomes particularly difficult to address in the presence of as many Fermi-surface sheets as is the case in the iron-based materials, which commonly possess up to five separate Fermi-surface sheets in the first Brillouin zone. This complexity arises mainly due to the fact that the optical conductivity represents an intertwined response of all electronic subsystems at once. To shed some light onto this issue detailed measurements in the far-infrared spectral range (i.e. in the range where the real part of the optical conductivity experiences a drastic suppression, as shown in Fig.~\ref{fig:sdw}) on a very fine temperature grid were carried out for three 122 parent materials in Ref.~\onlinecite{PhysRevB.88.184511}. The results are compiled in Fig.~\ref{fig:sdwgaps}, which shows the temperature dependence of the optical conductivity normalized to its value at $T=T_{\mathrm{0}}$ close to $T_{\mathrm{N}}$ ($T_{\mathrm{0}}=170\ \textrm{K},\ 200\ \textrm{K},\ \textrm{and}\ 140\ \textrm{K}$ for the Ca-, Sr-, and Ba-based compounds, respectively) at an equidistant set of wavenumbers from $180\ \textrm{cm}^{-1}$ (blue colors) to $520\ \textrm{cm}^{-1}$ (red colors) in steps of $20\ \textrm{cm}^{-1}$ with the exception of the spectral window of the $260\ \textrm{cm}^{-1}$ phonon for the Ca- and Sr-based compounds, as well as from $220\ \textrm{cm}^{-1}$ to $580\ \textrm{cm}^{-1}$ for $\textrm{BaFe}_2\textrm{As}_2$.

The temperature dependence of the normalized conductivity shows a very pronounced drop for all compounds at their respective N\'eel temperatures (right vertical dashed lines) due to the onset of spin-density--wave order, with a clear mean-field order-parameter--like temperature dependence at large wavenumbers (red colors). This temperature dependence changes, however, as one moves toward progressively smaller wavenumbers and in the case of $\textrm{CaFe}_2\textrm{As}_2$ reveals a clear second suppression feature at a temperature $T^*$ of about $80\ \textrm{K}$. This second feature becomes washed out in the Sr-based compound, although the temperature dependence of $\sigma_1$ at smallest wavenumbers remains markedly different from that at large wavenumbers and some sort of an analogue of $T^*$ can be sketched also in this case, although there is no clear second suppression feature present in any of the separate temperature dependences themselves. Finally, in the case of $\textrm{BaFe}_2\textrm{As}_2$, the change in the temperature dependence of $\sigma_1$ between large and small wavenumbers becomes hardly noticeable, with the exception of an upturn in its temperature dependence at intermediate frequencies due to the pile-up of the spectral weight transferred from the energies below the smaller spin-density--wave energy gap.

\begin{figure*}[t!]
\includegraphics[width=\textwidth]{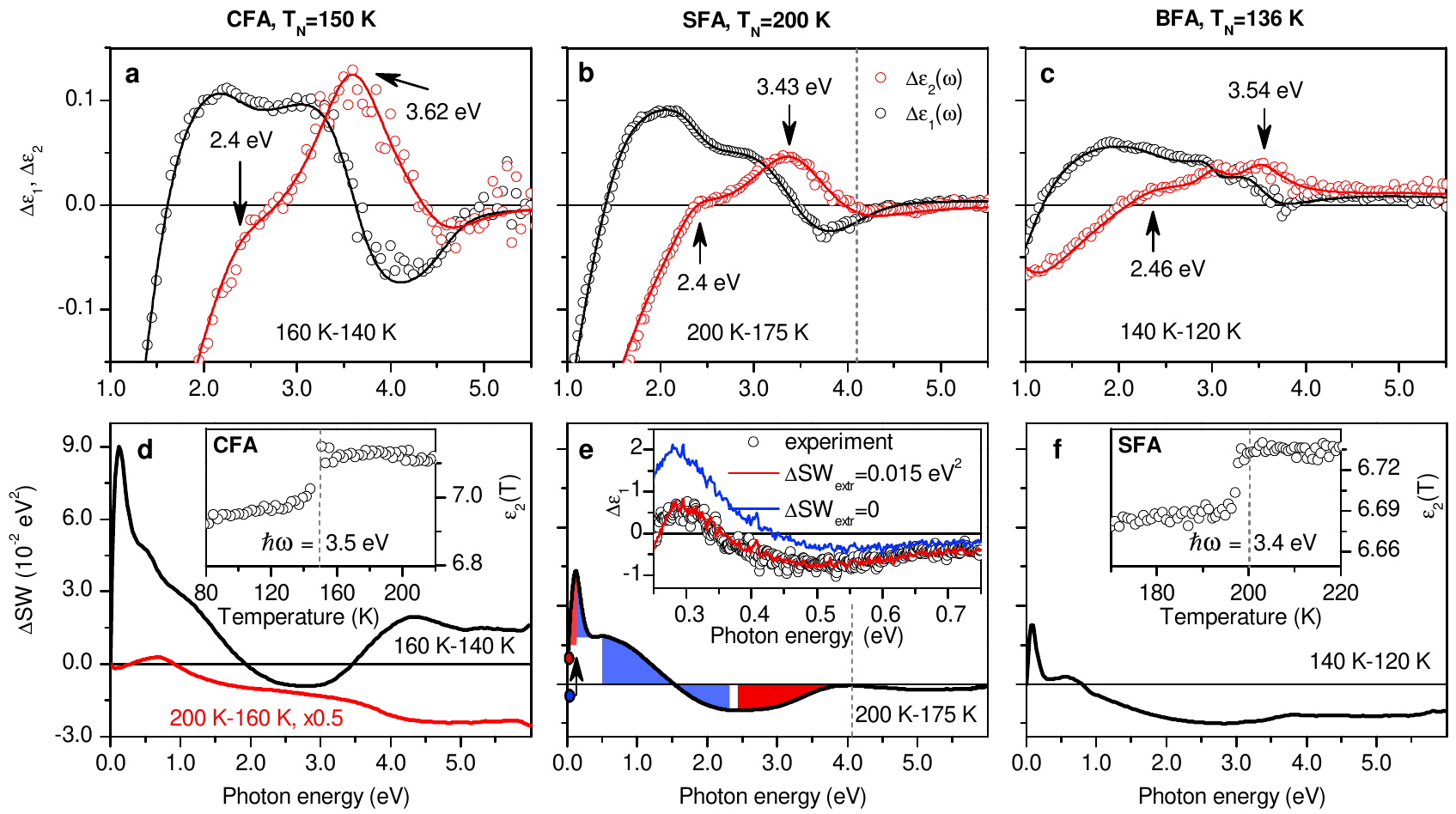}
\caption{\label{fig:anomalyvis}~\textbf{a}--\textbf{c}~Difference real and imaginary parts of the dielectric function $\Delta\varepsilon_1,\ \Delta\varepsilon_2$ in the visible spectral range between the temperatures above and below the N\'eel transition temperature, as specified in the panels. The arrows indicate the two spin-density--wave--suppressed absorption bands.~\textbf{d}--\textbf{f} Spectral-weight redistribution between the same temperatures and in the same compounds as in \textbf{a}--\textbf{c} as a function of photon energy (black solid lines). Blue and red areas in \textbf{e} indicate the regions of spectral-weight gain and loss, respectively, in the magnetic versus the normal state. Red solid line in \textbf{d} shows the spectral-weight redistribution between $200\ \textrm{K}$ and $160\ \textrm{K}$ scaled by a factor of $0.5$. [Inset in \textbf{e}], Real part of the dielectric function and the Kramers-Kronig transformations of the real part of the optical conductivity of $\textrm{SrFe}_2\textrm{As}_2$ with different amount of spectral weight contained in the extrapolation region at low frequencies (solid lines, colors match filled circles). [Insets in \textbf{d},\textbf{f}] Temperature scan of $\varepsilon_2$ in $\textrm{CaFe}_2\textrm{As}_2$ at $3.5\ \textrm{eV}$ and in $\textrm{SrFe}_2\textrm{As}_2$ at $3.4\ \textrm{eV}$, respectively. Figure, caption, and the corresponding discussion of the spectral-weight transfer in the text reprinted with permission from Ref.~\onlinecite{PhysRevB.88.184511}. Copyright (2013) by the American Physical Society. Panel \textbf{e}, the inset therein, as well as in \textbf{f} adapted by permission from Macmillan Publishers Ltd: Nature Communications Ref.~\onlinecite{CharnukhaNatCommun2011}, copyright (2011).}
\end{figure*}

The presence of the second suppression feature in the temperature dependence of the real part of the optical conductivity at small enough wavenumbers in $\textrm{CaFe}_2\textrm{As}_2$ indicates, that the electronic subsystem that develops the smaller spin-density--wave energy gap $2\Delta_{\mathrm{S}}^{\mathrm{SDW}}$ preserves some knowledge about its own N\'eel temperature that it {\it would have} if this subsystem were completely independent from the other(s). In any real material all electronic subsystems are coupled, even if weakly, so that all spin-density--wave energy gaps open at the same $T_{\mathrm{N}}$. However, in case of weak intersubsystem coupling, those with smaller gaps exhibit an anomaly below the real N\'eel temperature, as is the case in $\textrm{CaFe}_2\textrm{As}_2$, albeit this anomaly does not correspond to a true phase transition. Such an effect has been predicted theoretically for the analogous case of multiband superconductivity~\cite{PhysRevB.71.054501,PhysRevB.80.014507} and discovered experimentally in $\textrm{FeSe}_{1-x}$ in Refs.~\onlinecite{Andreev_two_gaps_FeSe_2011,PhysRevLett.104.087004,Andreev_two_gaps_FeSe_2013}. The corresponding non-BCS temperature dependence of the smaller superconducting gap resulting from weak interband coupling seems to be consistent with the scanning tunneling spectroscopy data obtained on the LiFeAs compound~\cite{PhysRevLett.109.087002}. 

This argument thus suggests that in the Ca-based material the two electronic subsystems developing the large and the small spin-density--wave energy gap are weakly coupled. Naturally, as the coupling between such electronic subsystems increases, the temperature dependence of the small gap would gradually approach that of the large gap via an intermediate state when it already does not show any anomaly but still has not matched the temperature dependence of the large gap. Figure~\ref{fig:sdwgaps}b provides evidence for such a behavior in the Sr-based iron pnictide. Finally, the temperature dependence of the normalized optical conductivity in $\textrm{BaFe}_2\textrm{As}_2$ is almost the same at all wavenumbers, both in the region of the large spin-density--wave gap and in that of the small one, indicating strongly coupled electronic subsystems in this compound. Thus, by monitoring the detailed temperature dependence of the optical conductivity in the far-infrared spectral range on a fine grid of wavenumbers, a gradual transition could be observed from weak coupling between the electronic subsystems developing the two different spin-density--wave energy gaps in $\textrm{CaFe}_2\textrm{As}_2$, via intermediate coupling in $\textrm{SrFe}_2\textrm{As}_2$, to strong coupling between them in $\textrm{BaFe}_2\textrm{As}_2$, systematically with increasing atomic number of the intercalant.

We have consistently emphasized throughout this review that, albeit originally believed to be purely itinerant, the antiferromagnetism of the iron-based compounds has been found to bear essentially dual, itinerant and local, character~\cite{NatPhys_Dai_Dagotto_ReviewMagnPnictides}. It suggests that the spectral-weight redistribution found in the optical conductivity of antiferromagnetic compounds might be driven by local-moment physics as well as by the opening of a spin-density--wave energy gap due to collective long-range order. This idea has indeed received significant attention. For instance, the authors of Ref.~\onlinecite{PhysRevB.81.205114} have assigned the smaller and the larger gaplike features in \efa, similar to those discussed above, to two different sets of optical transitions: related to transitions within the spin-minority bands formed by Fe ions with the same spin and, therefore, sensitive to the long-range magnetic order and between the spin-majority and the spin-minority bands formed by Fe sites with opposite spins, respectively. The latter is mostly defined by the exchange splitting of the local magnetic moment due to Hund's coupling and thus less sensitive to the long-range order. Calculations taking into account electronic correlations based on a combination of the density-functional and dynamic-mean-field theory have likewise emphasized that it is the gain of Hund’s-coupling energy rather than Hubbard repulsion energy that compensates the loss in kinetic energy and thus stabilizes the low-temperature antiferromagnetic phase~\cite{Yin_NatPhys_BFA_2011}. The same study has achieved a remarkable agreement in the overall shape between the calculated and experimentally obtained infrared conductivity of \bfa\ in the antiferromagnetic ground state, albeit with a substantially overestimated electronic background~\cite{PhysRevLett.108.147002}.

Theoretical calculations in both Ref.~\onlinecite{PhysRevB.81.205114} and Ref.~\onlinecite{Yin_NatPhys_BFA_2011} have shown that the significant role played by Hund's coupling in the magnetism of the iron-based compounds leads to a modification of the electronic band structure on the energy scale of several electron-volts in the antiferromagnetic ground state compared to the high-temperature paramagnetic state. A detailed ellipsometric study of the \sfa\ compound~\cite{CharnukhaNatCommun2011} and its subsequent systematic comparison to the \cfa and \bfa\ compounds~\cite{PhysRevB.88.184511} have revealed that energies even higher than expected, on the order of $4-5\ \textrm{eV}$ are in fact affected by the antiferromagnetic phase transition, as shown in Fig.~\ref{fig:anomalyvis}. The optical conductivity of all three compounds experiences a drastic suppression upon entering the antiferromagnetic state (Fig.~\ref{fig:anomalyvis}a--c). The magnetism-induced character of these modifications is confirmed by a sharp drop at the N\'eel temperature observed in the temperature dependence of the imaginary part of the dielectric function shown in the insets of Fig.~\ref{fig:anomalyvis}d~and~f as well as in Ref.~\onlinecite{PhysRevB.88.184511}. The magnitude of the suppression decreases systematically from the Ca- via the Sr- to the Ba-based compound, roughly proportionally to the square of the gap ratio determined from infrared-conductivity measurements on the same compounds~\cite{PhysRevB.88.184511}. Quite surprisingly, a similar although much smaller suppression has been also found to occur at the superconducting transition temperature in the non-magnetic \bkfaoptmpi\ compound at optimal doping level~\cite{CharnukhaNatCommun2011}. This discovery has suggested that the suppression likely results from a redistribution of the electronic population at the superconducting transition between several electronic bands.

Additional information about the energy scales involved in the antiferromagnetic phase transition in \cfa, \sfa, and \bfa\ can be inferred from a detailed analysis of the spectral-weight transfer in the optical conductivity at across the N\'eel temperature (see Fig.~\ref{fig:anomalyvis}d--f). Several common features can be identified in the energetics of the spin-density--wave transition: the spectral weight from below the optical spin-density--wave gap $2\Delta^{\mathrm{SDW}}$ (lowest-energy red area in Fig.~\ref{fig:anomalyvis}e and the corresponding areas in Figs.~\ref{fig:anomalyvis}d,f) is redistributed to energies directly above the optical gap (subsequent blue area in Fig.~\ref{fig:anomalyvis}e and the corresponding areas in Figs.~\ref{fig:anomalyvis}d,f). However, the spectral weight gained directly above the optical gap does not balance that lost within the gap, which implies that higher-energy processes beyond the characteristic magnetic energy scales proposed for the iron-based compounds are affected by the spin-density--wave transition, as discussed above, including the spin-density--wave--supressed band in the visible spectral range (higher-energy blue and red areas in Fig.~\ref{fig:anomalyvis}e and the corresponding areas in Figs.~\ref{fig:anomalyvis}d,f).

\begin{figure}[b!]
\includegraphics[width=\columnwidth]{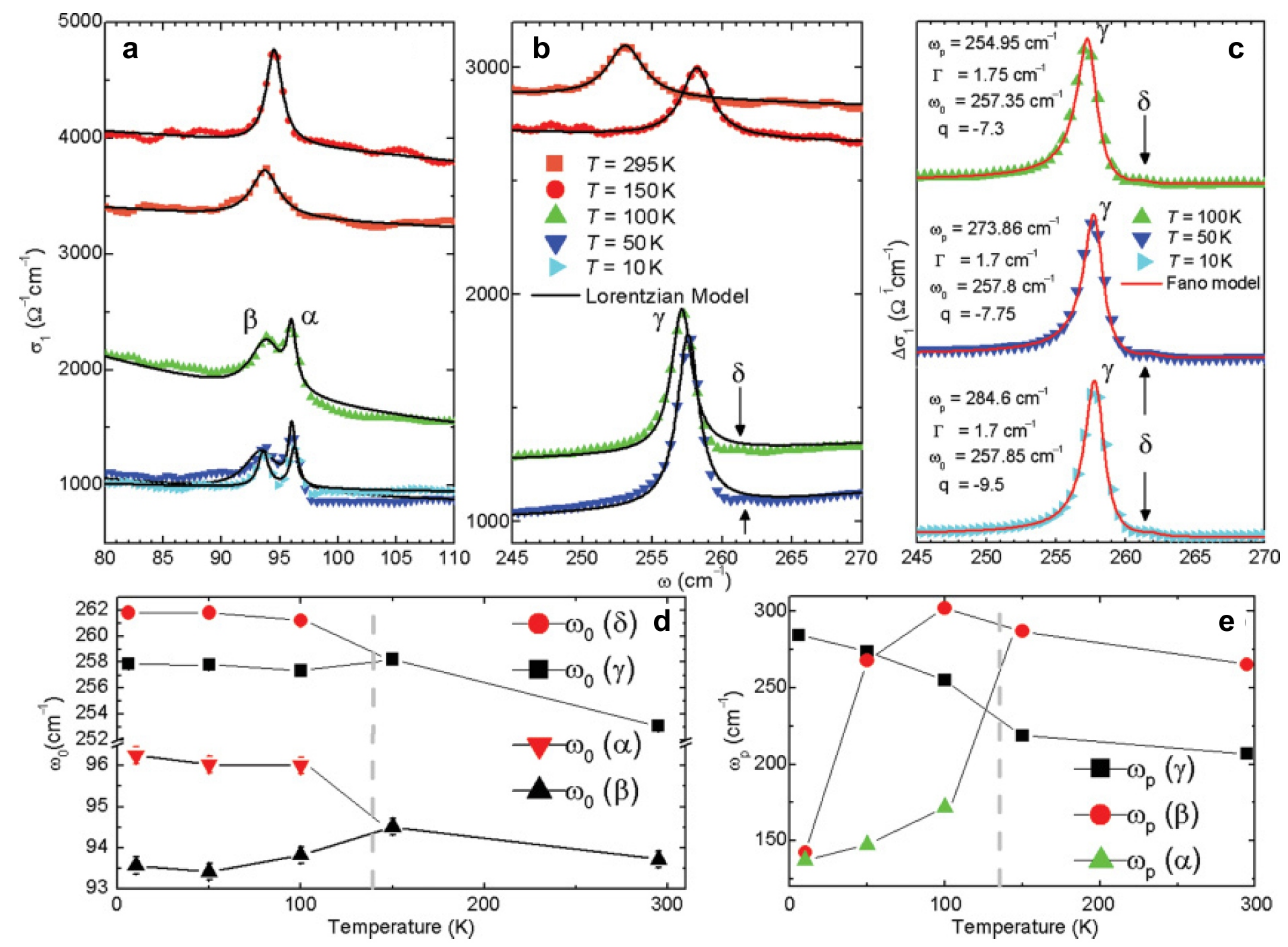}
\caption{\label{fig:bfaphonons}~\textbf{a}, \textbf{b}~Infrared-active phonons due to vibrations involving predominantly \textbf{a}~Ba and \textbf{b}~Fe-As ions at two temperatures above and below the magnetostructural phase transition showing the splitting due to the lowering of the crystallographic symmetry from tetragonal to orthorhombic.~\textbf{c} Analysis of the $260\textrm{-cm}^{-1}$ phonon using the Fano model.~\textbf{d}, \textbf{e} Temperature dependence of the position and strength of the infrared phonons shown in panels~\textbf{a} and~\textbf{b}. Figure reprinted with permission from Ref.~\onlinecite{PhysRevB.84.052501}. Copyright (2011) by the American Physical Society.}
\end{figure}

In contrast to the case of the phase transition, the temperature-induced spectral-weight transfer above the N\'eel temperature in $\textrm{CaFe}_2\textrm{As}_2$ (red solid line in Fig.~\ref{fig:anomalyvis}d) leaves the low-energy itinerant-charge-carrier response largely unaffected and is dominated by the temperature dependence of the $0.6\ \textrm{eV}$ absorption band, ubiquitous in the iron-based materials as can be seen in Fig.~\ref{fig:comp122interband}. The compensating effect due to the suppression of the $3.5\ \textrm{eV}$ absorption band is entirely absent in the normal state and only a very weak decrease in the spectral weight at higher energies takes place, comparable to that across the spin-density--wave transition. The absence of any significant temperature dependence of the $3.5\ \textrm{eV}$ absorption band above $T_{\mathrm{N}}$ is further confirmed by the temperature scan of the imaginary part of the dielectric function at $3.5\ \textrm{eV}$ shown in the inset of Fig.~\ref{fig:anomalyvis}d.

\subsection{Infrared phonons}

The general features of the infrared- and Raman-active phonons allowed by the tetragonal symmetry of the iron-based compounds and observed experimentally have been discussed in detail in Ref.~\onlinecite{Hu2009545}. Therefore, here we will concentrate on the phonon anomalies induced by the antiferromagnetically ordered phase.

The lowering of the crystal symmetry at the virtually simultaneous structural and antiferromagnetic phase transition in the \thcrsi-type iron-based parent compounds must results in the splitting of the two in-plane infrared-active phonon modes of the high-temperature tetragonal phase~\cite{Hu2009545,PhysRevB.80.180502} and has indeed been observed~\cite{PhysRevB.84.052501,PhysRevB.88.184511}. While the splitting of the lower-energy phonon mode involving the intercalating ion can be seen very clearly in the optical conductivity of the parent \bfa\ compound (Fig.~\ref{fig:bfaphonons}a), that of the higher-energy Fe-As vibrational mode is anomalous and features an extreme asymmetry in the intensity of the two split branches, with one of them being hardly distinguishable over the noise floor, as shown in Fig.~\ref{fig:bfaphonons}b. Very careful infrared reflectance measurements reported in Ref.~\onlinecite{PhysRevB.84.052501} and shown in Fig.~\ref{fig:bfaphonons} allowed one to distinguish all in-plane infrared-active phonon modes and trace their splitting upon entering the low-temperature orthorhombic phase. These measurements suggested that the large asymmetry as well as the unusual temperature dependence of the phonon intensities below the N\'eel temperature can only be explained by assuming a highly anisotropic electronic state in the antiferromagnetic phase due to strong Hund's-coupling correlations predicted theoretically~\cite{Yin_NatPhys_BFA_2011}, which would result in a different electronic background and thus screening of ionic vibrations by itinerant charge carries along the direction of the ferro- and antiferromagnetically ordered magnetic moments. This conclusion has been directly confirmed by the polarization-dependent measurements of the optical conductivity of mechanically detwinned \bfca\ samples, reviewed in the next section.

\begin{figure}[b!]
\includegraphics[width=\columnwidth]{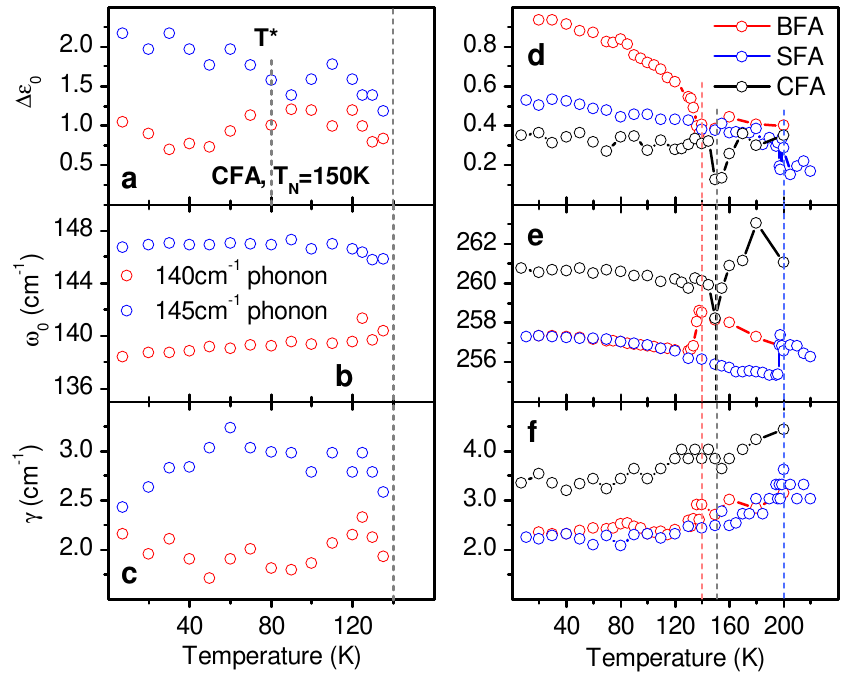}
\caption{\label{fig:phonons}~\textbf{a}--\textbf{c}~Temperature dependence of the strength $\Delta\varepsilon$, position $\omega_0$, and the width $\gamma$ of the Ca-related phonon below the N\'eel transition temperature of $\textrm{CaFe}_2\textrm{As}_2$, split due to the lowering of the crystallographic symmetry from tetragonal to orthorhombic at the concomitant structural transition (right vertical dashed line). The temperature $T^*$ of the small optical spin-density--wave gap $2\Delta_{\mathrm{S}}^{\mathrm{SDW}}$ (left vertical dashed line in~\textbf{a}) as inferred from Fig.~\ref{fig:sdwgaps}\textbf{a}.~\textbf{d}--\textbf{f} Temperature dependence of the strength $\Delta\varepsilon$, position $\omega_0$, and the width $\gamma$ of the phonon due to the vibrations of Fe and As ions. Vertical dashed lines indicate the N\'eel transition temperatures of the Ca-, Sr-, and Ba-based compounds (grey, light blue, and pink, respectively). Figure, caption, and parts of the corresponding discussion in the text reprinted with permission from Ref.~\onlinecite{PhysRevB.88.184511}. Copyright (2013) by the American Physical Society.}
\end{figure}

A very detailed study of the temperature dependence of the intensity, position, and width of the infrared-active phonons of \cfa, \sfa, and \bfa\ compounds and their comparison has been reported in Ref.~\onlinecite{PhysRevB.88.184511} and is presented in Fig.~\ref{fig:phonons}. The low-energy phonon due to the vibrations of the intercalating ions is reported only for the Ca-based parent compound (located at $\approx140\ \textrm{cm}^{-1}$) but not in the Sr- and Ba-based counterparts because for the latter it is located outside the experimentally accessible spectral range. The phonon arising from vibrations of Fe and As is seen at approximately the same position of $260\ \textrm{cm}^{-1}$ in all three materials. In the $\textrm{CaFe}_2\textrm{As}_2$ compound, the splitting of the Ca-related phonon at the magnetostructural transition can be clearly resolved and amounts to about $8\ \textrm{cm}^{-1}$ (Fig.~\ref{fig:phonons}b). Within the noise floor the temperature dependence of the width of these two phonons, plotted in Fig.~\ref{fig:phonons}c, does not appear to display any anomalies. That of the phonon strength $\Delta\varepsilon$, on the other hand, seems to slightly change at the temperature $T^*$ (left vertical dashed line in a) inferred from Fig.~\ref{fig:sdwgaps}a, albeit this change is quite close to the limit of the fit uncertainty.

The temperature dependence of all Lorentz parameters of the Fe-As phonon in all three compounds, Fig.~\ref{fig:phonons}d--f, shows noticeable anomalies at the respective N\'eel temperatures (grey, light blue, and pink vertical dashed lines for \hbox{Ca-}, \hbox{Sr-}, and Ba-based compounds, respectively). In $\textrm{SrFe}_2\textrm{As}_2$ and $\textrm{BaFe}_2\textrm{As}_2$ the phonon intensity $\Delta\varepsilon_0$ and position $\omega_0$ change abruptly at the magnetostructural transition temperature, as shown in Fig.~\ref{fig:phonons}d,~\ref{fig:phonons}e, whereas in $\textrm{CaFe}_2\textrm{As}_2$, quite surprisingly, the spin-density--wave--induced changes seem to set in already at somewhat higher temperatures (Fig.~\ref{fig:phonons}e). This observation indicates the existence of incipient critical lattice strain at temperatures higher than the magnetostructural transition temperature in this compound, which may be related to the electronic nematicity recently discovered in a doped compound of this class~\cite{Davis_Nematicity_BFCA_2013} and reviewed in more detail in the next section. In addition, while the width and the position of the Fe-As phonon in both Sr- and Ba-doped compounds [Fig.~\ref{fig:phonons}e,~\ref{fig:phonons}f] display essentially identical magnitude and behavior, those of the Ca-doped material are significantly larger. While the somewhat broader phonon feature could, in principle, be traced back to the quality of the sample surface or the sample itself, the harder Fe-As phonon of Ca-based compound compared to its Sr- and Ba-based counterparts must be intrinsic. Such a hardening most likely results from the shorter Ca-As and Fe-As bond lengths compared to those in the \hbox{Sr-} and Ba-based compounds and has also been observed in inelastic-neutron-scattering measurements on $\textrm{CaFe}_2\textrm{As}_2$, $\textrm{BaFe}_2\textrm{As}_2$ and predicted by ab-initio calculations~\cite{PhysRevB.79.144516}.

\subsection{Electronic anisotropy and nematicity}

Already very early theoretical investigations of the electronic band structure of the iron-based compounds predicted a strongly anisotropic electronic state in the low-temperature antiferromagnetic phase due to the importance of Hund's-coupling correlations~\cite{Yin_NatPhys_BFA_2011}. Experimentally, however, this issue had received limited attention until the discovery of a pronounced anisotropy in the dc resistivity of detwinned underdoped \bfca\ compounds significantly above the N\'eel transition temperature~\cite{Jiun-HawChu08132010}. This observation has led to an explosion of research into the electronic anisotropy of the antiferromagnetic ground state and the symmetry-breaking fluctuations, which lead to the occurrence of anisotropy already in the high-temperature tetragonal phase above the magnetostructural transition upon application of detwinning pressure/strain, reviewed in this section.

\begin{figure}[t!]
\includegraphics[width=\columnwidth]{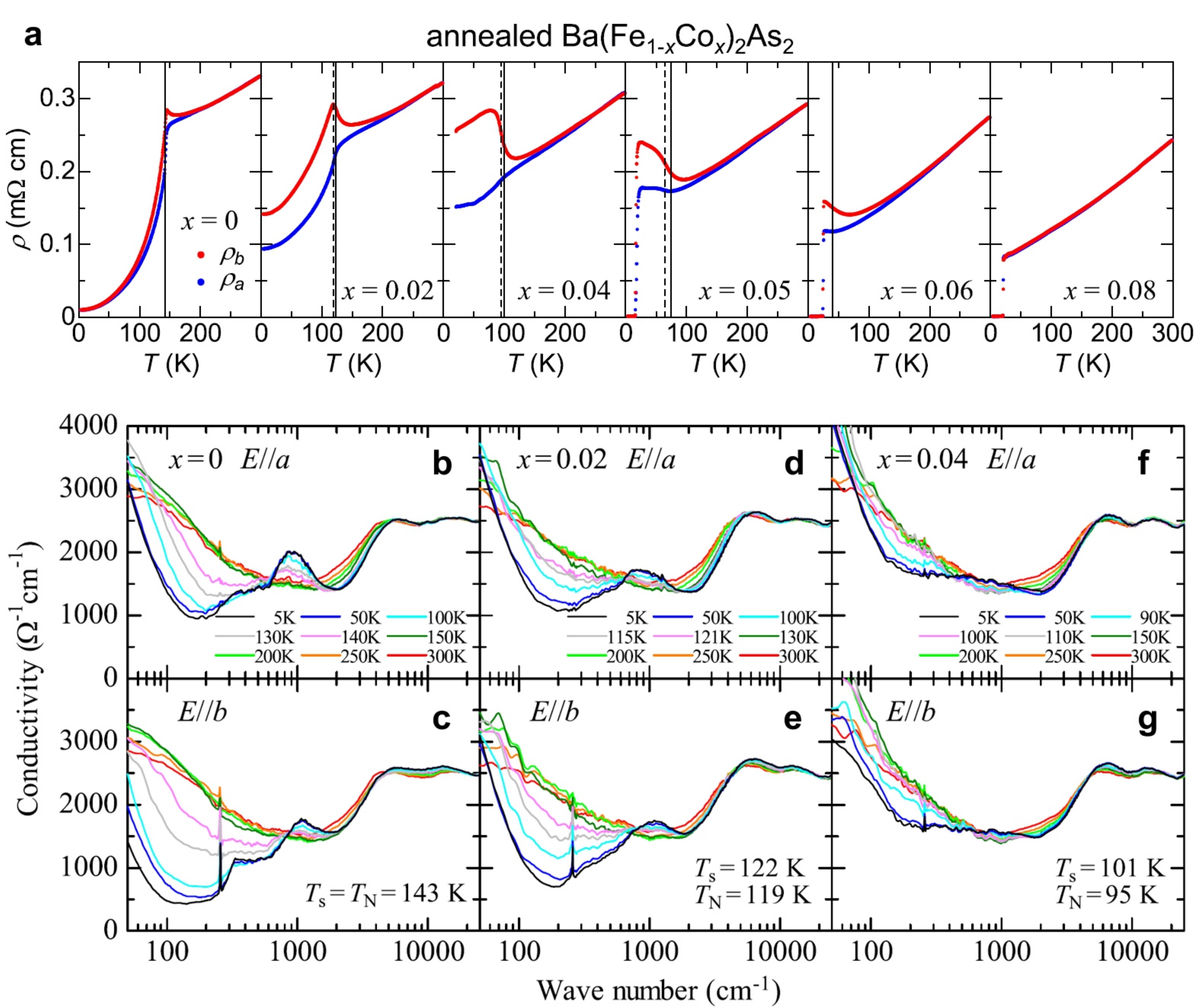}
\caption{\label{fig:anisotropy}~\textbf{a}~Temperature dependence of the dc resistivity along the two in-plane orthorhombic axes ($\rho_a$ and $\rho_b$) in annealed and detwinned single crystals of \bfca\ at different doping levels.~\textbf{b}--\textbf{g}~Real part of the optical conductivity of \bfca\ at several representative temperatures and different doping levels, with the electric field vector of the probing radiation aligned along the two in-plane orthorhombic axes. Panels~\textbf{a} and~\textbf{b}--\textbf{g} reprinted with permission from Refs.~\onlinecite{PhysRevLett.110.207001,PhysRevLett.109.217003}, respectively. Copyright (2012, 2013) by the American Physical Society.}
\end{figure}

The breaking of the fourfold rotational symmetry of the high-temperature tetragonal state at the magnetostructural transition leads to the formation of domains with mutually orthogonal orientation of the now inequivalent in-plane crystallographic axes. Any experimental technique that probes several of such domains at once will only be able to access an average of the inherent electronic response in the two inequivalent directions, producing an essentially isotropic result irrespective of the in-plane orientation. In order to overcome this difficulty, the formation of the antiferromagnetic domains must be hindered by imposing a preferred direction already in the tetragonal phase through the application of stress or strain. This procedure is fraught with several potential difficulties. First of all, the application of strain/stress by itself breaks the fourfold rotational symmetry of the high-temperature tetragonal phase and thus all measurements carried out on thus detwinned samples will not probe the inherent anisotropic electronic response but rather introduce pressure-related effects. Therefore, strain/stress must be released before any measurements are carried out on the detwinned samples. This requirement presents an additional complication for many techniques because of the low N\'eel transition temperatures of the iron-based compounds. Secondly, if too much strain/stress is applied to the sample, irreversible changes might occur, which would permanently alter the properties of the material. It has also been found that annealing drastically reduces the resistivity anisotropy in \bfca\ and makes the low-temperature in-plane charge transport of undoped \bfa\ essentially isotropic~\cite{PhysRevB.84.184514,PhysRevLett.110.207001} and Fig.~\ref{fig:anisotropy}a.

In order to confirm the intrinsic character of their results and to prove that the observed resistive anisotropy is electronic and not driven by the structural transition enhanced by applied strain, the authors of Ref.~\onlinecite{Jiun-HawChu08132010} have designed and successfully implemented a fully electrically controlled detwinning device based on a piezo stage with an additional reference piezo attached to the top surface of the sample. This device enables easy application and removal of strain in a controlled fashion and thus allows one to avoid irreversible damage to the sample as well as complete remove any residual strain~\cite{Fisher_BFCA_Nematicity_2012}. Armed with this elaborate device the authors have revisited their original study and discovered the divergence of the nematic susceptibility under constant strain, i.e. the sensitivity of the electronic anisotropy to the applied stress, which indicates the existence of inherently anisotropic fluctuations in the \bfca\ material. The authors further found that these fluctuations appear strongest near the optimal doping in this compound, where the long-range antiferromagnetic order disappears and the superconducting transition temperature is the highest, which strongly suggests the existence of a quantum critical point beneath the superconducting dome in these compounds.

A different method to avoid averaging of the anisotropic signal due to the existence of twin domains is to use a sample with the dimensions of the order of the size of a single domain, which will create a natural imbalance of domains with a certain orientation in some samples and thus give an overall anisotropic signal. This approach was used in the thermodynamic study of Ref.~\onlinecite{Kasahara_nematicity_BFAP_2012}, which observed a finite anisotropy in the magnetic-torque oscillations above the magnetostructural transition in micron-sized \bfap\ samples.

While the existence of nematic fluctuations in various iron-based compounds has been reliably established, their nature remains unclear. They might originate in the anisotropy of the relative orientation of neighboring spins (anisotropic spin fluctuations) or in the occupation imbalance of the iron $d_{\mathrm{xz}}$ and $d_{\mathrm{yz}}$ orbitals (anisotropic orbital fluctuations)~\cite{PhysRevLett.111.137001}. The former scenario is supported by the observation of a clear scaling relation between nuclear-magnetic-resonance and shear-modulus data, probing magnetic and orthorhombic fluctuations, respectively~\cite{PhysRevLett.111.137001,PhysRevB.87.174507}. On the other hand, a sizable orbital anisotropy has recently been discovered by means of x-ray linear dichroism absorption spectroscopy~\cite{PhysRevLett.111.217001} and angle-resolved photoemission spectroscopy~\cite{Yi26042011} on detwinned \bfca\ samples. This ambivalence indicates an intimate relationship between the spin and orbital degrees of freedom in the iron-based materials~\cite{PhysRevB.80.180418,PhysRevB.83.020505}.

Polarization-dependent measurements of the optical conductivity on mechanically detwinned \thcrsi-type samples have also revealed a strong anisotropy of the electronic state in the antiferromagnetic phase~\cite{Nakajima26072011,PhysRevLett.109.217003,detwinned_BFCA_lobo_2011}, which persists up to unusually high energies of the order of several electron-volts~\cite{Nakajima26072011}. Figure~\ref{fig:anisotropy}b--g shows the infrared conductivity of \bfa\ along the two perpendicular in-plane directions in the low-temperature orthorhombic phase and its evolution upon gradual substitution of cobalt for iron. These polarized infrared measurements, similarly to the unpolarized measurements discussed in the previous sections, reveal the existence of two spin-density-wave--induced gaplike features in the infrared conductivity of the parent \bfa\ compound along {\it both} orthorhombic directions, thus strongly suggesting that the difference in the energy scale of these gaplike features is unrelated to the anisotropy of the magnetic interactions. These measurements further find a coherent and an incoherent contribution to the infrared conductivity along both directions at all investigated doping levels~\cite{Nakajima26072011,PhysRevLett.109.217003,detwinned_BFCA_lobo_2011}. 

Given that in the underdoped regime the residual dc resistivity of \bfca\ is dominated by the coherent component (see Fig.~\ref{fig:anisotropy}), it is instructive to compare their respective anisotropies as a function of cobalt concentration. Such a detailed analysis reveals that the doping dependence of the anisotropy of both the residual dc resistivity and the quasiparticle scattering rate of the coherent term in the optical conductivity is linear up to the doping level of $x=0.04$ and shows a pronouced anisotropy, whereas the spectral weight of the coherent term is isotropic and only weakly depends on the cobalt concentration~\cite{PhysRevLett.110.207001,PhysRevLett.109.217003}. This result implies that the anisotropy of the dc resistivity in the antiferromagnetic phase is extrinsic and is likely driven by the anisotropic impurity potential of randomly distributed cobalt dopants~\cite{PhysRevLett.109.217003,PhysRevLett.110.207001}, polarized by the anisotropic antiferromagnetic environment~\cite{PhysRevB.83.214502}. Such anisotropic impurity states have indeed been directly visualized by means of scatting tunneling spectroscopy in a related isostructural \cfca\ system~\cite{Davis_Nematicity_BFCA_2013}.

The doping level of $x=0.04$ represents a clear turning point in the doping dependence of the dc-resistivity anisotropy: below this level the latter increases with cobalt concentration and decreases above it~\cite{PhysRevLett.110.207001,Jiun-HawChu08132010}. The very same doping level seems to be singular also in the doping dependence of the incoherent contribution to the infrared conductivity, which becomes isotropic by $x=0.04$ and virtually ceases to be affected by the antiferromagnetic phase transition. It is noteworthy that this doping level is very close to the zero-temperature boundary of the superconducting region of the phase diagram~\cite{Jiun-HawChu08132010}. The decrease of the anisotropy of the residual dc resistivity beyond the doping level of $x=0.04$ is consistent with the weakening of the orthorhombic antiferromagnetic phase and the splitting of the iron $d_{\mathrm{xz}}$ and $d_{\mathrm{yz}}$ orbitals~\cite{Yi26042011} in the absence of external pressure towards the doping level at which the structural and N\'eel temperatures reduce to zero. It is, however, worth noting that a recent study of the linear dichroism in the x-ray absorption of the same \bfca\ family of compounds has revealed the existence of anisotropic orbital fluctuations stabilized to long-range orbital order under applied pressure well beyond this doping level (at least up to $10\%$)~\cite{PhysRevLett.111.217001}. Additionally, the divergence of the nematic susceptibility mentioned above was found to be strongly enhanced towards the doping level of $x=0.08$ at which the long-range antiferromagnetic order disappears~\cite{Fisher_BFCA_Nematicity_2012}.

Detailed transport measurements on \bfca, \bfna, and \bfcua\ have attributed the nonmonotonic behavior of the anisotropy of the residual dc resistivity with a peak at $4\%$ cobalt doping and its virtual absence in the parent \bfa\ compound to the existence of a very mobile isotropic subsystem of itinerant charge carriers at zero doping, quickly suppressed upon substitution thus revealing the underlying anisotropy in the transport of the remaining charge carriers, significant even in the parent compound~\cite{PhysRevB.84.054540}. This interpretation is consistent with the spectroscopic observation of several contributions to the coherent component of the optical conductivity in \bfa, one of which was found to be extremely narrow and suggested to originate in the topologically protected from backscattering sheet of the Fermi surface formed by bands with a Dirac-cone--shape dispersion~\cite{Nakajima26072011}.

The cobalt concentration of $4\%$ also marks the reversal in the spin-density-wave--induced modification of the dc resistivity, which goes from experiencing a suppression at the N\'eel temperature below this doping level to being enhanced at higher doping levels, as can be clearly seen in Fig.~\ref{fig:anisotropy}a for both directions. This behavior can be easily explained with the help of the spectroscopic data in Figs.~\ref{fig:anisotropy}b--g. To understand the origin of this phenomenon one needs to remember that the suppression/enhancement of the dc resistivity ($\rho_0=4\pi\gamma/\omega_{\mathrm{pl}}^2\sim\gamma/n$, where $\gamma$ is the total elastic scattering rate, $\omega_{\mathrm{pl}}$ is the plasma frequency of itinerant charge carriers, and $n$ is the electron density) at a spin-density--wave transition is governed by two competing effects, both due to the partial gapping of the Fermi surface: a decrease in the quasiparticle density, which increases the dc resistivity, and a reduction of the quasiparticle scattering rate due to the elimination of some scattering channels, which decreases the dc resistivity. It is quite clear from Fig.~\ref{fig:anisotropy}a that at low doping levels the latter effect dominates and the dc resistivity is significantly reduced below the N\'eel temperature. The corresponding reduction of the quasiparticle scattering rate below the antiferromagnetic transition can be clearly seen in the spectroscopic data in Figs.~\ref{fig:anisotropy}b,c. The aforementioned reversal of the suppression in Figs.~\ref{fig:anisotropy}a can now be explained by the gradual decrease in the intensity of the reduction of the quasiparticle scattering rate at the phase transition upon cobalt doping so that at $x=0.04$ the two factors in the dc resistivity are comparable and at higher doping levels the decrease in the quasiparticle density starts to dominate and the dc resistivity is enhanced below the N\'eel temperature. The assumption that the gapping of the Fermi surface below the spin-density--wave transition temperature indeed changes significantly from $x=0$ to $x=0.04$ is further confirmed by the dramatic reduction of the overall suppression in the optical conductivity with increasing cobalt concentration: at $x=0.04$ not only does the quasiparticle scattering rate not show a significant reduction but the overall temperature dependence of the infrared conductivity is rather weak (see Figs.~\ref{fig:anisotropy}f,g) as compared to that at $x=0$ (Figs.~\ref{fig:anisotropy}b,c). 

The infrared phonon due to the relative in-plane displacement of Fe and As ions located at about $260\ \textrm{cm}^{-1}$ and discussed in the previous section also shows a pronounced anisotropy in the conductivity spectra along the two perpendicular directions in the antiferromagnetic state. Its intensity in the direction of antiferromagnetically ordered ion magnetic moments [$\sigma_{\mathrm{b}}(\omega)$] is strongly enhanced compared with the high-temperature paramagnetic tetragonal phase, whereas that along the ferromagnetic axis [$\sigma_{\mathrm{a}}(\omega)$] reduces almost to the noise level, consistent with the previous observations in twinned single crystals~\cite{PhysRevB.84.052501}. One of the possible explanations for this behavior is the difference in the degree of electronic screening of ionic vibrations in the ferro- and antiferromagnetic directions due to the anisotropic optical conductivity originating in significant Hund's coupling and has been outlined in the previous section. The authors of Ref.~\onlinecite{PhysRevLett.109.217003} put forward an alternative explanation suggesting that this lattice vibration produces anisotropic electronic polarization in the antiferromagnetic state, which almost completely cancels out the dipolar field due to the in-plane displacement of Fe and As ions in the ferromagnetic direction and simultaneously enhances it in the perpendicular direction. This anisotropy of the phonon intensity, most pronounced in the parent \bfa\ compound~\cite{PhysRevLett.109.217003,Nakajima26072011}, gradually decreases upon cobalt substitution, which provides additional evidence that the electronic state in the antiferromagnetic phase becomes progressively less polarizable towards the optimal doping level.

Finally, we would like to note that the $0.6\ \textrm{eV}$ absorption band, discussed in detail above and renormalized by strong Hund's-coupling correlations~\cite{Yin_NatPhys_BFA_2011}, exhibits a pronounced anisotropy in its temperature dependence in the antiferromagnetic state, as can be seen in Figs.~\ref{fig:anisotropy}b--g.

\section{\label{sec:superconductors}Superconducting compounds}

\subsection{Superconducting state}

Superconductivity in the iron-based materials is achieved, in most cases, by iso- or aliovalent substitution into or application of external pressure to the parent antiferromagnetic compounds~\cite{Johnston_Review_2010,0953-8984-22-20-203203,RevModPhys.83.1589}. The superconducting phase occurs once the antiferromagnetic correlations have been sufficiently weakened by one of the aforementioned means and the superconducting transition temperature appears to be maximized at the doping level at which long-range antiferromagnetic order completely disappears and thus the corresponding fluctuations of the iron magnetic moments are the largest. This quite general observation suggests that the exchange of antiferromagnetic spin fluctuations between electrons might mediate their binding into Cooper pairs, similarly to the scenario proposed for the mechanism of superconducting pairing in the copper-based high-temperature superconductors~\cite{0034-4885-66-8-202}. This conclusion has been supported indirectly by the early {\it ab-initio} calculations, which demonstrated that the attractive electron-phonon interaction, responsible for the formation of the superconducting state in all conventional superconductors~\cite{RevModPhys.62.1027}, is too weak in the iron-based compounds to be able to account for their relatively high superconducting transition temperatures~\cite{boeri:026403,Boeri2009628,PhysRevB.82.020506}. Since the interaction due to the exchange of antiferromagnetic spin fluctuations is repulsive, the general form of the gap equation~\cite{Tinkham_superconductivity_1995_articlestyle} maintains that the formation of Cooper pairs can only take place if the superconducting order parameter has opposite signs on different portions of the Fermi surface, in stark contrast to the sign-constant superconducting order parameter in all conventional superconductors~\cite{Superconductivity_Conventional_Unconventional_articlestyle}. One of the clear manifestations of such an unconventional electronic state is the existence of a resonance peak in the energy spectrum of neutron scattering intensity at the wave vector connecting the portions of the Fermi surface with the opposite signs of the superconducting order parameter~\cite{PhysRevB.47.3419,Tinkham_superconductivity_1995_articlestyle}. This resonance peak has indeed been observed in virtually all known iron-based superconductors~\cite{0953-8984-22-20-203203}.

In the presence of only one large sheet of the Fermi surface the alternating-sign character of the superconducting order parameter implies the existence of nodes, i.e. zeros in the superconducting energy gap, as is the case in the $d$-wave copper-based superconductors~\cite{RevModPhys.75.473}. The inherently multiband nature of the iron-based materials, on the other hand, makes a realization of nodeless sign alternation across the Fermi surface possible: the so-called extended $s$-wave or $s_\pm$ symmetry~\cite{2009PhyC469614M}, which complies with the tetragonal symmetry of the underlying lattice, features a different sign of the superconducting order parameter on different sheets of the Fermi surface but requires no nodes. Given that the superconducting energy gap has been found to be nodeless in the majority of the iron-based compounds~\cite{Johnston_Review_2010,1674-1056-22-8-087406,1674-1056-22-8-087407}, it is now widely believed that this extended $s$-wave symmetry is indeed realized in most of the iron-based superconductors, albeit several alternative scenarios have been proposed~\cite{PhysRevLett.104.157001,PhysRevB.81.060504,PhysRevB.83.140512,2013arXiv1307.6119O}. Due to the preservation of the tetragonal symmetry of the lattice, a simple phase-sensitive test similar to the one devised to experimentally establish the $d$-wave symmetry of the order parameter in the cuprate superconductors~\cite{RevModPhys.67.515} cannot be carried out. Nevertheless, several ingenious proposals for the determination of the pairing symmetry in the iron-based compounds based on the multiband semimetallic character of these materials have been put forward~\cite{golubov032601,2013arXiv1312.5930B} and even carried out on one member of the iron-chalcogenide family~\cite{Science328.474}.

\begin{figure*}[t!]
\includegraphics[width=\textwidth]{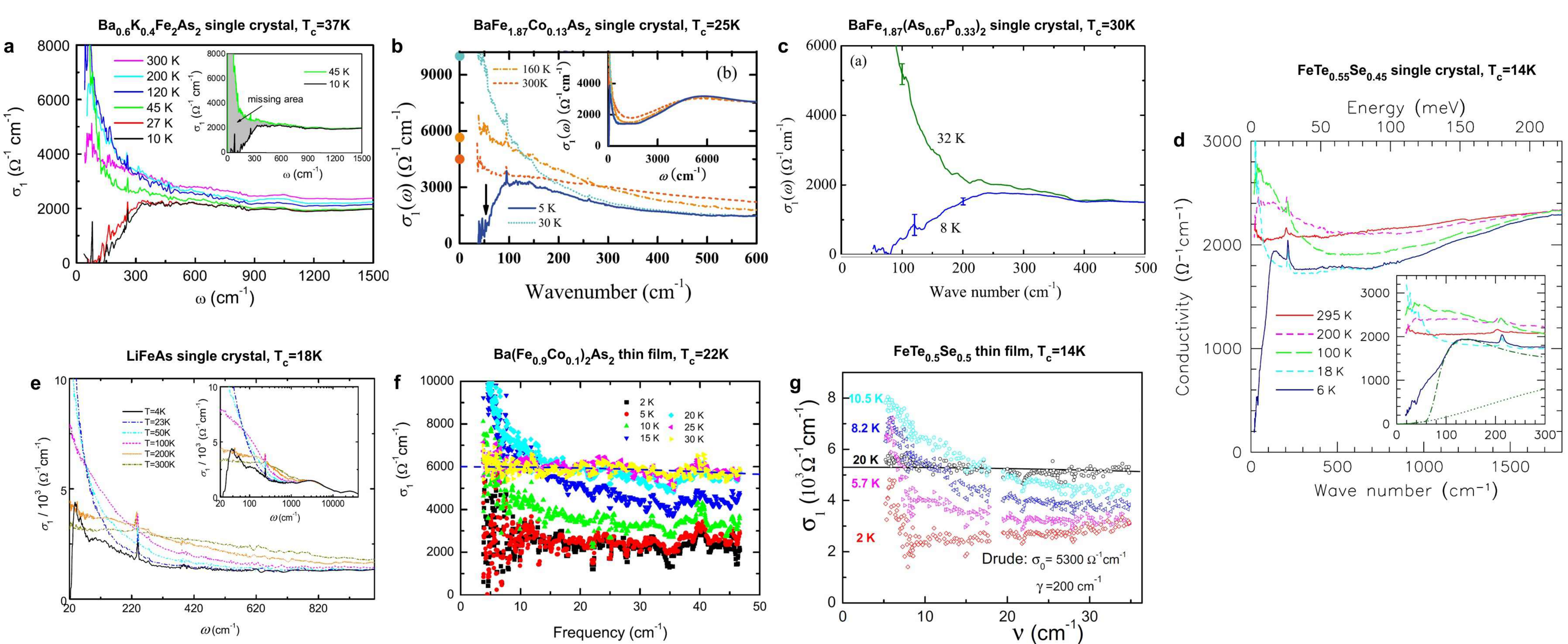}
\caption{\label{fig:superconductivity} Real part of the optical conductivity of single-crystalline \textbf{a}~\bkfaopt, \textbf{b}~$\textrm{BaFe}_{1.87}\textrm{Co}_{0.13}\textrm{As}_2$, \textbf{c}~$\textrm{BaFe}_2(\textrm{As}_{0.67}\textrm{P}_{0.33})_2$, \textbf{d}~$\textrm{FeTe}_{0.55}\textrm{Se}_{0.45}$, \textbf{e}~\lfa\ and thin-film \textbf{f}~$\textrm{Ba}(\textrm{Fe}_{0.9}\textrm{Co}_{0.1})_2\textrm{As}_2$ and \textbf{g}~$\textrm{FeTe}_{0.5}\textrm{Se}_{0.5}$ at several representative temperatures in the far-infrared \textbf{a}--\textbf{e} and terahertz \textbf{f},~\textbf{g} spectral range. The insets show: \textbf{a}~the missing spectral weight between the normal and the superconducting state, \textbf{b},~\textbf{e} the same optical conductivity as in the main panel in a broader spectral range, and \textbf{d} in a narrower spectral range. Panels~\textbf{a}--\textbf{d},~\textbf{f} reprinted with permission from Refs.~\onlinecite{PhysRevLett.101.107004},~\onlinecite{PhysRevB.81.214508},~\onlinecite{PhysRevLett.109.027006},~\onlinecite{PhysRevB.81.180508},~\onlinecite{PhysRevB.82.224507}. Copyright (2008, 2010, 2012) by the American Physical Society. Panels~\textbf{e},~\textbf{g} reprinted with permission from Refs.~\onlinecite{1367-2630-15-7-073029},~\onlinecite{1367-2630-15-1-013032}. Copyright (2013) by the Institute of Physics Publishing Ltd. and Deutsche Physikalische Gesellschaft.}
\end{figure*}

Superconductivity is an electronic instability of the entire Fermi surface, which implies that all quasiparticle excitations are gapped in the superconducting ground state with the exception of the nodal regions, if any. In the case of the iron-based superconductors it amounts to the presence of as many superconducting energy gaps as there are sheets of the Fermi surface of either electronic character. In all materials, on which reliable angle-resolved photoemission measurements have been carried out to date, the existence of up to five superconducting energy gaps has indeed been identified~\cite{1674-1056-22-8-087406,1674-1056-22-8-087407,Johnston_Review_2010}. Quite surprisingly, the specifics of the individual band structure of each particular material notwithstanding, all superconducting gaps have been found to cluster in two groups with approximately the same magnitude (but possibly different signs) within each group but very disparate between them~\cite{PhysRevB.83.214520}. Such clustering has been observed with a variety of experimental probes~\cite{Johnston_Review_2010,0953-8984-22-20-203203,RevModPhys.83.1589} but its origin remains unclear. It appears likely that orbital-specific interactions might play a key role in this phenomenon.

The clustering of the superconducting energy gaps suggests that it may be possible to distill the essence of the complex multiband (up to five bands) electronic structure of the iron-based superconductors into an effective two-band model by identifying the most important electronic degrees of freedom in these materials. Such efforts have indeed borne some fruit the iron-isospin effective model recently argued to capture essential low-energy physics~\cite{PhysRevX.2.021009,1742-6596-449-1-012017,0295-5075-104-5-57007}. Similarly, the analysis of the optical conductivity of a prototypical iron-based superconductor in the framework of the Eliashberg theory has shown that a good semi-quantitative description of the experimental data can be achieved in an effective two-band model~\cite{PhysRevB.84.174511}, obtained via a controlled reduction from a microscopically more accurate four-band model constrained by the aforementioned clustering of the superconducting energy gaps.

Opening of a superconducting gap dramatically affects the low-energy electrodynamics of solids: quasiparticle excitations become eliminated within the energy window of approximately one gap value above and below the Fermi level in all nodeless regions of the Fermi surface. This effect can be clearly identified in the frequency dependence of the complex optical conductivity: due to the conservation of the total number of electrons, the total spectral weight (the area under the real part of the optical-conductivity curve) is temperature independent~\cite{Mahan_Many_particle_Physics_articlestyle} and any loss of it due to the elimination of quasiparticle excitations must be compensated by its gain elsewhere. In conventional superconductors the low-energy spectral weight lost due to the opening of the superconducting gap is transferred into the coherent dc response of the superconducting condensate at zero frequency and can be used to quantify the London penetration depth of a superconductor~\cite{PhysRev.104.844}. Such a redistribution of the overall conserved infrared spectral weight is called the Ferrell-Glover-Tinkham sum rule~\cite{PhysRev.109.1398,Tinkham_superconductivity_1995_articlestyle}. In some exotic mechanisms of superconductivity, the spectral-weight redistribution has been suggested to involve not only infrared frequencies on the order of the superconducting energy gap but also energies much higher than those of the itinerant-charge-carrier response~\cite{PhysRevB.39.11515}. The suppression of the in-plane optical conductivity in the superconducting state due to the gapping of quasiparticles has been identified in all investigated iron-based superconductors with a variety of optical techniques, as illustrated in Fig.~\ref{fig:superconductivity}: single-crystalline \bkfa~\cite{PhysRevLett.101.107004,CharnukhaNatCommun2011,PhysRevB.84.174511,PhysRevB.86.100501,PhysRevLett.111.117001}, single-crystalline~\cite{PhysRevB.81.214508,PhysRevB.82.174509,1367-2630-12-7-073036,PhysRevB.87.014506,PhysRevB.82.184527} and thin-film~\cite{PhysRevB.81.060509,PhysRevB.82.180514,PhysRevB.82.224507,Nakamura2011634} \bfca, single-crystalline \bfap~\cite{PhysRevLett.109.027006}, single-crystalline \lfa~\cite{1367-2630-15-7-073029}, single-crystalline~\cite{PhysRevB.81.180508,Homes2011505} and thin-film~\cite{Nakamura2011634,1367-2630-15-1-013032} \fetese, as well as a static and pump-probe investigation of the infrared reflectance of a \lfaof\ thin film~\cite{PhysRevB.87.180509}. All of these materials have been found to exhibit signatures of multiple superconducting gaps in their infrared optical conductivity, consistent with the inherently multiband character of the iron-based materials and in most cases only two different magnitudes of the superconducting gap have been identified despite the presence of five distinct sheets of the Fermi surface~\cite{Johnston_Review_2010}. The infrared conductivity of the iron-based superconductors in the out-of-plane direction has only been investigated in two compounds, \fetese~\cite{PhysRevLett.106.217001} and \bfca~\cite{PhysRevLett.110.097003}. In a tour-de-force optical study, the latter work demonstrated that although it is notoriously difficult to obtain a fresh surface of good optical quality for infrared measurements in any iron-based material, such a surface, as opposed to its {\it polished} counterpart, is indispensable for the observation of the inherent out-of-plane optical response and the signatures of the superconducting state.

In a single-band $s$-wave weakly coupled conventional superconductor, no infrared absorption is possible at energies smaller than the optical superconducting energy gap $2\Delta$ due to the finite binding energy of the Cooper pairs making up the superconducting condensate~\cite{PhysRev.111.412,Zimmermann199199}, which provides a simple experimental means for the determination of its value from the energy at which the optical conductivity first reduces to zero as a function of decreasing frequency (absorption edge). In many single-crystalline iron-based superconductors such an absorption edge has indeed been observed, as demonstrated in Figs.~\ref{fig:superconductivity}\textbf{a}--\textbf{d}, and the corresponding optical superconducting energy gap $2\Delta$ extracted. Its values for the optimally doped \bkfaopt, $\textrm{BaFe}_{1.87}\textrm{Co}_{0.13}\textrm{As}_2$, $\textrm{BaFe}_2(\textrm{As}_{0.67}\textrm{P}_{0.33})_2$, and~$\textrm{FeTe}_{0.55}\textrm{Se}_{0.45}$ have been found to be approximately $19,\ 6.6,\ 10$, and $<3\ \textrm{meV}$ (in the latter case the absorption edge could not be reached) and agree very well with the values of the superconducting energy gap obtained using other experimental techniques~\cite{Johnston_Review_2010,PhysRevB.83.214520}. Yet in multiband superconductors, such as the iron-based materials, the situation is significantly complicated by the fact that several (up to five) bands contribute to the experimentally observed ac optical conductivity and the latter, strictly speaking, only reaches zero below the smallest optical superconducting energy gap and remains non-zero above it. This complication makes the accurate determination of the superconducting gaps more difficult and model-dependent, thus imparting additional significance to the development of a minimal effective low-energy model with clear justified microscopic underpinnings~\cite{PhysRevB.84.174511}.

\begin{figure}[t!]
\includegraphics[width=\columnwidth]{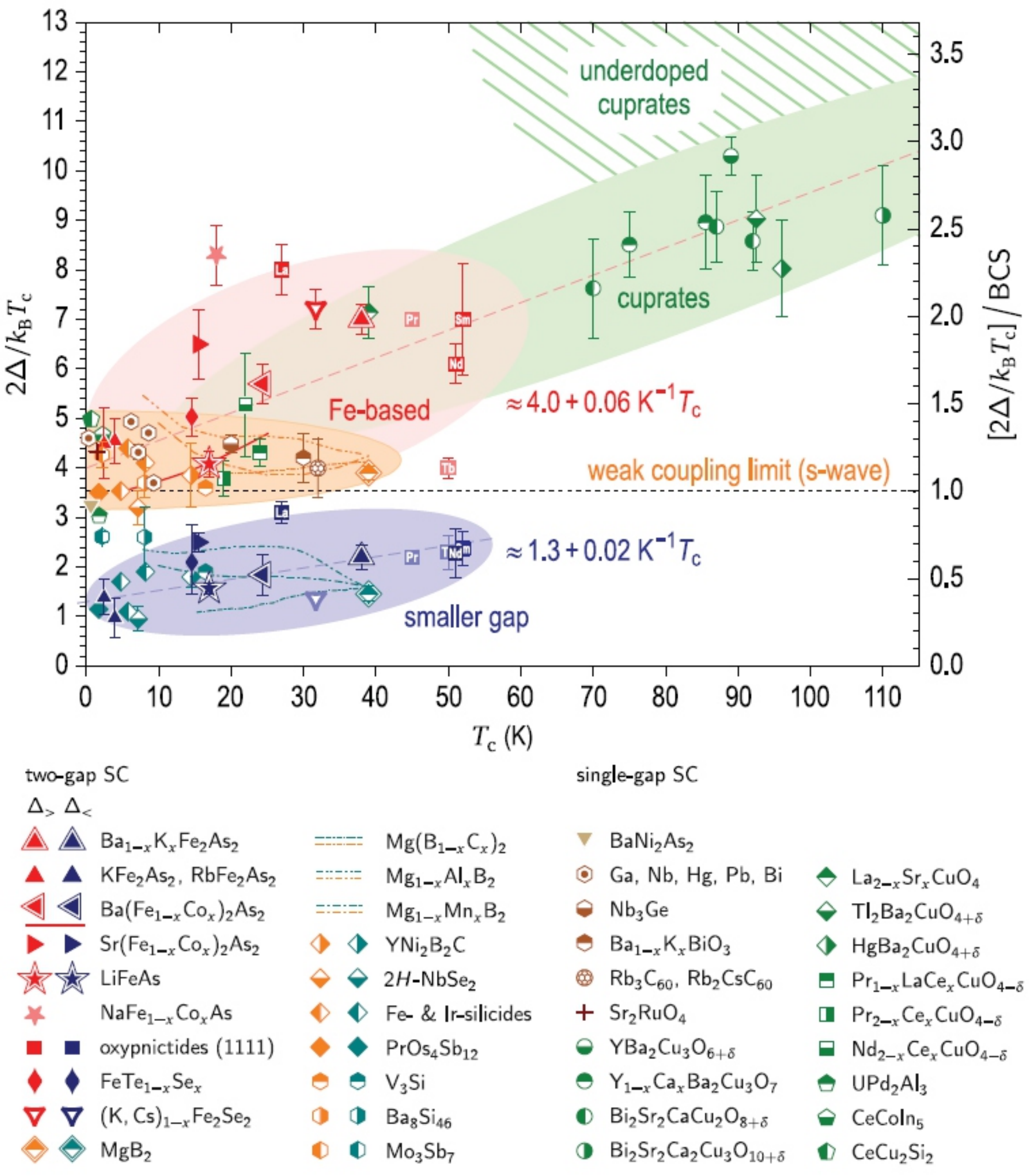}
\caption{\label{fig:superplot}The gap ratios, $2\Delta/k_{\mathrm{B}}T_{\mathrm{c}}$, for different families of single- and two-gap superconductors vs their critical temperatures $T_{\mathrm{c}}$ at ambient pressure. The data points summarize most of the recent energy-gap measurements in ferropnictides, high-$T_{\mathrm{c}}$ cuprates, and some conventional superconductors. Each data point is an average of all the available measurements of the corresponding compound by various complementary techniques (see Tables I--III in Ref.~\onlinecite{PhysRevB.83.214520}). The error bars represent one standard deviation of this average for repeatedly measured compounds or the experimental errors of single measurements, whenever averaging could not be performed. Such unconfirmed points are shown in lighter colors. Points confirmed in a considerable number of complementary measurements are additionally outlined. The weak-coupling limit, predicted for s-wave superconductors by the BCS theory, is shown by the dotted line. For weakly coupled d-wave superconductors, a slightly higher value of $4.12$ is expected (not shown). Figure and caption reprinted with permission from Ref.~\onlinecite{PhysRevB.83.214520}. Copyright (2011) by the American Physical Society.}
\end{figure}

One of the most common ways to extract the values of superconducting energy gaps from optical-conductivity data is to use the weak-coupling Mattis-Bardeen expression for the optical conductivity of a single-band superconductor~\cite{PhysRev.111.412,Zimmermann199199} obtained in the framework of the Bardeen-Cooper-Schrieffer theory of superconductivity~\cite{PhysRev.108.1175} or a linear combination thereof for multiband materials. The values of the superconducting energy gaps of various iron-based superconductors extracted using this approach can be found in the original works cited above and have also been listed and compared with those obtained using various experimental techniques in numerous superconducting compounds including conventional, heavy-fermion, and high-temperature copper-based superconductors in a comprehensive review presented in Ref.~\onlinecite{PhysRevB.83.214520} and summarized in Fig.~\ref{fig:superplot}.

In the case of the iron-based superconductors, however, the application of the Mattis-Bardeen expression or a combination thereof to describe the optical response in the superconducting state appears questionable. First of all, as already mentioned, this expression has been derived based on the weak-coupling microscopic theory of superconductivity developed by Bardeen, Cooper, and Schrieffer and, therefore, is strictly speaking only applicable to the description of the optical response of weakly coupled superconductors. The coupling strength in the iron-based materials has, on the other hand, been shown to span a wide range from relatively weak to strong coupling, with a clear trend towards intermediate/strong coupling~\cite{PhysRevB.83.214520}. Secondly, the Mattis-Bardeen expression applies only to a single-band superconductor. It can, nevertheless, be readily generalized to the case of several bands, as has been demonstrated in Ref.~\cite{PhysRevLett.3.552}. However, this generalization is based on the same assumptions as the BCS theory itself, thus remaining in the weak-coupling regime also with respect to the coupling {\it between} different bands, while it is now widely accepted that in many iron-based materials the interband coupling between at least two of the bands is quite strong, as evidenced by the peak in the magnetic susceptibility at the wave vector connecting bands in the center and in the corner of the Brillouin zone, probed by inelastic neutron scattering~\cite{Johnston_Review_2010}, as well as an enhanced quasiparticle relaxation between analogous bands identified by means of time- and angle-resolved photoemission spectroscopy~\cite{2010arXiv1008.1561R}. Finally, the Mattis-Bardeen expression is only applicable to the case of impurity scattering much larger than the corresponding coupling strength, i.e. in the limit of high impurity concentration. Some of the iron-based materials, like \bfca, can indeed be expected to have relatively high levels of elastic impurity scattering, while others, like \bkfa\ and \bfap, are considered to possess quite long quasiparticle mean free paths at least in some of the bands~\cite{PhysRevLett.104.057008,PhysRevLett.105.267002,PhysRevB.84.174511,0034-4885-74-12-124507,PhysRevLett.110.257002}. The above arguments show that while the application of the Mattis-Bardeen expression to some of the iron-based superconductors (namely, for weakly coupled compounds with weak interband scattering and a relatively high impurity concentration) could be partially justified, any adequate description of others requires a more complete theory of superconductivity, which overcomes the above limitations~\cite{PhysRevB.84.174511}.

\begin{figure}[t!]
\includegraphics[width=\columnwidth]{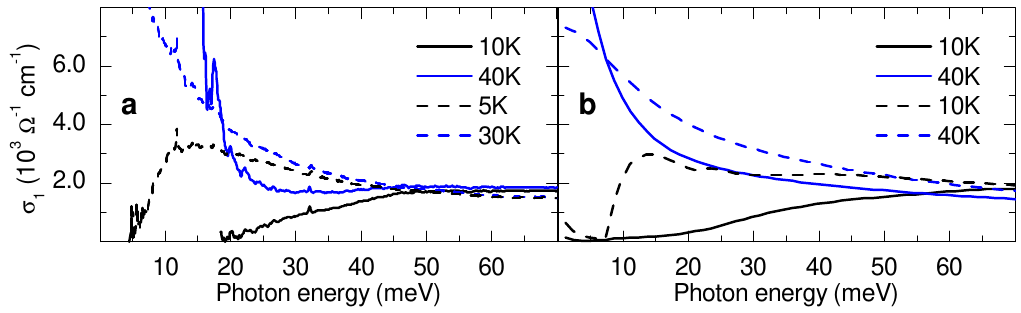}
\caption{\label{fig:compbkfabfca}~\textbf{a} Real part of the optical conductivity of the optimally doped \bkfaoptmpi\ (solid lines) and $\textrm{BaFe}_{1.87}\textrm{Co}_{0.13}\textrm{As}_2$ (dashed lines) deep in the superconducting state at $10$~and $5~\textrm{K}$, respectively (black lines) and just above the superconducting transition temperature at $40$~and $30~\textrm{K}$, respectively (blue lines).~\textbf{b} The corresponding prediction of the Eliashberg theory of superconductivity as described in the text (line style and color same as in~\textbf{a}; black lines at $10~\textrm{K}$, blue lines at $40~\textrm{K}$). The data for \bkfaoptmpi\ and $\textrm{BaFe}_{1.87}\textrm{Co}_{0.13}\textrm{As}_2$ in panel~\textbf{a} reprinted with permission from Refs.~\onlinecite{PhysRevB.84.174511,PhysRevB.81.214508}. Copyright (2011, 2010) by the American Physical Society. Panel~\textbf{b} reprinted with permission from Ref.~\onlinecite{PhysRevB.84.174511}. Copyright (2011) by the American Physical Society.}
\end{figure}

The most sophisticated theory of superconductivity to date is the so-called Eliashberg theory of superconductivity, applicable at arbitrary coupling strengths both within and between all of the bands and at arbitrary impurity concentrations~\cite{PhysRev.156.470,PhysRev.156.487}. Figure~\ref{fig:compbkfabfca} shows a comparison of the experimental infrared conductivity of optimally doped \bkfa\ and \bfca\ superimposed onto the prediction of an effective two-band Eliashberg theory of superconductivity obtained in a reduction from a microscopically more justified four-band theory constrained by the aforementioned clustering of the superconducting energy gaps into two groups by magnitude~\cite{PhysRevB.84.174511}. It can clearly be seen that the qualitative differences in the shape of the infrared conductivity between the two compounds are captured in the same theory, with only two parameters adjusted: the intraband impurity scattering rate and the plasma frequencies of the bands. The former change can be justified by the different location of the dopants, namely, in the superconducting iron planes in the case of \bfca, which is expected to significantly disturb the coherent superconducting transport taking place in these planes, as compared to the substitution of the intercalating ion relatively far from the iron planes in \bkfa~\cite{Johnston_Review_2010}. The change of the plasma frequencies of the bands is to be expected due to the doping with charge carriers of opposite signs in these two cases. The above theory assumes that superconductivity in \bkfaoptmpi\ is mediated by a boson with a characteristic energy of about $13\ \textrm{meV}$, consistent with the peak energy of the resonance mode in the magnetic susceptibility observed by means of inelastic neutron scattering~\cite{Osborn_INS_BKFA_2008} and thus with superconducting pairing mediated by antiferromagnetic spin fluctuations. The effect of various model parameters on the infrared conductivity predicted by the effective two-band Eliashberg theory used above can be analyzed in an \href{http://prb.aps.org/epaps/PRB/v84/i17/e174511/bkfafir.jar}{interactive simulation} in the Supplementary Material of Ref.~\onlinecite{PhysRevB.84.174511}.

We can now apply the above reasoning to compare the shape of the optical conductivity across the entire family of the iron-based superconductors for which spectroscopic data in the superconducting state are available, shown in Fig.~\ref{fig:superconductivity}. It is immediately evident that the shape of the optical conductivity of optimally doped \bfap\ (Fig.~\ref{fig:superconductivity}\textbf{c}) in the superconducting state compared to that in the normal state just above the superconducting transition temperature is very similar to the case of \bkfaopt\ shown in Fig.~\ref{fig:superconductivity}\textbf{a}: down to a certain characteristic energy the superconductivity-induced suppression of the optical conductivity is small, while below that energy there is a gradual quasilinear fall-off in the low-temperature infrared conductivity until it reaches essentially zero, to the accuracy of the corresponding reflectance measurements. Based on this comparison one can conclude that optimally doped \bfap\ must likewise be at least moderately coupled superconductor with very low levels of elastic impurity scattering. The coupling strength can be estimated by the ratio of the value of the largest superconducting energy gap obtained, e.~g., by means of angle-resolved photoemission spectroscopy $2\Delta\approx8\ \textrm{meV}$~\cite{Feng_BFAP_ARPES_2012} and the superconducting transition temperature of about $30\ \textrm{K}$ at optimal doping: $2\Delta/k_{\mathrm{B}}T_{\mathrm{c}}\approx6.5$ --- very close to that of optimally doped \bkfa\ but with even lower superconducting transition temperature, see Fig.~\ref{fig:superplot}. Low levels of impurity scattering are eloquently confirmed by the observation of de~Haas--van~Alphen quantum oscillations, notoriously sensitive to disorder, starting from the terminal composition \bfp\ down to almost optimally doped \bfap~\cite{PhysRevLett.104.057008,0034-4885-74-12-124507,PhysRevLett.110.257002} as well as by the analysis of the collective vortex pinning in this material at optimal doping~\cite{PhysRevLett.105.267002}. Additionally, and similarly to the comparison of \bkfa\ and \bfca\ above, substitution in \bfap\ occurs at the arsenic sites away from the iron planes in which the coherent superconducting transport takes place.

The shape of the optical conductivity of \fetese\ in the superconducting state compared to that in the normal state (Fig.~\ref{fig:superconductivity}\textbf{d}), on the other hand, is of the Mattis-Bardeen type and closely resembles that of optimally doped \bfca. In view of the discussion above this suggests that impurity scattering significantly perturbs coherent quasiparticle transport in the superconducting state. This conclusion is consistent with the very high doping levels required to achieve the maximal superconducting transition temperature, at which virtually every second tellurium atom is substituted by selenium. Sizable quasiparticle scattering is further confirmed by the clear observation of quasiparticle-interference patterns by means of scanning tunneling microscopy in this material~\cite{Science328.474}.

In the case of \lfa, yet again, the shape of the optical conductivity is of the Mattis-Bardeen type (Fig.~\ref{fig:superconductivity}\textbf{e}). At first glance, this observation might lead one to conclude that, similarly to \bfca\ and \fetese, impurity scattering rate is relatively high. However, even taking into account the well-known difficulty to control the lithium stoichiometry in \lfa~\cite{ZAAC:ZAAC19683610107,Chu2009326}, recent scanning tunneling microscopy studies have shown that in the best samples the concentration of impurities can be very low~\cite{PhysRevB.85.214505}. Therefore, it is more likely that in the case of \lfa\ (assuming that the samples investigated by the authors of Ref.~\onlinecite{1367-2630-15-7-073029} were of comparable purity) the Mattis-Bardeen, rather than quasilinear, shape of the optical conductivity in the superconducting state results from the overall much weaker coupling suggested by the relatively small gap ratio $2\Delta/k_{\mathrm{B}}T_{\mathrm{c}}$ for the largest superconducting gap, located close to the BCS limit, as shown in Fig.~\ref{fig:superplot}.

\begin{figure}[t!]
\includegraphics[width=\columnwidth]{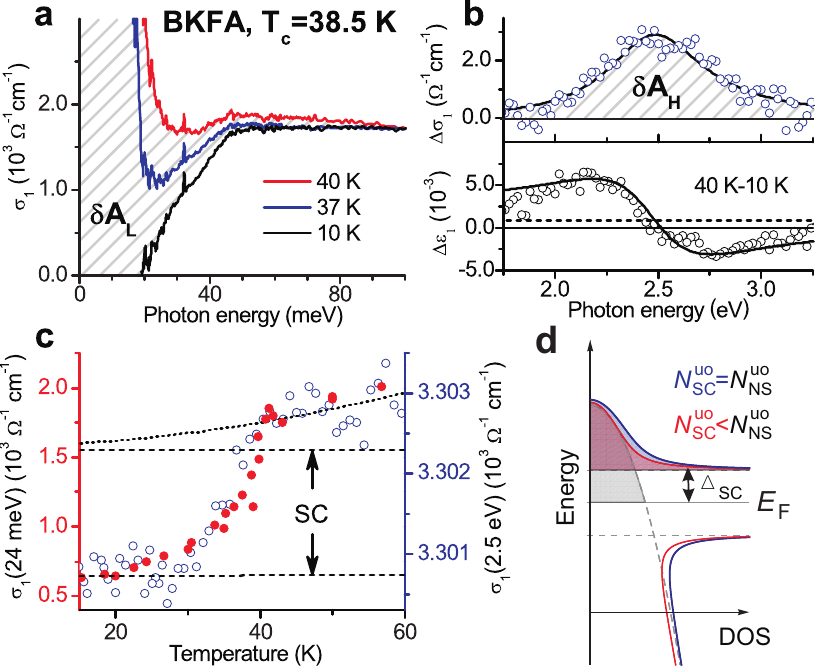}
\caption{\label{fig:bkfacoloring}~\textbf{a}~Real part of the far-infrared optical conductivity of \bkfaoptmpi\ and the corresponding missing area.~\textbf{b}~Difference spectra of the real part of the optical conductivity (top panel) and dielectric function (bottom panel) between $40$ and $10~\textrm{K}$, with a small background shift (horizontal dashed line) detected by temperature modulation measurements. Lorentzian fit to both spectra (black solid lines).~\textbf{c}~Temperature scan of $\sigma_1$ at $2.5~\textrm{eV}$ (blue) and $24\ \textrm{meV}$ (red). Contribution of the normal-state dynamics (dotted line) was estimated to determine the magnitude of the SC-induced jump (dashed lines).~\textbf{d}~Density of states in the normal (NS; grey dashed line), conventional superconducting state (SC; blue line), and a superconducting state with a depletion of the unoccupied states (UO; red line). Filled areas of respective colors represent the total number of unoccupied states. Figure and caption adapted by permission from Macmillan Publishers Ltd: Nature Communications Ref.~\onlinecite{CharnukhaNatCommun2011}, copyright (2011).}
\end{figure}

Another set of instructive spectroscopic optical-conductivity data complementary to those discussed above has been obtained on thin films of two representative iron-based superconductors, \bfca\ and \fetese, as shown in Fig.~\ref{fig:superconductivity}f,g, by means of terahertz transmission spectroscopy. This technique extends the spectral range covered by the traditional reflectance techniques and ellipsometry down to $\sim5\ \textrm{cm}^{-1}$ and is sensitive to both the amplitude and phase of the transmitted signal, thus providing essential information about low-energy itinerant charge transport in the normal and superconducting state. The terahertz optical conductivity of both compounds shows several important common features. First of all, the superconductivity-induced suppression of the optical conductivity due to the opening of the superconducting energy gap is evident in both materials but, unlike in the case of single-crystalline compounds discussed above, the optical conductivity of thin-film samples has never been found to reach zero at any achievable terahertz frequency and always features some virtually frequency-independent contribution on the order of $1500$--$2500\ \Omega^{-1}\textrm{cm}^{-1}$~\cite{PhysRevB.81.060509,PhysRevB.82.180514,PhysRevB.82.224507,Nakamura2011634,1367-2630-15-1-013032}. This residual conductivity has been argued to originate either in strong pair-breaking impurity scattering~\cite{PhysRevB.82.180514} or in the presence of nodes in the superconducting energy gap~\cite{PhysRevB.82.224507}. As pointed out by the authors of Ref.~\onlinecite{PhysRevB.82.224507}, the former mechanism appears inconsistent with the clear observation of a coherence peak in the optical conductivity of both compounds~\cite{PhysRevB.81.060509,PhysRevB.82.180514,PhysRevB.82.224507,1367-2630-15-1-013032}, which is expected to be strongly suppressed by pair-breaking impurity scattering~\cite{PhysRevB.44.7741}. While the existence of nodes in the \bfca\ compound cannot be fully ruled out at present, it seems to be at odds with the experimental data obtained by means of angle-resolved photoemission spectroscopy~\cite{Terashima05052009} as well as some terahertz transmission measurements~\cite{PhysRevB.81.060509,PhysRevB.82.180514}. In \fetese\, the existence of nodes is incompatible with the clear observation of a U-shaped rather than a V-shaped tunneling spectrum in the superconducting state by scanning tunneling spectroscopy~\cite{Science328.474}, unambiguously proving the absence of quasiparticle excitations below the superconducting energy gap beyond thermal excitation~\cite{Science328.474}. This evidence indicates that neither the strong pair-breaking impurity scattering nor the existence of nodes provides a consistent explanation of the aforementioned residual conductivity and that it could, in fact, be extrinsic. For instance, such residual optical conductivity could come from absorption in an intermediate non-superconducting metallic layer between the substrate and the superconducting thin film. Indeed, strong evidence for the spontaneous formation of an intermediate several-nanometer-thick layer most likely composed of Fe atoms (along with possible inclusions in the bulk) in thin films grown using pulsed laser deposition has been reported for many different growth conditions~\cite{PhysRevB.81.100507,PhysRevB.83.134514,APEX.3.043102,0034-4885-77-4-046502}. Given that all of the above terahertz-transmission studies have been carried out on the thin films of iron-based superconductors grown by this method, the extrinsic character of the residual conductivity appears likely.

Notwithstanding the significant residual conductivity observed in transmission spectroscopy of the thin films of \bfca\ and \fetese, several features of superconductivity could be clearly identified in these measurements. The authors of Ref.~\onlinecite{PhysRevB.81.060509} succeeded in directly observing one of the optical superconducting energy gaps $2\Delta\approx3.7\ \textrm{meV}$ as the energy at which the real part of the terahertz conductivity drops to the background level. Furthermore, virtually all terahertz transmission studies have reported the presence of a coherence peak in the real part of the optical conductivity, which manifests itself as a slight enhancement of the optical conductivity at low enough energies just below the superconducting transition temperature as compared to its overall suppression at higher energies and lower temperatures. This behavior is a hallmark of superconductivity, very well understood in the conventional BCS and Eliashberg theory of superconductivity, and stems from the formation of highly coherent superconducting pairs~\cite{Tinkham_superconductivity_1995_articlestyle} and/or the elimination of quasiparticle scattering channels below the optical superconducting energy gap~\cite{PhysRevB.44.7741}. It must be noted, however, that the existence of a coherence peak in the low-energy optical conductivity is a sufficient but not a necessary condition for superconductivity as in certain circumstances it can be partially or fully suppressed~\cite{PhysRevB.44.7741,PhysRevB.80.174526}.

According to the Ferrell-Glover-Tinkham sum rule~\cite{PhysRev.109.1398,Tinkham_superconductivity_1995_articlestyle}, the missing area under the infrared conductivity curve as compared to that in the normal state is transferred into the coherent response of the superconducting condensate at zero frequency. It manifests itself in the most fundamental properties of superconductors: zero dc resistivity and Meissner effect. The latter is characterized by the London penetration depth $\lambda_{\mathrm{L}}$ for the magnetic field, which is related to the missing area $SW$ via $\lambda^{-2}_{\mathrm{L}}=8SW/c^2$ (in CGS units), where $c$ is the speed of light. The missing area obtained from the integration of the optical conductivity of the iron-based superconductors typically gives a London penetration depth on the order of $2000$--$3500$~\AA, consistent with other measurements~\cite{Johnston_Review_2010}.

Intriguingly, the superconductivity-induced suppression of the optical conductivity with non-zero missing area has also been discovered in optimally doped \bkfaoptmpi\ in the visible spectral range~\cite{CharnukhaNatCommun2011}, at energies two orders of magnitude larger than the superconducting energy gap, as demonstrated in Figs.~\ref{fig:bkfacoloring}\textbf{a}--\textbf{c}. This observation challenges the conventional theories of superconductivity, which only allow for the occurrence of the missing area at energies of the order of the superconducting energy gap~\cite{PhysRev.111.412,PhysRev.156.487,Tinkham_superconductivity_1995_articlestyle}. Some exotic theories of superconductivity have proposed a mechanism by which the superconductivity-induced suppression of the optical spectral weight could extend to the spectral range of interband transitions~\cite{PhysRevB.16.2437,Hirsch1992305,PhysRevB.62.15131}. The authors of Ref.~\onlinecite{CharnukhaNatCommun2011} have demonstrated, however, that the observed effect can be understood phenomenologically in conventional terms by utilizing the inherently multiband nature of the iron-based superconductors. The suppression of an interband transition over its entire width with a local non-conservation of the spectral weight implies a change in the oscillator strength of this transition, which can take place due to the modification of the occupation in either initial or final states that its spans. Since all conventional superconductivity-induced modifications of the electronic structure and dynamics occur in the vicinity of the Fermi level within several values of the superconducting energy gap, one may conclude that the only conventional way for an interband transition to feel the effect of superconductivity is for its initial or final states to be largely concentrated in this vicinity. However, this mechanism by itself would not entail any loss of spectral weight in the considered interband transition because the opening of the superconducting energy gap leads to the expulsion of electronic states from within the gap to above it with the total number of states conserved (illustrated in Fig.~\ref{fig:bkfacoloring}\textbf{d} with gray and blue lines and areas). Therefore, the corresponding modifications in the interband optical conductivity would be limited to small shifts/corrugations on the order of the superconducting energy gap~\cite{Dobryakov1994309} with the total spectral weight conserved. In the present case, on the other hand, it is then required that the number of states in the superconducting state be unequal to that in the normal state, as shown in Fig.~\ref{fig:bkfacoloring}\textbf{d} with the red line and area. In a conventional single-band superconductor such a scenario would be impossible and exotic mechanisms would have to be invoked to explain the observed suppression. The situation changes drastically in the case of multiband superconductivity. In the iron-based superconductors, due to the presence of multiple sheets of the Fermi surface, some of which are relatively strongly coupled to each other~\cite{Johnston_Review_2010}, a population redistribution could and does indeed occur dynamically between the strongly coupled sheets of the Fermi surface~\cite{PhysRevLett.108.097002,2010arXiv1008.1561R} and could lead to a static population imbalance in the superconducting state should it be energetically favorable. This latter condition can be fulfilled in a multiband superconductor due to the well-known, albeit very small, effect of the change in the chemical potential upon entering the superconducting state~\cite{parks_superconductivity_1969_vol1_articlestyle,PhysRevB.46.14245}, proportional to the square of the superconducting energy gap. In case a material features different magnitudes of the superconducting energy gap on different sheets of the Fermi surface and/or a different character of the charge carriers (electrons vs. holes), there will be a band- or orbital-specific shift of the chemical potential leading to a population redistribution between different bands. None of the existing theories of superconductivity takes this effect into account self-consistently. Albeit the missing area in the visible spectral range shown in Fig.~\ref{fig:bkfacoloring}\textbf{b} constitutes only a small fraction of that in the far-infrared (Fig.~\ref{fig:bkfacoloring}\textbf{a}), a simple estimate in the framework of the tight-binding nearest-neighbour approximation~\cite{PhysRevB.16.2437,Hirsch1992305} shows that it would entail a reduction of the electronic kinetic energy of $0.60\ \textrm{meV/unit cell}$ in the superconducting state~\cite{CharnukhaNatCommun2011}, should it contribute to superconducting pairing. This is close to the condensation energy $\Delta F(0)=0.36\ \textrm{meV/unit cell}$ obtained from specific-heat measurements on the same sample~\cite{PhysRevLett.105.027003}. It thus remains to be seen whether the physics behind this phenomenon is of consequence for superconductivity of multiband materials.  

\subsection{Spin fluctuations and quantum criticality}

Arguably the most remarkable feature of the phase diagram of the iron-based superconductors is the proximity of the superconducting phase to antiferromagnetic order~\cite{0953-8984-22-20-203203,Johnston_Review_2010}. With the exception of very few compounds (such as 1111-type materials and \efa), the superconducting transition temperature is maximized close to the zero-temperature end point of the antiferromagnetic/structural phase transition line underneath the superconducting dome, at which the long-range antiferromagnetic order continuously disappears and the corresponding antiferromagnetic spin fluctuations are expected to be strongest. This observation immediately suggested that superconductivity in the iron-based superconductors might be mediated by the quantum-critical fluctuations of the antiferromagnetic order parameter, similarly to the mechanism proposed for their high-temperature copper-based~\cite{0034-4885-66-8-202,RevModPhys.75.913,KeimerSachdev_QC_2011,Sachdev_QPT_2011_articlestyle} and heavy-fermion counterparts~\cite{Gegenwart_HeavyFermions_review}. One of the expected manifestations of the quantum-critical regime is the non-Fermi-liquid linear temperature dependence of the dc resistivity in a funnel-shaped region of the phase diagram originating at the quantum critical point, as compared to the quadratic low-temperature behavior of the conventional Fermi liquid~\cite{KeimerSachdev_QC_2011}, due to quasiparticle scattering dominated by antiferromagnetic spin fluctuations (marginal Fermi liquid)~\cite{PhysRevLett.63.1996}. Such behavior has indeed been identified in \bfap~\cite{HusseyFisher_TransportQCP_Tlinear_2014,2013arXiv1308.6133N} along with a sharp peak in the zero-temperature London penetration depth near the doping level of the quantum critical point~\cite{Hashimoto_Prozorov_QCPinBFAP_2012}, which can be traced back to the enhancement of the quasiparticle effective mass, previously observed by means of de~Haas--van~Alphen measurements in the same material~\cite{PhysRevLett.104.057008,0034-4885-74-12-124507,PhysRevLett.110.257002}, by critical antiferromagnetic spin fluctuations~\cite{PhysRevLett.110.177003}. 

\begin{figure}[t!]
\includegraphics[width=\columnwidth]{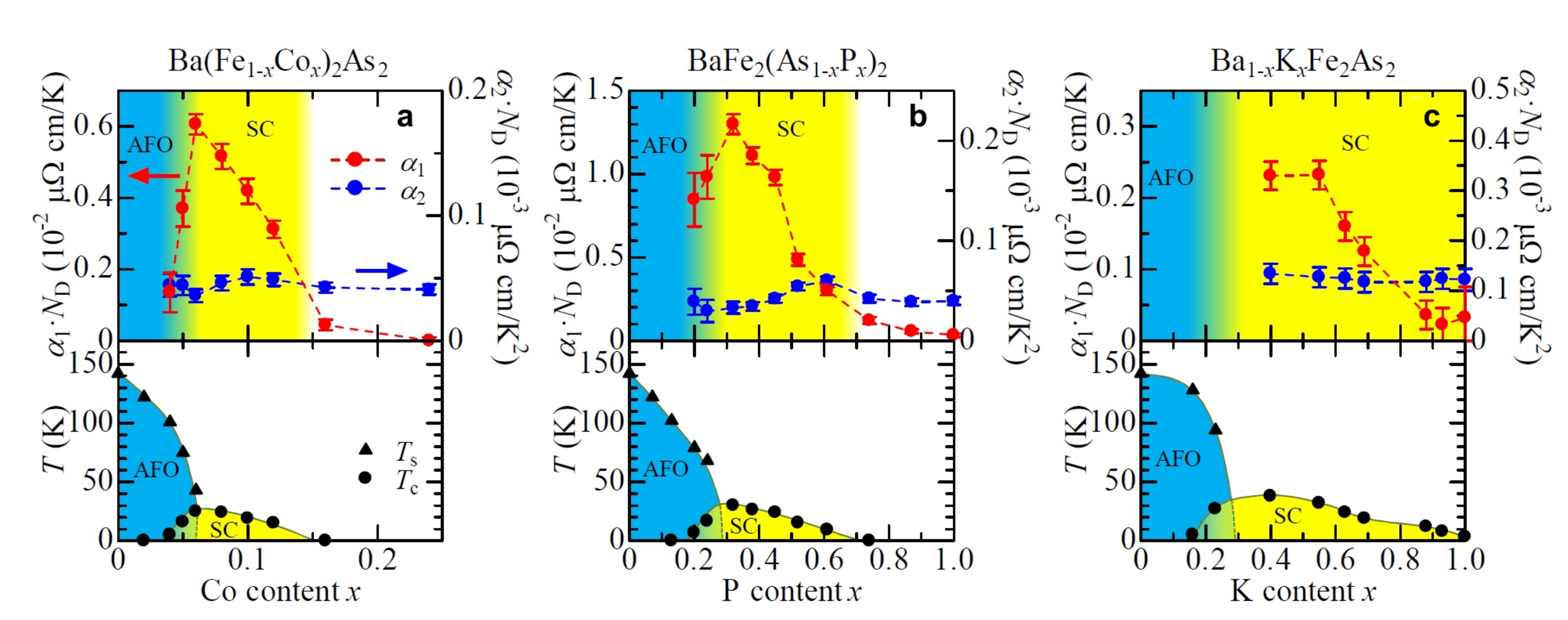}
\caption{\label{fig:tlinearcomp} Doping evolution of the coefficients of the $T$-linear and $T^2$ components ($\alpha_1\cdot N_{\mathrm{D}}$ and $\alpha_2\cdot N_{\mathrm{D}}$, respectively) for~\textbf{a} Co-
Ba122,~\textbf{b} P-Ba122, and~\textbf{c} K-Ba122 determined from the fitting shown in Fig.~4 of Ref.~\onlinecite{2013arXiv1308.6133N}. The $T= 0$ electronic phases are color-coded: AFO in blue, SC in yellow, AFO-SC coexisting region in gradation, and non-SC metallic in white. Lower panels show the electronic phase diagrams for each system. Figure and caption reprinted with permission from Ref.~\onlinecite{2013arXiv1308.6133N}. Copyright (2013) by the American Physical Society.}
\end{figure}

Similar $T$-linear dc resistivity has been observed near optimal doping in \bfca\ and \bfra\ in Refs.~\onlinecite{PhysRevB.80.140508,PhysRevB.81.224503}, respectively. Despite an overall very similar phase diagram of the hole-doped \bkfa\, the temperature dependence of its dc resistivity at optimal doping was found to display a more complex behavior: quadratic at low temperatures with an inflection point and a trend towards saturation at higher temperatures~\cite{PhysRevLett.101.107006,PhysRevB.78.014507,0953-2048-21-12-125014}, thus seemingly inconsistent with quasiparticle scattering dominated by antiferromagnetic spin fluctuations. However, similarly to the decomposition of the infrared conductivity of the 122 iron-based compounds into two distinct contributions~\cite{PhysRevB.81.100512} discussed in the previous section, it has been shown that this unusual temperature dependence of the dc resistivity can be described within the Eliashberg theory of normal metals by considering two electronic subsystems with very different elastic scattering rates and electron-boson interactions, one of which dominates the charge transport at low temperatures, while the other one --- at high temperatures, giving rise to an inflection point in the temperature dependence~\cite{Golubov_BKFA_2011}. A subsequent very detailed spectroscopic investigation of the temperature dependence of the infrared conductivity in \bkfaopt\ carried out in Ref.~\onlinecite{PhysRevLett.111.117001} uncovered, using an analogous two-component decomposition, that the quasiparticle scattering rate of the broad component is virtually temperature independent, while that of the narrow, coherent, contribution exhibits a linear temperature dependence, which, in turn, translates into the linear temperature dependence of the partial dc resistivity associated with this electronic subsystem. Thus it has been established that the temperature dependence of the dc resistivity of the coherent itinerant response in optimally doped \bkfaopt\ is, in fact, consistent with quasiparticle scattering dominated by marginal-Fermi-liquid spin fluctuations but is masked by multiband effects in raw transport data. Eliminating the dc value of the temperature-independent incoherent term obtained from the two-component analysis of the infrared-conductivity spectra one may extract the doping dependence of the quantum-critical $T$-linear and the conventional Fermi-liquid $T^2$ contributions to the dc resistivity based on comprehensive transport data. This analysis has been carried out in Ref.~\onlinecite{2013arXiv1308.6133N} on a representative set of electron-doped \bfca, isovalently substituted \bfap, and hole-doped \bkfa\ based on the same parent compound, \bfa, and is juxtaposed with the phase diagrams of these materials in Fig.~\ref{fig:tlinearcomp}. A recent magneto-transport investigation of the parent \efa\ compound under high pressure~\cite{PhysRevB.88.224510} has revealed the presence of a similar $T$-linear dependence of the in-plane dc resistivity and the recovery of the conventional quadratic temperature dependence upon application of an external magnetic field, which strongly suggests that the linear temperature dependence in the nearly optimally doped superconducting compounds indeed originates in strong coupling to spin fluctuations.

\begin{figure}[t!]
\includegraphics[width=\columnwidth]{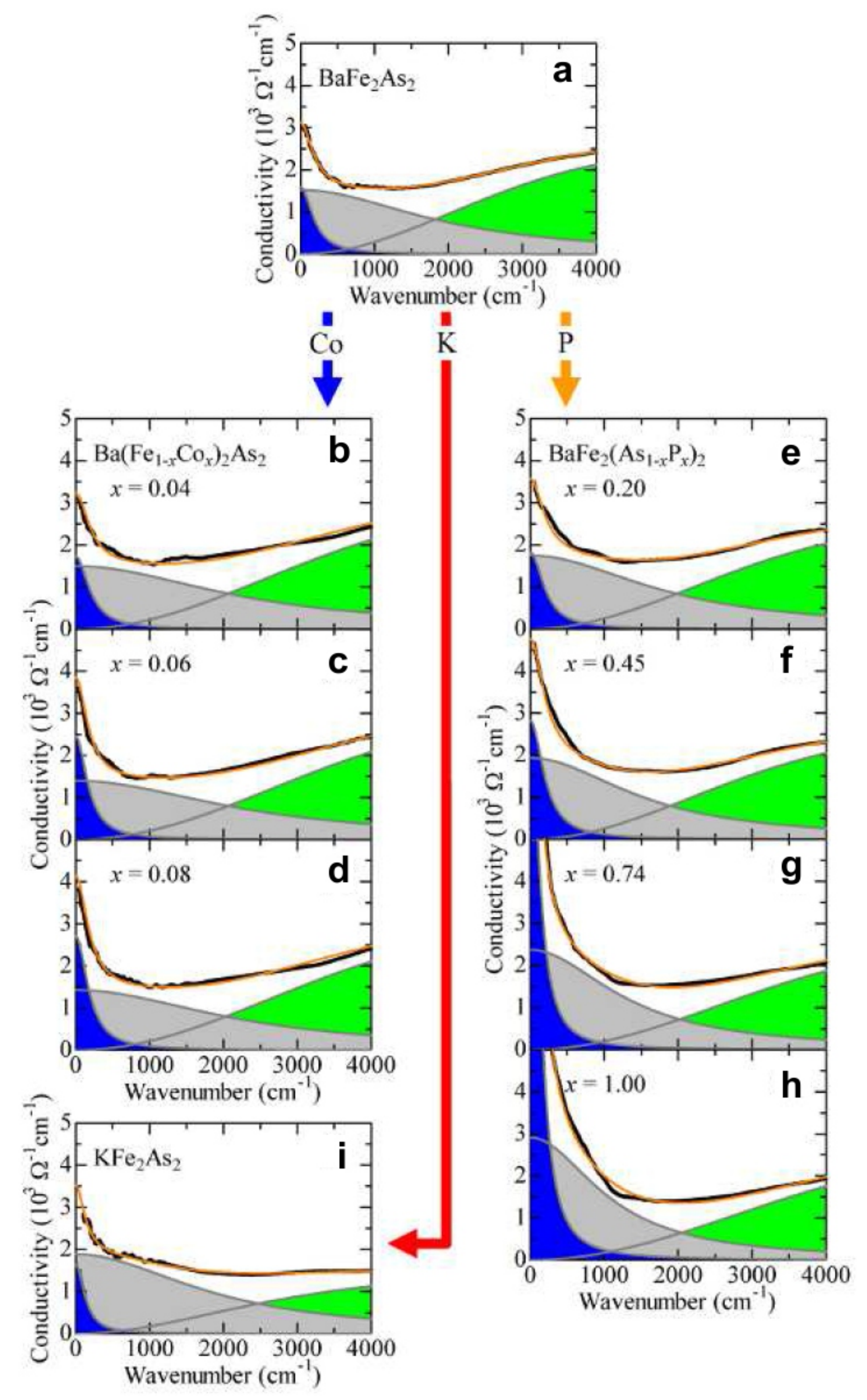}
\caption{\label{fig:dopingdep122} Doping evolution of the room-temperature optical conductivity and its decomposition into a broad and a narrow charge-carrier response for \textbf{a}~\bfa, \textbf{b}--\textbf{d}~\bfca, \textbf{e}--\textbf{h}~\bfap, and \textbf{i}~$\textrm{KFe}_2\textrm{As}_2$. Figure reprinted with permission from Ref.~\onlinecite{2013arXiv1308.6133N}. Copyright (2013) by the American Physical Society.}
\end{figure}

While the occurrence of the linear temperature dependence of the in-plane dc resistivity near optimal doping has been reliably established in the majority of iron-based superconductors and thus strongly suggests the universal relevance of underlying quasiparticle interaction for the mechanism of superconductivity, the possible existence and importance of the antiferromagnetic quantum critical point underneath the superconducting dome near optimal doping is under debate. As discussed above, in the case of \bfap\ significant experimental evidence has been furnished by means of various experimental techniques in favor of the existence of the quantum critical point, manifested in the enhancement of the quasiparticle effective mass upon approaching the optimal doping, as well as of a funnel-shaped region of the phase diagram with the linear temperature dependence of the dc resistivity. Similarly, magnetoelastic studies of \bfca\ across its phase diagram have revealed the divergence of the nematic susceptibility near the optimal doping and thus strongly suggest the existence of a nematic quantum critical point at optimal doping~\cite{Fisher_BFCA_Nematicity_2012}. Nuclear-magnetic-resonance investigations on the same material further corroborate the existence of a quantum critical point~\cite{PhysRevLett.104.037001,PhysRevB.87.174507}. On the other hand, the aforementioned investigation of the magnetotransport in the parent \efa\ compound under pressure reported the absence of any detectable enhancement of the quasiparticle effective mass across the optimal doping level~\cite{PhysRevB.88.224510}. Finally, the phase diagram of the 1111-type iron-based materials shows a rich variety of qualitatively different phase diagrams. The \lfaof\ and \nfaof\ materials feature a sharp disappearance of antiferromagnetism at the border with the superconducting phase in a first-order phase transition~\cite{Luetkens_La1111_phasediagram_2009,Malavasi_Nd1111_phasediagram_2010}. The phase diagram of \cfaof\ was found to exhibit the complete continuous disappearance of antiferromagnetism with doping before the onset of superconductivity~\cite{Dai_TcvsPnictogenBondAngle}, while that of the Sm-based compound --- to possess a large region of coexistence of superconductivity and antiferromagnetism, with strong spin fluctuations extending to the very overdoped region~\cite{Drew_Bernhard_coexistence_Sm1111_2009}, thus providing evidence for both the existence of a quantum critical point underneath the superconducting dome (see also Ref.~\onlinecite{PhysRevLett.101.087001}) and the importance of spin fluctuations for superconductivity. A recent nuclear-magnetic-resonance investigation of the phase diagram of isovalently substituted $\textrm{LaFeAs}_{1-x}\textrm{P}_x\textrm{O}$ provided another example of an 1111-type material in which quantum criticality seems to bear upon superconductivity~\cite{Kitagawa_PressureNMR_LaFeAsPO_2014}. These contrasting data across different classes of iron-based superconductors indicate that quantum criticality in these compounds is material-dependent and in some of them might not bear directly on the occurrence of superconductivity.

A recent systematic analysis of the optical conductivity spectra of several representative 122-type iron-based compounds across their respective phase diagrams in the framework of the two-component free-charge-carrier response, shown in Fig.~\ref{fig:dopingdep122}, has revealed several interesting trends~\cite{2013arXiv1308.6133N}. First of all, it has discovered the presence of a relatively weak dependence of the broad incoherent contribution on doping. Secondly, and more importantly, it found that superconductivity in these materials only occurs when the spectral weight of the narrow Drude contribution to the optical response is relatively small (the so-called ``bad-metal'' behavior) and fades away as the latter becomes much larger and the material --- more metallic. Based on the comparison of their results with the existing experimental data the authors conclude that the decrease of the spectral weight of the narrow component in the superconducting compounds most likely originates in the corresponding enhancement of the quasiparticle mass renormalization and thus strengthening of electronic correlations.

Strong coupling to spin fluctuations strongly suggested by the observation of the $T$-linear temperature dependence of the in-plane dc resistivity discussed above implies a sizable amount of inelastic scattering at the characteristic energies of these excitations in the normal state. It is tempting to associate the optical-conductivity contribution resulting from such scattering with the broad incoherent electronic background observed in all investigated iron-based superconductors, particularly in view of the unphysically small values of the quasiparticle mean free path obtained under the assumption of this electronic background entirely originating in elastic scattering, as discussed in the previous section. On the other hand, the broad component's origin in the inelastic scattering due to spin fluctuations seems to be at odds with the rather weak  temperature and doping dependence of its scattering rate~\cite{PhysRevLett.111.117001,PhysRevB.81.104528,PhysRevB.88.094501,2013arXiv1308.6133N,PhysRevB.88.094501,2013arXiv1308.6133N}, compared to the corresponding temperature dependence of the optical conductivity predicted in the framework of the marginal-Fermi-liquid theory~\cite{PhysRevB.44.7741}. This issue requires further systematic experimental investigation. 

\section{\label{sec:selenides}Iron-selenide superconductors}

The early successful growth of large high-quality single crystals of the 122-type iron-arsenide superconductors and their extensive investigation by a variety of experimental techniques stimulated significant interest in other compounds that could be obtained by substitution of either iron or arsenic while preserving the characteristic Fe-As layered structure~\cite{Johnston_Review_2010}. Late in 2010 the condensed-matter community got stirred up once again by the discovery of Fe-Se--based superconducting materials \afs\ with relatively high transition temperatures of about $32\ \textrm{K}$~\cite{PhysRevB.82.180520,PhysRevB.83.212502,APL10.10631.3549702,PhysRevB.83.060512}. They were first believed to crystallize in the same I4/mmm symmetry of $\textrm{ThCr}_2\textrm{Si}_2$ type as their iron-arsenide predecessors but soon it became clear that there is an inherent iron-deficiency order present in these materials with a chiral $\sqrt{5}\times\sqrt{5}\times1$ superstructure, which reduces the symmetry to I4/m and makes it more appropriate to classify these materials into the 245 stoichiometry~\cite{C1SC00070E}. The Fe-defect and antiferromagnetic orders occur at rather high transition temperatures of $400-550\ \textrm{K}$. Neutron-scattering studies showed that these compounds possess a magnetic moment on iron atoms of about $3.3\ \mu_{\textrm{B}}$~\cite{Bao_KFS_2011}, which is unusually large for iron pnictides. At the same time a resonance peak has been observed by the inelastic neutron scattering below $T_{\mathrm{c}}\approx32~\textrm{K}$ at an energy of $\hbar\omega_{\mathrm{res}}=14\ \textrm{meV}$ and the $\mathbf{Q}$-vector $(0.5,0.25,0.5)$ in the unfolded Fe-sublattice notation~\cite{PhysRevLett.107.177005,PhysRevB.86.094528,0295-5075-99-6-67004}, which is also unprecedented for the iron pnictides.

\begin{figure}[t!]
\includegraphics[width=\columnwidth]{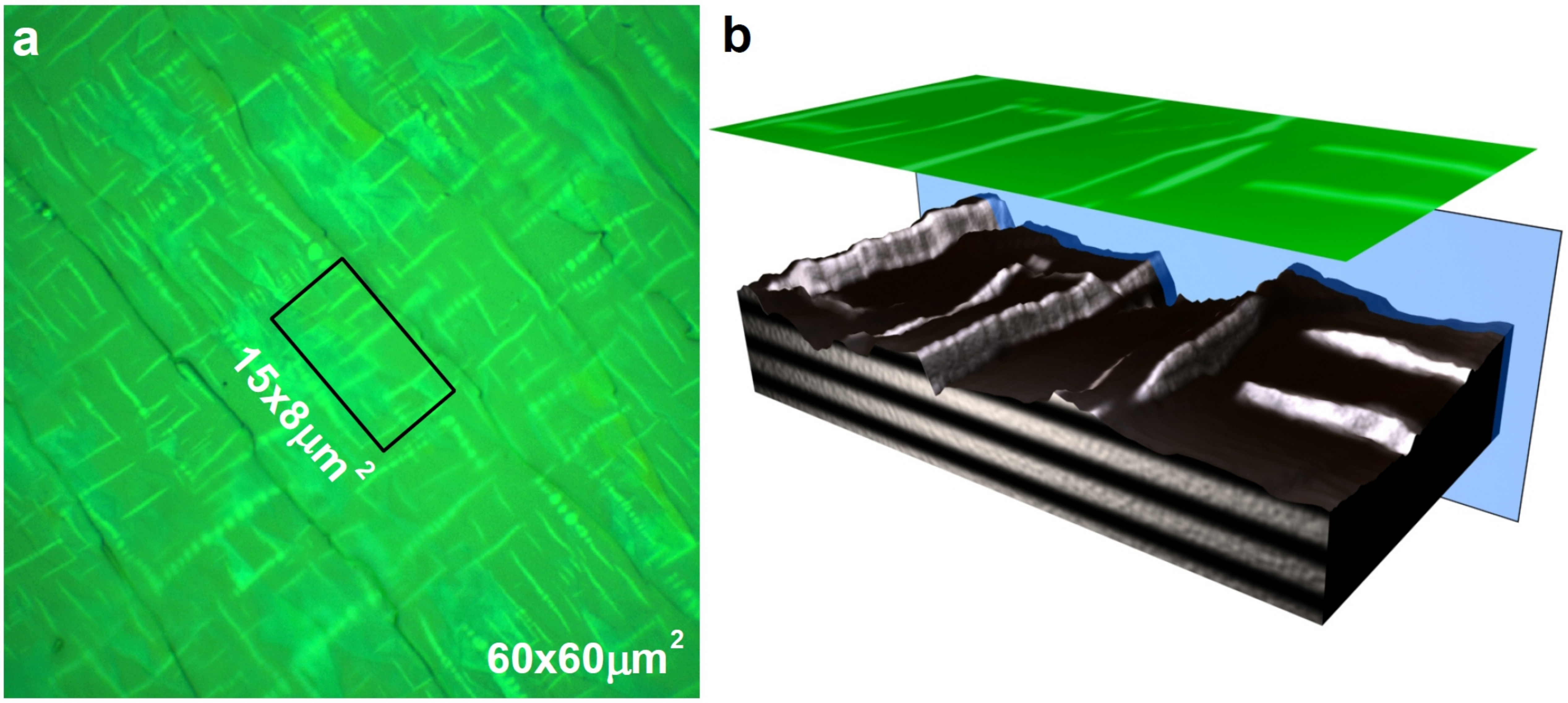}
\caption{\label{fig:phaseseparation}~\textbf{a}~Microscope image of a $60\times60\ \mu\textrm{m}^2$ surface patch of a freshly cleaved superconducting \rfs\ single crystal. Typical rectangular $15\times8\ \mu\textrm{m}^2$ area studied via near-field microscopy.~\textbf{b} Superposition of the topography of a $15\times8\ \mu\textrm{m}^2$ rectangular area (terrain) and the optical signal (brightness) normalized to that of silicon. Glossy areas indicate high silicon-\rfs\ contrast and thus metallicity, while the matt areas are semiconducting. Figure reprinted with permission from Ref.~\onlinecite{PhysRevLett.109.017003}. Copyright (2012) by the American Physical Society.}
\end{figure}

For a short time it remained under debate how superconductivity with such a high transition temperature could survive on such a strong magnetic background. Gradually, experimental evidence indicating an inherent phase separation in these iron-chalcogenide materials started to appear and as of today the phase separation of the superconducting and antiferromagnetic phases has been observed by or found consistent with essentially all experimental probes.~\cite{PhysRevLett.107.137003,PhysRevB.83.140505,PhysRevB.84.060511,PhysRevB.84.180508,KFS_MBE_thinfilm_NatPhys_2011,PhysRevLett.108.237002,PhysRevLett.109.017003,PhysRevLett.109.057003,PhysRevLett.109.077003,PhysRevB.85.100501,PhysRevB.85.100504,PhysRevB.85.140511,PhysRevB.85.180510,PhysRevB.85.214503,PhysRevB.85.214519,PhysRevB.85.224510,PhysRevB.85.224515,PhysRevB.86.054503,PhysRevB.86.134502,PhysRevB.86.134530,PhysRevB.86.144530,PhysRevB.86.174107,PhysRevB.86.184511,PhysRevB.86.224502,PhysRevX.1.021020,SciRep_Wang2012,0295-5075-99-6-67004,0034-4885-75-11-112501,0953-8984-24-45-455702,PhysRevLett.110.137003,PhysRevB.87.100501,PhysRevB.87.094508,PhysRevB.88.134501}. The superconducting phase has been found to be a minority phase with a fraction of $12$--$20\%$~\cite{PhysRevB.84.180508,PhysRevB.85.100501,PhysRevB.85.214503,PhysRevB.85.214519,PhysRevLett.109.017003,RevModPhys.85.849}. Importantly, this fraction is above the percolation limit~\cite{0295-5075-96-3-37010} as $100\%$ superconducting shielding fraction and zero dc resistivity have been clearly observed~\cite{PhysRevB.84.144520}. Based on the available experimental evidence the geometry of the phase separation illustrated in Fig.~\ref{fig:phaseseparation} appears the most likely. Figure~\ref{fig:phaseseparation}\textbf{a} shows a microscope image of a freshly cleaved surface of the \rfs\ compound exhibiting a pattern of bright stripes at an angle of $45^\circ$ with respect to the in-plane crystallographic axes. Such a percolating network has been widely observed in \afsgen\ compounds~\cite{PhysRevB.83.104526,PhysRevLett.109.017003,PhysRevB.86.144507,0953-8984-25-27-275701} and shown to be very sensitive to heat treatment and result from the superconducting phase precipitating during the formation of the iron-vacancy--ordered phase~\cite{PhysRevB.86.144507}. No experimental route has so far been found to stabilize the stoichiometric 122 phase. Consistent with the precipitation-driven formation of the stable phase of a percolating superconducting network, the superconducting transition temperature of $\approx32\ \textrm{K}$ has been found essentially independent of the {\it nominal} material composition in the entire region of the phase diagram in which superconductivity occurs~\cite{PhysRevB.84.144520}. Furthermore, virtually all electronic properties of these materials have been found to be independent of the intercalating atom~\cite{RevModPhys.85.849}.

Investigation of the \kfs\ and \rfs\ compounds by means of transmission electron~\cite{PhysRevB.83.140505} and near-field microscopy~\cite{PhysRevLett.109.017003}, respectively, has revealed an unprecedented nanoscale phase separation in \afsgen\ materials in the out-of-plane direction, whereby the ``layers'' of the superconducting phase only several nanometers thick are embedded into a matrix of semiconducting $A_2\textrm{Fe}_4\textrm{Se}_5$ ($A=$K, Rb), as shown schematically in Fig.~\ref{fig:phaseseparation}\textbf{b}. A complementary detailed low-energy muon-spin--rotation study has further shown that both the magnetic properties of the semiconducting matrix and its overall volume fraction strongly depend on the distance from the sample surface~\cite{PhysRevLett.109.017003}.

While the exact composition of the semiconducting iron-vacancy--ordered matrix in the phase-separated \afsgen\ compounds was reliably determined to be of the 245 type shortly after their discovery~\cite{NatPhys_Dai_Dagotto_ReviewMagnPnictides,RevModPhys.85.849}, that of the superconducting phase long eluded quantification. Suggestions ranged from a vacancy-disordered 245 phase to the stoichiometric 122 phase, similar to the iron-arsenide superconductors. It is worth noting, however, that due to the different valency of selenium the stoichiometric 122 phase of the \afsgen\ compounds would be heavily electron-overdoped, as compared to the doping levels of the superconducting iron-pnictides, obtained by moderate substitution in compensated semimetalic parents~\cite{Johnston_Review_2010}. Based on the respective volume fractions of the superconducting and antiferromagnetic phase, detailed nuclear-magnetic-resonance investigation of the superconducting \rfs\ compound succeeded in pinpointing the elusive chemical composition of the superconducting phase to iron-vacancy--free $\textrm{Rb}_{0.3}\textrm{Fe}_2\textrm{Se}_2$ and found general agreement with the nanoscale phase-separation picture discussed above.

\begin{figure}[t!]
\includegraphics[width=\columnwidth]{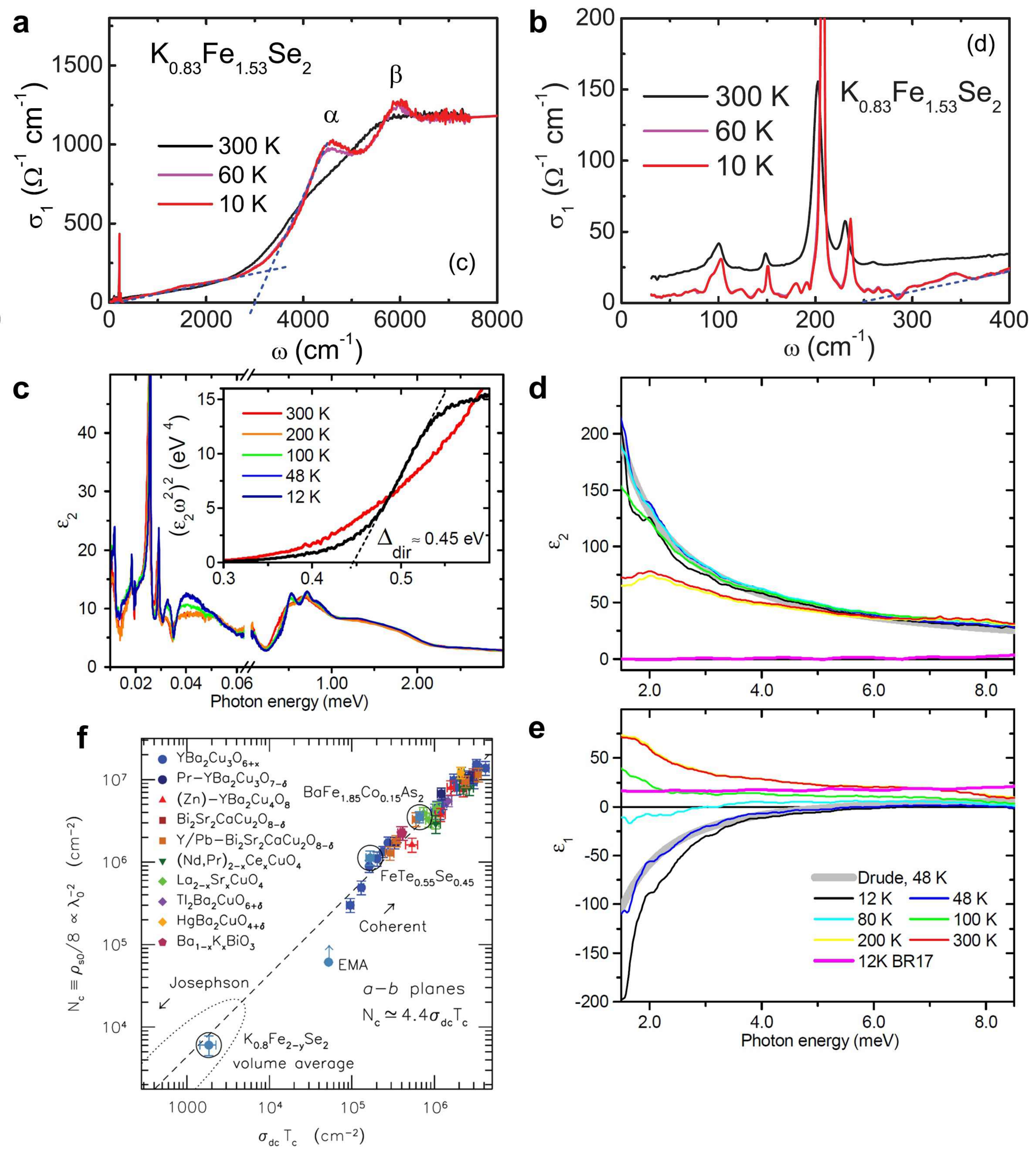}
\caption{\label{fig:optics245} Real part of the optical conductivity of $\textrm{K}_{0.83}\textrm{Fe}_{1.53}\textrm{Se}_2$ in the \textbf{a}~middle- and \textbf{b}~far-infrared spectral range.~\textbf{c} Real part of the optical conductivity of superconducting \rfs . Inset: Plot of $(\varepsilon_2(\omega)\omega^2)^2$ near the absorption edge. The intersection of the dashed line with the energy axis defines the direct energy gap $\Delta_{\mathrm{dir}}=0.45\ \textrm{eV}$ at $12\ \textrm{K}$.~\textbf{d},\textbf{e} Imaginary and real part of the dielectric function of the same \rfs\ material in the terahertz spectral range.~\textbf{f} Log-log plot of the spectral weight of the superfluid density $N_{\mathrm{c}}\equiv\rho_{s0}/8$ vs $\sigma_{\mathrm{dc}}T_{\mathrm{c}}$ in the a-b planes for a variety of cuprate superconductors as well as several iron-based superconductors compared with the volume-average and EMA results for $\textrm{K}_{0.8}\textrm{Fe}_{2-y}\textrm{Se}_2$. The dashed line corresponds to the general result for the cuprates $\rho_{s0}/8\simeq4.4\sigma_{\mathrm{dc}}T_{\mathrm{c}}$, whereas, the dotted line denotes the region of the scaling relation typically associated with Josephson coupling along the c axis. Although the volume-average result signaled a Josephson phase, the EMA result now lies very close to the coherent regime. Panels~\textbf{a},~\textbf{b}; \textbf{c}--\textbf{e}; and~\textbf{f} as well as the caption of~\textbf{f} reprinted with permission from Refs.~\onlinecite{PhysRevB.83.220507},~\onlinecite{PhysRevB.85.100504},~\onlinecite{PhysRevB.86.144530}, respectively. Copyright (2011, 2012) by the American Physical Society.}
\end{figure}

Optical studies of the \afsgen\ compounds began with the investigation of the purely semiconducting iron-vacancy--ordered 245 phase in $\textrm{K}_{0.83}\textrm{Fe}_{1.53}\textrm{Se}_2$, as shown in Fig.~\ref{fig:optics245}\textbf{a},\textbf{b}~\cite{PhysRevB.83.220507}. In the lowest infrared spectral range the optical conductivity manifests a very low level of electronic background, consistent with the semiconducting character of this material. It also features numerous infrared-active phonon modes (as compared to only two infrared-active phonon modes allowed by the tetragonal symmetry of their iron-arsenide counterparts), which result from the symmetry lowering due to the formation of the $\sqrt{5}\times\sqrt{5}\times1$ superstructure at the iron-vacancy ordering temperature~\cite{C1SC00070E}. In the mid-infrared spectral range the optical conductivity of this material exhibits a clear direct band gap of about $0.37\ \textrm{eV}$, comparable with the prediction of the {\it ab initio} calculations of its electronic structure in the antiferromagnetic checkerboard ground state~\cite{PhysRevB.83.233205}. Above the direct band gap and at low temperatures the optical-conductivity spectrum features two distinct absorption bands ($\alpha$ and $\beta$ in Fig.~\ref{fig:optics245}\textbf{b}), as compared to the single absorption band previously observed in all iron-arsenide compounds, as discussed in the previous section. Therefore, this splitting must likewise be associated with the antiferromaganetic checkerboard ground state in this system.

Due to the large volume fraction of the antiferromagnetic semiconducting phase in the superconducting \afs\ compounds ($80$--$88\%$)~\cite{RevModPhys.85.849}, their optical conductivity is dominated by the features related to the 245 semiconducting phase discussed above down to energies as low as $8\ \textrm{meV}$, as demonstrated for the case of \rfs\ in Fig.~\ref{fig:optics245}\textbf{c}. Similarly to the semiconducting case, it shows a large number of infrared-active phonons as well as a strongly temperature-dependent absorption band at about $300\ \textrm{cm}^{-1}$, clearly absent in the semiconducting compound. The mid-infrared conductivity of \rfs\ further reveals a somewhat larger direct band gap of about $0.45\ \textrm{meV}$ than that of the K-based semiconductor. Quite interestingly, instead of two absorption bands just above the absorption edge, the optical conductivity of \rfs\ exhibits three distinct peaks at low temperatures. The temperature dependence of these peaks is very similar to that of the magnetic Bragg peak, strongly suggesting a spin-controlled character of these interband transitions~\cite{PhysRevB.85.100504} and further confirming the close relation of the doublet/triplet feature to the antiferromagnetic order in this material.

The predominantly semiconducting character of the optical conductivity of the superconducting compound down to $8\ \textrm{meV}$ made it possible to investigate the charge dynamics of these materials at terahertz frequencies in a transmission geometry on sufficiently thin slabs of $\approx12\ \mu\textrm{m}$~\cite{PhysRevB.85.100504} --- a study impossible in any good metal due to their near-unity reflectivity well below the plasma frequency. The results of such an investigation using terahertz time-domain spectroscopy are shown in Figs.~\ref{fig:optics245}\textbf{d}--\textbf{e}. The simultaneous determination of both the real and imaginary parts of the dielectric function $\varepsilon(\omega)=\varepsilon_1(\omega)+i\varepsilon_2(\omega)$ using this phase-sensitive technique enabled the spectroscopic observation of the semiconductor-metal crossover apparent in the temperature dependence of the dc resistivity of all \afsgen\ materials~\cite{PhysRevB.82.180520,PhysRevB.83.060512,PhysRevB.83.212502,0295-5075-94-2-27009,PhysRevB.84.144520}. Figures~\ref{fig:optics245}\textbf{d}--\textbf{e} demonstrate that although the level of the electronic background in $\varepsilon_2$ is sizable even at room temperature, $\varepsilon_1(\omega)$ remains positive down to lowest achievable energies of about $1\ \textrm{meV}$, which implies a semiconducting character of the charge dynamics. As the temperature is lowered, however, $\varepsilon_1(\omega)$ gradually decreases until it crosses zero between $100$ and $80\ \textrm{K}$, indicative of a crossover into a metallic state with a finite plasma frequency. It is important to note that this zero crossing develops at temperatures much larger than the superconducting transition temperature of $32\ \textrm{K}$ and is thus unrelated to superconductivity. The characteristic temperature of the semiconductor-metal crossover in the dc resistivity and the initial slope of the latter have been found to be correlated with the connectivity of the superconducting domains in the \afs\ single crystals, strongly affected by the annealing temperature and duration used in their post-processing~\cite{2014arXiv1401.6906S,PhysRevB.86.144507}.

Even at the lowest temperatures the zero crossing in $\varepsilon_1(\omega)$, which corresponds to the screened plasma frequency, occurs at energies on the order of $6.5\ \textrm{meV}$, as can be seen in Fig.~\ref{fig:optics245}\textbf{e}, --- much lower than the corresponding values in the single-phase iron-based superconductors~\cite{Hu2009545,Johnston_Review_2010}. The value of the unscreened plasma frequency in \rfs\ was found to be approximately $100\ \textrm{meV}$, also much smaller than the corresponding values in the single-phase compounds. These observations clearly indicate that the optical response of the \afsgen\ materials represents an effective-medium response of the superconducting phase embedded into a semiconducting matrix and must be so modeled in order to extract the inherent characteristics of the superconductor. Such analysis has indeed been carried out in Refs.~\onlinecite{PhysRevB.85.214503,PhysRevB.86.144530} and revealed that the extracted inherent optical response of the superconducting phase is, in fact, quite close to that of the single-phase iron-based superconductors, as shown by the scaling plot in Fig.~\ref{fig:optics245}\textbf{f}. The effective-medium model in Ref.~\onlinecite{PhysRevB.85.214503} could further reproduce the absorption band at $300\ \textrm{cm}^{-1}$, mentioned above, as well as its temperature dependence. Application of the effective-medium approach to the optical conductivity of both \rfs\ and \kfs\ superconducting compounds in Refs.~\onlinecite{PhysRevLett.109.017003,PhysRevB.86.144530} lead to an estimate for the London penetration depth of the superconducting condensate on the order of $2\ \mu\textrm{m}$, much smaller than a direct estimate from the raw data without taking into account the phase separation in these materials but still about $8$ times larger than the value of $\approx250\ \textrm{nm}$ obtained by means of low-energy and conventional muon-spin--rotation studies on the same compounds~\cite{PhysRevLett.109.017003,PhysRevB.85.214503}. The effective-medium models applied to the experimental data in Ref.~\onlinecite{PhysRevB.85.214503} extracted a value of the London penetration depth of $\approx400\ \textrm{nm}$, even closer to the muon-spin--rotation estimate. The above discussion implies that both the normal and the superconducting properties of the \afsgen\ superconductors are, in fact, quite close to those of all other iron-based superconductors once the phase separation in these materials is carefully taken into account.

The similarity of many of the properties of the \afsgen\ compounds to those of their iron-arsenide and iron-selenide counterparts notwithstanding, there are several features of interest that do distinguish this new class of materials. One of such features is the symmetry of the superconducting order parameter. As already mentioned above, extensive inelastic-neutron-scattering studies carried out on the \afs\ compounds revealed the existence of a resonance peak below $T_{\mathrm{c}}\approx32~\textrm{K}$ at the $\mathbf{Q}$-vector $(0.5,0.25,0.5)$ in the magnetic Brillouin zone (one-iron unit cell~\cite{Johnston_Review_2010})~\cite{PhysRevLett.107.177005,PhysRevB.86.094528,0295-5075-99-6-67004}, which requires a sign change in the superconducting order parameter on the portions of the Fermi surface connected by this $\mathbf{Q}$-vector. Based on the results of angle-resolved photoemission spectroscopy measurements~\cite{PhysRevB.85.220504} and theoretical calculations~\cite{PhysRevB.83.100515}, the $\mathbf{Q}$-vector of the neutron resonance was found compatible with momentum transfer between two Fermi surface of the same electronic character located at the M point of the magnetic Brillouin zone, which led to a natural conjecture of a nodeless $d$-wave pairing symmetry of the superconducting order parameter in this {\it unfolded} Brillouin zone~\cite{PhysRevB.83.100515}. However, when the transition to the real two-iron unit cell is carried out, the Fermi surfaces hosting superconducting order parameters of different sign overlap and hybridize, giving rise to nodes (zeros) in the momentum-dependent superconducting energy gap around the resulting sheets of the Fermi surface~\cite{PhysRevB.83.140512}. From the same argument it immediately follows that the small electronic sheet of the Fermi surface observed around the $Z$ point of the Brillouin zone~\cite{PhysRevB.85.220504,PhysRevLett.109.037003} must, too, exhibit zeros in the superconducting energy gap in the nodal directions. This scenario is, however, at odds with the aforementioned angle-resolved photoemission studies, which have reported a nodeless superconducting gap on all observed sheets of the Fermi surface~\cite{PhysRevB.85.220504,NatMatZhangFengARPESnohole2011}. A recent theoretical proposal appears to have reconciled these disparate experimental results by demonstrating that a nodeless extended $s$-wave (sing-changing) pairing symmetry in which the superconducting order parameter changes its sign between the hybridized electron pockets in the corners of the real Brillouin zone does supports a spin resonance at the experimentally observed $\mathbf{Q}$-vector~\cite{PhysRevB.88.224505}.

It has further been argued that the antiferromagnetic checkerboard ground state of the semiconducting iron-selenides has a Mott-insulator character~\cite{PhysRevLett.106.186401}. It then appears to be of interest how the crossover from this Mott-insulating state to the metallic superconducting state occurs and what interactions are responsible for it. Ref.~\onlinecite{0953-8984-24-38-385702} has explored theoretically the interplay between antiferromagnetism and superconductivity in the iron-vacancy--ordered configuration and revealed a complex phase diagram with both microscopically coexisting and phase-separated blocked-checkerboard antiferromagnetic and superconducting phases at different electron doping levels. In another theoretical work it has been suggested that for intermediate values of the Hund's coupling this crossover occurs via an orbital-selective Mott phase, in which the iron $3d_{xy}$ orbital is Mott localized, while the other $3d$ orbitals remain itinerant~\cite{PhysRevLett.110.146402}. Intriguingly, precisely such an orbital-selective localization has indeed bee observed experimentally in both \kfs\ and \rfs\ as a function of temperature~\cite{PhysRevLett.110.067003}. This orbital selectivity seems to be supported by the recent terahertz-transmission measurements in superconducting \rfs~\cite{2013arXiv1309.6084W}. The aforementioned theoretical predictions and experimental observations together with the prediction and observation of orbital-selective quasiparticle mass renormalization and band shifts with respect to the results of the density-functional theory~\cite{PhysRevLett.109.177001,PhysRevLett.110.067003,2013arXiv1307.1280M} further support the importance of the Hund's-coupling correlations~\cite{PhysRevB.87.045122} for superconductivity in the iron-based superconductors.

Finally, a recent theoretical investigation of the interplay between superconductivity and antiferromagnetism in phase-separated iron selenides suggested a long-sought rationale for maximizing the superconducting transition temperature in the iron-based superconductors: it has identified, based on a two-layer model, that the transition temperature in these phase-separated materials could be suppressed by interlayer hopping~\cite{PhysRevB.85.104506} between the superconducting and antiferromagnetic phase at the interface and minimization of this hopping could lead to superconducting transition temperatures on the order of $65\ \textrm{K}$~\cite{PhysRevB.88.174510}. In this regard it appears remarkable that in the class of binary iron selenides, the simplest iron-based superconductors, a significant increase in the transition temperature to $\approx45\ \textrm{K}$ with respect to that of bulk FeSe ($\approx8\ \textrm{K}$) has been demonstrated by intercalating the iron-selenium layers with large molecular complexes~\cite{MolecularIntercalated_FeSe_2013} and to $65\ \textrm{K}$ by isolating a single monolayer of FeSe on a $\textrm{SrTiO}_3$ substrate~\cite{MonolayerFeSe_65K_2013}, in which the interlayer hopping is either minimized or, in the latter case, completely absent.

\section{\label{sec:conclusions}Conclusions}

In summary, we have reviewed the developments in the field of high-temperature unconventional iron-based superconductivity since their discovery in 2006, largely through the prism of itinerant-charge-carrier dynamics investigated by means of optical spectroscopy. In the parent compounds, the effect of the antiferromagnetic phase transition on the optical conductivity has been clearly demonstrated by the observed redistribution of the optical spectral weight from low to high frequencies in the vicinity of the corresponding characteristic energies (optical energy gaps) due to the modification of the electronic structure at the phase transition. Some evidence for the inherently dual, simultaneously itinerant and local, character of antiferromagnetism in the iron-based compounds has been derived from the analysis of the infrared conductivity across the transition. Detailed investigation of the temperature dependence of the infrared conductivity of various parent iron pnictides has demonstrated the evolution of coupling between the low-energy and intermediate-energy electronic subsystems affected by the antiferromagnetic phase transition from weak in \cfa\ via intermediate in \sfa\ to strong in \bfa, systematically with the atomic number of the intercalant. All infrared-active modes expected in the high-temperature tetragonal phase have been observed in all compounds along with their modification at the concomitant structural transition, with some evidence for sizable Hund's-coupling correlations derived from the anomalous line shape of some of the phonons.

The dispersion analysis of iron-based materials has revealed that the overall structure of the interband absorption bands in the optical conductivity at energies above $1\ \textrm{eV}$ can be well reproduced by standard {\it ab initio} density-functional calculations in the local-density approximation. We have emphasized that the interband transitions in the iron-based compounds make an unusually large contribution on the order of $60$ to the zero-frequency permittivity and, therefore, must be explicitly taken into account lest significant artifacts arise in the analysis of the itinerant-charge-carrier response by means of the extended-Drude theory. Likewise, the multiband character of the iron-based superconductors must be considered in order to account for the frequency dependence of the quasiparticle scattering rate and mass renormalization due to the contribution of multiple bands to the optical conductivity before other sources of this frequency dependence such as the interaction with bosonic excitations are involved. In the infrared spectral range, the optical conductivity of the iron-based compounds has been found to exhibit two well-discernible contributions according to the regime of quasiparticle coherence distinguished based on the strength of the quasiparticle scattering rate: a broad incoherent and a narrow coherent term. This observation is consistent with the established inherently multiband character of the iron-based materials, with several (up to five) sheets of the Fermi surface of both electron and hole character contributing to the itinerant dynamics. The general shape of the infrared conductivity both in the normal and in the antiferromagnetic state has been demonstrated to be well-captured by dynamical-mean-field-theory theoretical calculations explicitly taking into account sizable Hund's-coupling correlations between quasiparticles. The occurrence of superconductivity has been shown to correlate with the degree of incoherence of the narrow contribution to the itinerant-charge-carrier response and the latter --- with the pnictogen bond angle and electron filling, which provide a measure of the degree of electronic correlations.

Parent compounds in the 122 class of the iron-based materials have further revealed the presence of pronounced electronic nematicity in the nominally tetragonal state above the antiferromagnetic and structural phase transition. The corresponding anisotropy, originally observed in the dc transport, has now been reliably established by means of many different techniques, including investigations of the optical conductivity, and traced back to the orbital anisotropy in the (stabilized by detwinning pressure or strain) orthorhombic state and corresponding large nematic/orbital susceptibility in the tetragonal state close to the phase transition. The existence of a nematic instability due to spin/orbital fluctuations and their strong coupling to the lattice demonstrate the intimate connection between these three fundamental degrees of freedom in the iron-based compounds.

In the superconducting compounds, the hallmark of superconductivity --- the missing area in the infrared conductivity --- has been observed in all investigated materials and has been shown to be consistent with a nodeless superconducting energy gap(s), supporting the established view of the extended $s$-wave symmetry (fully gapped with a sign change between different sheets of the Fermi surface in the Brillouin zone) of the superconducting order parameter in most materials of this family of superconductors. Angle-resolved photoemission measurements have revealed an unusual clustering of the superconducting gaps into two groups with respect to their magnitude, widely observed with other techniques, and their strong dependence on the orbital character of the underlying electronic structure.

It has been demonstrated that, based on an extensive comparison of various conventional and unconventional high-temperature superconductors, the iron-based materials appear to span a wide range of the coupling regime from weak to strong coupling. The applicability of the widely used Mattis-Bardeen theory for the description of the optical conductivity to the case of iron-based superconductors beyond the weak-coupling regime has been shown to be questionable at best and in the case of strongly coupled superconductors in the clean limit with respect to impurity scattering --- completely inadequate. Analysis of the optical conductivity in the framework of the Eliashberg theory of superconductivity has shown that it is largely consistent with pairing driven by antiferromagnetic spin fluctuations. This premise has been strongly supported by the observation of the inherent linear temperature dependence of the dc resistivity in the coherent channel (narrow contribution to the itinerant-charge-carrier response), associated with strong coupling to antiferromagnetic spin fluctuations, as well as the recovery of the conventional Fermi-liquid $T^2$ dependence upon the application of a spin-stabilizing magnetic field. In some but not all iron-based superconductors a quantum critical point and the associated finite-temperature quantum-critical regime have been identified. The relation of quantum-critical fluctuations to superconducting pairing, however, remains unclear.

Intriguingly, the changes in the optical conductivity induced by both the antiferromagnetic and superconducting phase transition have been shown to affect energies much higher than the characteristic energies of either, most pronouncedly superconducting, phase. This effect can be explained quite naturally by a population redistribution between different electronic bands due to a band/orbital-specific shift of the chemical potential in the symmetry broken state, an effect that up to now has not been accounted for self-consistently by any of the widely employed theories of superconductivity.

The recently discovered addition to the family of iron-based superconductors, iron-selenide materials of nominally 122 type, have been found to exhibit quite high superconducting transition temperatures of $32\ \textrm{K}$ simultaneously with exceptionally strong antiferromagnetism with N\'eel temperatures of about $550\ \textrm{K}$. This peculiar coexistence has been demonstrated to occur in the form of unusual nanoscale phase separation between the superconducting and antiferromagnetic phase in the out-of-plane direction. The transport properties of these materials show a clear temperature-induced semiconductor-metal crossover, likely attributable to the observed orbital-selective Mott transition, at which one of the electronic orbitals gets localized while the rest remain itinerant. The latter observation further emphasizes the aforementioned importance of orbital physics for the electronic properties of the iron-based materials and superconductivity therein.

While several important questions in the physics of the iron-based superconductors remain open, such as the symmetry of the superconducting order parameter and the precise pairing mechanism, it is clear that a synergistic consideration of their multiband electronic structure, the dual, itinerant and local, character of antiferromagnetism in the parent compounds, as well as the complex intimate interplay between the spin, orbital, and lattice degrees of freedom is key to a complete understanding of these fascinating materials.

\section{Acknowledgements}

This work was supported by the German Science Foundation under grant BO 1912/2-2 within SPP 1458. We would like to thank A.~V. Boris, D.~V.~Efremov, D.~V.~Evtushinsky, and S. Haindl for useful comments and suggestions.


\end{document}